\documentclass[twocolumn,aps,prc,tightenlines,floats,floatfix,nofootinbib]{revtex4}

\def\etal{{\it et.\ al.\/},$\,$}

\def\bea{\begin{eqnarray}}
\def\eea{\end{eqnarray}}
\def\etal{{\it et al.}\,}

\def\sfrac#1#2{{\textstyle \frac{#1}{#2}}}

\def\be{\begin{equation}}
\def\ee{\end{equation}}
\def\ba{\begin{eqnarray}}
\def\ea{\end{eqnarray}}

\usepackage{graphics}
\usepackage{graphicx}
\usepackage{epsf} 
\usepackage{amsmath}
\usepackage{amssymb}

\begin{document} 

\phantom{0}
\vspace{-0.2in}
\hspace{5.5in}
\parbox{1.5in}{ \leftline{JLAB-THY-10-1139}}

\vspace{-1in}

\title
{\bf Electromagnetic form factors of the $\Delta$ with
D-waves}
\author{G. Ramalho$^{1,2}$, M.T. Pe\~na$^{2,3}$ 
and 
Franz Gross$^{1,4}$
\vspace{-0.1in}  }

\affiliation{
$^1$Thomas Jefferson National Accelerator Facility, Newport News, 
Virginia 23606, USA \vspace{-0.15in}}
\affiliation{
$^2$Centro de F{\'\i}sica Te\'orica de Part{\'\i}culas, 
Av.\ Rovisco Pais, 1049-001 Lisboa, Portugal \vspace{-0.15in}}
\affiliation{
$^3$Department of Physics, Instituto Superior T\'ecnico, 
Av.\ Rovisco Pais, 1049-001 Lisboa, Portugal \vspace{-0.15in}}
\affiliation{
$^4$College of William and Mary, Williamsburg, Virginia 23185, USA
}

\vspace{0.2in}
\date{\today}

\phantom{0}

\begin{abstract}
The electromagnetic form factors of the $\Delta$ baryon
are evaluated within the framework of a covariant spectator quark model, where
S and D-states are included in the $\Delta$ wave function.
We predict all the four  
$\Delta$ multipole form factors: the
electric charge $G_{E0}$, the magnetic dipole 
$G_{M1}$, the electric quadrupole $G_{E2}$ 
and the magnetic octupole $G_{M3}$.
We compare our predictions with other theoretical calculations.
Our results are compatible with 
the available 
experimental data 
and recent lattice QCD data.
\end{abstract}

\vspace*{0.9in}  
\maketitle

\section{Introduction}

In the history of hadronic physics 
the
nucleon was one of the first particles to have its
internal structure experimentally uncovered.
The nonpointlike character of the nucleon was suggested by
the measurements of the proton anomalous magnetic moment.
Around 1950,
measurements disclosed an exponential falloff for the
charge distribution, and three decades later the 
SLAC measurements showed that the
corresponding charge form factor in momentum space
coincided, almost perfectly, with the magnetic dipole form factor.
This image of the proton proved to be a good 
approximation for low $Q^2$, the square of the four-momentum transfer, and was 
only superseded by the results from the Jlab polarization 
measurement in 1999 
\cite{Perdrisat06}.

The study of the next baryonic state (the $\Delta$) has been more challenging.
Although the pure spin 3/2 and 
isospin 3/2 structure of the 
$\Delta$ was clearly isolated from the 
background of the 
pion nucleon scattering experimental cross sections, 
its theoretical description is not yet totally clear.
Quantum chromodynamics (QCD) 
emphasizes the three-quark state component, and historically
the existence of the $\Delta^{++}$ 
led to the introduction of color 
as a quantum number. Still, the issue on 
how much of the $\Delta$ 
structure comes from
a three-quark state, and how much is a molecularlike state 
of a nucleon and a pion, remains an open question.

Since the $\Delta$ is a spin $3/2$ particle, its electromagnetic structure
can be characterized 
by four form factors, namely
the electric charge 
$G_{E0}$, magnetic dipole $G_{M1}$, 
the electric quadrupole $G_{E2}$
and electric octupole $G_{M1}$ form factors.
The first two describe
the charge and magnetic dipole distributions, while 
the last two measure the deviation of the 
first ones from the symmetric form \cite{Buchmann08}.
At zero transferred momentum squared, $Q^2=0$, 
the form factors define the $\Delta$ multipole moments: 
the charge $e_\Delta$, the magnetic moment $\mu_\Delta$, 
the electric quadrupole moment $Q_\Delta$ 
and the magnetic octupole moment ${\cal O}_\Delta$. 
If the $\Delta$ would simply be a completely symmetric state, 
with no quark D-states relatively to 
the quark pair (diquark), both $Q_\Delta$ 
and ${\cal O}_\Delta$ would vanish \cite{DeltaFF}.
 
The unstable character of the $\Delta$ 
(with a mean lifetime of 5.6$\times 10^{-24} s$) makes
difficult to probe the electromagnetic 
properties of this particle/resonance.
Until recently the available information on
the $\Delta$ structure came from 
the study of  the $\gamma N \to \Delta$ reaction only.
The dominance of the magnetic dipole 
moment in that reaction asserts 
the spin flip of a quark in the nucleon as the 
main process, but the 
magnitudes of the quadrupole contributions  
($G_E^\ast$ and $G_C^\ast$) 
indicate a small deformation of the $\Delta$ \cite{NDeltaD}.

The experimental information on
the $\Delta$ is restricted to the 
$\Delta^{++}$ and $\Delta^{+}$ magnetic moments 
\cite{Yao06,Kotulla02,Castro01,SIN,UCLA,Exp,Kotulla07,Chiang05}.
Even so, those results are affected by  
considerable error bars, due to the unstable nature 
of the $\Delta$ and theoretical model 
uncertainties. This is 
particularly true for the  $\Delta^+$ magnetic moment \cite{Kotulla02}.
See Refs.~\cite{Castro01,Kotulla07} for a review.
New experiments are in progress at MAMI \cite{Kotulla07} 
to extract $\mu_{\Delta^+}$ more accurately with 
the help of reaction models based on
hadronic degrees of freedom \cite{Chiang05,Machavariani99,Pascalutsa05}.
The scarce experimental information on $\mu_\Delta$
is compensated by extensive theoretical studies
within various frameworks
\cite{Chiang05,Pascalutsa05,Hikasa92,Beg64,Chao74,Kim89,Dillon94,Jenkins94,Hong94,Luty95,Banerjee96,Linde98,Ha98,Franklin02,Hayne82,Giannini90,Schlumpf93,Butler94,Buchmann97,Lee98,Wagner00,Aliev00,Kerbikov00,Sahu02,An06,Kim,Machavariani08,He05,Hashimoto08,Dahiya03,Diaz04,Yang04,Arndt,Slaughter09,Jiang09,Thakkar10,Geng09,Mendieta09}.
For a review see Ref.~\cite{DeltaFF}.
As for other $\Delta$ observables, 
there is no other information,
with the exception of the 
electric quadrupole moment  $Q_\Delta$, 
which was estimated using a connection to the $\gamma N \to \Delta$ 
transition quadrupoles \cite{Blanpied01}.
For $Q^2\ne 0$
there is no experimental information 
related to the $\Delta$ form factors.
Even a very basic property like the $\Delta$ 
electric charge radius
is at present rather unknown, contrarily to what 
happens for the nucleon, where 
that quantity is accurately measured \cite{Amsler08}.
Our knowledge on the $\Delta$ charge radius relies so far
only on diverse
theoretical model calculations
\cite{Giannini90,Buchmann97,Wagner00,He05,Schwesinger92,Gobbi92,Sahoo95,Barik95,Dillon99,Buchmann00b,Buchmann02,Buchmann02b,Buchmann03,Arndt,Ledwig08}.
There are also theoretical calculations 
of the $\Delta$ quadrupole moment $Q_\Delta$ 
\cite{Giannini90,Buchmann97,Buchmann02,Buchmann02b,Butler94,Wagner00,Ledwig08,Isgur82,Drechsel84,Leonard90,Krivoruchenko91,Kroll94,Geng09,Buchmann00,Azizi08,Lebed95,Tiburzi05,Tiburzi09}.

Recently, the nature of the $\Delta$ was 
explored in another direction, and under another light.
Lattice QCD, the discrete version 
of the fundamental nonperturbative 
theory of the hadrons, was used
to estimate the $\Delta$ elastic form factors 
\cite{Alexandrou07,Alexandrou08,Alexandrou09,Boinepalli09}, 
following the pioneer works of Bernard \etal~\cite{Bernard82} and 
Leinweber \etal~\cite{Leinweber92}.
The works \cite{Alexandrou07,Boinepalli09,Leinweber92} 
are based on the quenched approximation.
The works \cite{Alexandrou08,Alexandrou09} 
are unquenched calculations, which include 
also sea quark effects.
The results of the MIT-Nicosia group 
\cite{Alexandrou07,Alexandrou08,Alexandrou09} 
were obtained for a wide range of $Q^2$, while
the calculations of the Adelaide group \cite{Boinepalli09} 
are performed for only one small value 
of $Q^2$ (the limit case $Q^2=0$ is
impossible to reach for technical reasons) 
and extrapolated for $Q^2=0$ to obtain 
the $\Delta$ multipoles and the 
charge and magnetic radii.
In particular for $Q^2=0$ the background 
field method was used to 
calculate with a good accuracy the $\Delta$ magnetic moment
\cite{Bernard82,Lee05,Aubin}.
These methods were applied to pion mass values 
within the 0.3$-$1 GeV region
and consequently some extrapolations 
are required \cite{Pascalutsa05,Cloet03}.
In the absence of direct 
experimental information lattice QCD 
provides the best reference for 
theoretical calculations
for finite $Q^2$.

Motivated by the recent publication of lattice QCD data for the
$\Delta$ form factors
\cite{Alexandrou07,Alexandrou08,Alexandrou09,Boinepalli09,Aubin}, 
several models 
were also extended to determine those physical quantities.
Calculations of the $\Delta$ form factors 
were performed within the covariant spectator theory
\cite{DeltaFF,Letter} and also within a chiral Quark-Soliton model 
($\chi$QSM) \cite{Ledwig08}.
Also, the magnetic octupole moment ${\cal O}_\Delta$ 
was estimated by Buchmann \cite{Buchmann08},  using
a deformed pion cloud model, and by Geng  \cite{Geng09},
using  $\chi$PT.
Both $Q_\Delta$ and ${\cal O}_\Delta$ 
were estimated using QCD sum rules \cite{Azizi08}
and a covariant spectator quark model \cite{Letter}.

In this work we use a constituent quark model
based on the covariant spectator formalism 
\cite{Nucleon,NDelta,NDeltaD,LatticeD}
to predict the $Q^2$ dependence of all four 
$\Delta$ multipole form factors.
Our results will be compared with the available
experimental data, and the results from other
models in the literature, as well as with 
recent lattice data.
The $\Delta$ electromagnetic form factors 
give us an insight into the $\Delta$ internal structure.
In particular the 
determination of the $\Delta$ 
electric quadrupole and magnetic octupole moments
gives information on how the
$\Delta$ is deformed, and on its size and shape. 
And this information
is valuable, as we will conclude, to adequately constrain quark models. These in turn can
be used to extrapolate and interpret lattice results.

The nucleon, as
a spin 1/2 particle, has no electrical quadrupole moment,
and the $\Delta$, a spin 3/2 state, 
emerges as the first candidate for a deformed baryon. 
A deformation (compression or expansion 
along the direction of the spin projection)
would imply $Q_\Delta \ne 0$.
The interpretation of $Q_\Delta \ne 0$ as 
a signature of distortion
matches well with our intuitive notion of deformation, 
based on the nonrelativistic limit 
of the Breit frame form factors.
In that limit the Fourier transformation 
of the form factors gives the spatial 
distribution of charge or magnetic moment.
We will therefore identify this notion as 
{\it classical deformation}.
Recently, an alternative interpretation of distortion, 
based on transverse densities
defined in the infinite moment frame,
was introduced by other authors 
\cite{Alexandrou09,Transverse1,Transverse2,Miller08}.
Our calculations in this paper provide the basis needed for 
a forthcoming analysis of
the $\Delta$ deformation 
under the {\it classical} perspective and 
also in the newer perspective \cite{Deform}.

It is clear that the study of the $\Delta$ form factors
is a very interesting topic, 
even if experimentally arduous.
The only experimental source of information about the  
$\Delta$ deformation comes today  
from the $\gamma N \to \Delta$ quadrupole transition.
However, this reaction
depends on orbital angular momentum components in {\it both\/} the nucleon
and of the 
$\Delta$ 
\cite{Buchmann00,Pascalutsa07}.
It is then very important to have an 
independent 
estimate of the 
angular momentum components in the $\Delta$ (leading to a deformation) \cite{AlexandrouDeform,Alexandrou08,Alexandrou09}.
This is the main motive
for our focus in this paper on the direct $\gamma \Delta \to \Delta$ reaction.

In our previous work a first covariant spectator 
constituent quark model was developed for the 
nucleon, the $\gamma N \to \Delta$ 
transition and the $\Delta$
\cite{Nucleon,NDelta,FixedAxis, NDeltaD,LatticeD,Lattice}. 
In the covariant spectator quark 
model a baryon is described 
as a system of three constituent quarks, 
with a off-shell quark free to interact 
with a electromagnetic field, while 
the quark-pair (diquark) acts as 
a spectator on-shell particle.
Confinement is described
effectively, with a quark-diquark vertex assumed 
to have a zero at the singularity of the three-quark propagator.
The quarks are dressed, having an electromagnetic
form factor with a behavior consistent with vector meson dominance. 
Within this first model we calculated the $\Delta$ 
form factors considering the $\Delta$
wave function parameterized  simply by 
an S-wave state for the quark-diquark system.
In this case only $G_{E0}$ and $G_{M1}$ 
have nonzero contributions because
of the symmetry assumed for the system.
In the work reported here we extend the previous model 
to the inclusion of the D-states, following Ref.~\cite{NDeltaD}.
It becomes then possible to predict, without further parameter fitting,
nonzero results also for $G_{E2}$ and $G_{M3}$.
Although our results regard only the valence quark contributions, they
compare well with a variety of lattice QCD data, 
besides the available experimental
data.

We organize this paper in the following main sections:
In Sec.~\ref{secII} we introduce the basic 
definitions used to calculate the $\Delta$ 
electromagnetic form factors.
In Sec.~\ref{secWF} we review the covariant 
spectator quark model and present the $\Delta$ 
wave function.
In Sec.~\ref{secFF} we derive the $\Delta$
form factors within our model.
In Sec.~\ref{secResults} we show the numerical results. 
Finally, in Sec.~\ref{secConclusions} 
we draw some conclusions.

\section{$\Delta$ electromagnetic form factors}
\label{secII}

The interaction of a photon with an on-mass-shell $\Delta$ isobar, 
with initial four-momentum $P_-$ and final four- momentum $P_+$, 
can be parameterized in terms of the electromagnetic current 
\cite{Nozawa90,Pascalutsa07}:
\ba
J^\mu&=&
-\bar w_\alpha (P_+) 
\left\{
\left[ F_1^\ast(Q^2) g^{\alpha \beta} 
+ F_3^\ast (Q^2)\frac{q^\alpha q^\beta}{ 4 M_\Delta^2} \right] \gamma^\mu  
\right. 
\nonumber \\
& &
\left.
+ 
 \left[ F_2^\ast(Q^2) g^{\alpha \beta} 
+ F_4^\ast (Q^2)\frac{q^\alpha q^\beta}{ 4 M_\Delta^2} \right]
\frac{i \sigma^{\mu \nu} q_\nu}{2M_\Delta}    
\right\} w_\beta(P_-), \nonumber \\
& &
\label{eqJgen}
\ea
where $M_\Delta$ is the $\Delta$ mass, 
$w_\alpha$ is the Rarita-Schwinger 
spin 3/2 state, $F_1^\ast(0) =e_\Delta$, with
$e_\Delta=0,\pm 1,+2$ the four possible $\Delta$ charge values.
The four functions
 $F_i^\ast$ ($i=1,...,4$) of $Q^2$ 
define the four $\Delta$ form factors
[in appendix \ref{apGeneric},
we discuss the general form of (\ref{eqJgen})]. 
They can 
be combined to form the two electric and two magnetic 
multipoles, which have a direct 
physical interpretation: 
the electric charge (E0)  and quadrupole (E2), and the
magnetic dipole (M1) and octupole (M3) form factors, 
defined as \cite{DeltaFF,Pascalutsa07,Nozawa90}
\ba
G_{E0}(Q^2) &=& 
\left[ F_1^\ast -\tau F_2^\ast
\right] \left( 1+ \frac{2}{3} \tau \right) \nonumber \\
& & - \frac{1}{3}\left[ F_3^\ast -\tau F_4^\ast
\right] \tau \left( 1+ \tau \right) 
\label{eqGE0}\\
G_{M1}(Q^2) &=& 
\left[
F_1^\ast + F_2^\ast\right]
\left(   
1+ \frac{4}{5} \tau
\right)\nonumber \\
& & 
-\frac{2}{5}
\left[
F_3^\ast  + F_4^\ast\right] \tau
\left( 1 + \tau \right) \label{eqGM1}
\\
G_{E2}(Q^2) &=&
\left[ F_1^\ast -\tau F_2^\ast
\right] \nonumber \\
& & - \frac{1}{2}\left[ F_3^\ast -\tau F_4^\ast
\right] \left( 1+ \tau \right)  \label{eqGE2} \\
 G_{M3}(Q^2) &=&
\left[
F_1^\ast + F_2^\ast\right] \nonumber \\
& & 
-\frac{1}{2}
\left[
F_3^\ast  + F_4^\ast\right]
\left( 1 + \tau \right) 
\label{eqGM3}
\ea 
where the dimensionless factor
\be
\tau = \frac{Q^2}{4 M_\Delta^2}
\label{eqtau}
\ee
was introduced.
The static magnetic dipole ($\mu_\Delta$), 
electric quadrupole ($Q_\Delta$) and 
magnetic octupole (${\cal O}_\Delta$) moments are defined 
in the $Q^2=0$ limit, as
\ba
& &
\mu_\Delta = \frac{e}{2 M_\Delta} G_{M1}(0) \nonumber \\
& &
Q_\Delta = \frac{e}{M_\Delta^2} G_{E2}(0) \nonumber \\
& &
{\cal O}_\Delta = \frac{e}{2 M_\Delta^3} G_{M3}(0).
\label{eqMoments}
\ea

\section{Delta wave functions}
\label{secWF}

In our model the $\Delta$ wave function is
described by an mixture of an S-state and two D-state components
for the quark-diquark system
\be
\Psi_\Delta=
N \left[
\Psi_S + a \Psi_{D3} + b \Psi_{D1} \right],
\label{eqPsiDel}
\ee
where $a$ and $b$ are respectively the admixture coefficients 
for the two D-states:  D3 (quark core spin $3/2$) 
and D1 (quark core spin $1/2$).
$N$ is the normalization constant.
To characterize $\Psi_{\Delta}$ we took the 
covariant spectator theory for a quark and a on-mass-shell diquark system,
as proposed in Ref.~\cite{NDeltaD}. 

From its symmetry properties,
the $J=3/2$ quark-diquark S-state 
(quark core ${\cal S}=3/2$ and 
orbital angular momentum $L=0$) 
has the general form 
\be
\Psi_S (P,k) =
- \psi_S(P,k) \left(T \cdot \xi^{1 \ast} \right) 
 \varepsilon_P^\alpha \;w_\alpha(P).
\label{eqPsiS}
\ee
as proposed in Refs.~\cite{NDelta,NDeltaD}.
In this equation $\psi_S$ 
is a scalar function describing the momentum distribution 
of the quark-diquark system in terms 
of the $\Delta$
and the diquark moment, $P$ and $k$ respectively;  
$\varepsilon_{P}^\ast$ is the polarization state 
associated with the diquark spin \cite{FixedAxis}
(the polarization index was omitted in this short-hand notation)
and  $w_\beta$ is the Rarita-Schwinger 
state.
The isospin-$\frac{3}{2}$ 
to isospin-$\frac{1}{2}$ transition operators are represented by
$T_i$ ($i=x,y,z$)
\cite{NDelta,DeltaFF,Pascalutsa05,Pascalutsa07},
and $\xi^{1\ast}(I_z)$ is the 
spin-1 diquark state (a combination of $u$ and $d$ quarks). 
The isospin operator $(T\cdot \xi^{1\ast})$
acts on the $\Delta$ isospin states.
(See Ref.~\cite{NDelta} for details.)
The normalization of the wave function (\ref{eqPsiS}) 
to unity implies for the scalar wave function
the normalization condition 
\be
\int_k  \left[ \psi_S (\bar P,k) \right]^2 =1,
\ee
where $\bar P= (M_\Delta,0,0,0)$ and the covariant integral over the momentum $k$ is defined in Eq.~(\ref{eq:kint}) below.

The $\Delta$ D-states in the spectator formalism 
were introduced and explained in detail 
in Ref.~\cite{NDeltaD}. Their definition is made through
the ${\cal D}$-state operator defined as
\be
{\cal D}^{\alpha \beta}(P_\pm,k_\pm) 
= \tilde k_\pm^\alpha \tilde k_\pm^\beta -
\frac{\tilde k_\pm^2}{3} \tilde g_\pm^{\alpha \beta},
\ee
where
\be
\tilde k_\pm^\alpha = k^\alpha - \frac{P_\pm \cdot k}{M_\Delta^2} P_\pm^\alpha,
\label{eqKtil}
\ee
is the covariant generalization of the three-momentum in 
$\Delta$ rest frame, and
\be
\tilde g_\pm^{\alpha \beta}= g^{\alpha \beta} - 
\frac{P_\pm^\alpha P_\pm^\beta}{M_\Delta^2},
\label{eqGtil}
\ee
generalizes the unity operator.
The variables $P_-$ ($P_+$)
describe the initial (final) momentum, and in the same way
$\tilde k_-$ ($\tilde k_+$) 
refer to the initial (final) state.

Using the operator ${\cal D}$ a generic spin 3/2 D-state is
\be
{\cal W}^\alpha (P,k; \lambda) 
= 
{\cal D}^{\alpha \beta} (P,k) w_\beta(P,\lambda).
\ee 
It decomposes into a sum of two D-states, associated with the two possible 
spin states of the quark core, ${\cal S}= 1/2$ 
or   ${\cal S}= 3/2$ (core-spin states  ).
To separate the two core-spin states
 we consider the two projection projectors:
\ba
& &({\cal P}_{1/2})^{\alpha \beta}= 
\frac{1}{3} \tilde \gamma^\alpha \tilde \gamma^\beta \\
& &  ({\cal P}_{3/2})^{\alpha \beta}= 
\tilde g^{\alpha \beta}-
\frac{1}{3} \tilde \gamma^\alpha \tilde \gamma^\beta,
\ea
where
\be
\tilde \gamma^\alpha=
\gamma^\alpha - \frac{\not \! P P^\alpha}{M_\Delta^2}.
\ee
The properties of the core-spin projectors 
are known in the literature \cite{Benmerrouche89,Haberzettl98}.

The amplitudes for the states associated to the core-spin ${\cal S}=1/2$ 
(D1-state) and ${\cal S}=3/2$ (D3-state) 
are defined as
\ba
\hspace{-.5cm}
& &\Psi_{D1}(P,k)= 
-3  \psi_{D1} (T \cdot \xi^{1 \ast})
 \varepsilon_{\lambda P}^{\alpha \ast}
W_{D1\, \alpha} (P,k) 
\label{eqPsiD1}
\\
\hspace{-.5cm}
& &
\Psi_{D3}(P,k)= 
-3 \psi_{D3} (T \cdot \xi^{1 \ast}) \varepsilon_{\lambda P}^{\alpha \ast}
W_{D3\, \alpha} 
(P,k), 
\label{eqPsiD3}
\ea
where
\ba
\hspace{-.5cm}
& &
W_{D1\, \alpha} (P,k;\lambda_\Delta)
= \left({\cal P}_{1/2} \right)_{\alpha \beta}
{\cal W}^\beta (P,k;\lambda_\Delta) 
\label{eqW1} \\
\hspace{-.5cm}
& &
W_{D3\, \alpha} (P,k;\lambda_\Delta)
= \left({\cal P}_{3/2} \right)_{\alpha \beta}
{\cal W}^\beta (P,k;\lambda_\Delta).
\label{eqW3}
\ea
As before, $(T\cdot \xi^{1\ast})$
acts on the $\Delta$ isospin states. 
For simplicity, in our notation  the diquark polarization 
($\lambda$) and $\Delta$ spin projections ($\lambda_\Delta$) are suppressed 
in the wave functions.
Also $\psi_{D1}= \psi_{D1}(P,k)$,  $\psi_{D3}= \psi_{D3}(P,k)$
are scalar wave functions.
The factor $-3$ was introduced for convenience.
The normalization of the wave functions is 
\be
\int_k \left\{
\tilde k^4 \left[ \psi_{D \, 2S} (\bar P,k) \right]^2 \right\}=1,
\ee
and ensures that
the wave function $\Psi_\Delta$ satisfies the 
charge condition,
\be
3 \sum_\lambda \int_k \bar \Psi_\Delta (\bar P,k) j_1 (0) \gamma^0 
\Psi_\Delta(\bar P,k)= e_\Delta. 
\ee  
The variable $\bar P$ represents $P=P_+=P_-$ in the 
$\Delta$ rest frame ($Q^2=0$) and  
$j_1(0)= \frac{1}{6}+ \frac{1}{2} \tau_3$ is the (quark) charge operator
at $Q^2=0$. 
See Ref.~\cite{NDeltaD} for more details about the D-states.

\section{Current and Form Factors}
\label{secFF}

Within the covariant spectator theory, the electromagnetic current   in 
impulse approximation, can be written as in  
Refs.~\cite{Nucleon,NDelta,NDeltaD}
\be
J^\mu = 3 \sum_\lambda 
\int_k
\bar \Psi_\Delta(P_+,k) j_I^\mu \Psi_\Delta (P_-,k),
\label{eqJen}
\ee
which sums in all intermediate
diquark polarization  and integrates over the diquark three-momentum
\be
\int_k \equiv \int \frac{d^3 k}{(2\pi)^3 2 E_k}.\label{eq:kint}
\ee
It also includes the symmetrization 
of the quark current.
Following Ref.~\cite{Nucleon}
the photon-quark interaction is represented 
by
\ba
j_I^\mu&=& \left(\frac{1}{6}f_{1+} + \frac{1}{2} f_{1-}
\tau_3 \right) \gamma^\mu  +
\nonumber \\
      & & \left(\frac{1}{6}f_{2+} + \frac{1}{2}f_{2-} 
\tau_3 \right) \frac{i \sigma^{\mu \nu} q_\nu}{2M_N},
\label{eqJI}
\ea
where $M_N$ is the nucleon mass 
and the isospin ($\tau_3$) and spin ($\gamma^\mu$) 
operators acts on the $\Delta$ initial and 
final states. 
The coefficients $f_{i\pm}$ with $i=1,2$ 
are the quark form factors and function of $Q^2$.
The explicit form of $f_{i\pm}$ will 
be introduced later in Sec.~\ref{secResults}.

For numerical applications we define
the isospin dependent functions $\tilde e_\Delta$ 
and $\tilde \kappa_\Delta$ as
\ba
& &\tilde e_\Delta (Q^2)= 
\frac{1}{2} \left[ f_{1+}(Q^2) + f_{1-}(Q^2)  \bar T_3  \right] 
\label{eqEd}
\\
 & &\tilde \kappa_\Delta (Q^2)= 
\frac{1}{2} \left[ f_{2+}(Q^2) + f_{2-}(Q^2)  \bar T_3  \right]
\frac{M_\Delta}{M_N},
\label{eqKd}
\ea
and $\bar T_3$ is the isospin-$\frac{3}{2}$ matrix defined as
\be
\overline T_3=
3 \sum_{i} T_i^\dagger \tau_3 T_i=
\left[
\begin{array}{rrrr}
\; 3 & \; 0 & 0 & 0 \cr 
         \; 0 & \;1 & 0 & 0 \cr
         \; 0 & \; 0 &-1 & 0 \cr
         \;0 & \;0 & 0 &-3 \cr 
\end{array}
\right].
\ee
We can also write $\bar T_3$ as $\bar T_3= 2 M_T$, 
where $M_T$ is the diagonal matrix with 
the $\Delta$ isospin projections:
$M_T=\mbox{diag}(+3/2,+1/2,-1/2,-3/2)$ \cite{Buchmann97}.
In the limit $Q^2=0$ we use
\be
e_\Delta= \sfrac{1}{2}(1+ \bar T_3), 
\hspace{.5cm}
\kappa_\Delta= \sfrac{1}{2}(\kappa_+ + \kappa_- \bar T_3)
\frac{M_\Delta}{M_N}.
\ee
The values of $\kappa_\pm $ were fixed 
in Ref.~\cite{Nucleon} by the nucleon 
magnetic moments.

It is convenient also to define
\be
\tilde g_\Delta = \tilde e_\Delta - \tau \tilde \kappa_\Delta, \hspace{.5cm}
\tilde f_\Delta = \tilde e_\Delta + \tilde \kappa_\Delta.
\label{eqfDelta}
\ee
Similarly as for $\tilde e_\Delta$ and $\tilde \kappa_\Delta$ 
we suppress the tilde for \mbox{$Q^2=0$.}

As both the initial and final states 
depend of the diquark polarization  with 
the same polarization index, the product of both 
polarization vectors factors out in the current (\ref{eqJen}).
Then, since the initial 
and final state have the same mass ($M_\Delta$) 
 \cite{Nucleon,NDelta,FixedAxis},
\be
\Delta^{\alpha \beta} \equiv 
\sum_{\lambda} \varepsilon_{\lambda P_+}^\alpha 
\varepsilon_{\lambda P_-}^{\beta \ast},
\ee
becomes
\ba
\Delta^{\alpha \beta} &=& -g^{\alpha \beta} - 
\frac{P_+^\alpha P^\beta_-}{M_\Delta^2} 
\nonumber \\
& &+ 
2 \frac{(P_++P_-)^\alpha (P_++ P_-)^\beta}{4M_\Delta^2 + Q^2}. 
\label{eqProp}
\ea

Using Eq.~(\ref{eqPsiDel}) we can decompose 
the transition current into the contributions from the 
different components of the wave function: 
\ba
J^\mu &=& 3 \sum_\lambda \bar \Psi_\Delta j_I^\mu \Psi_\Delta  \nonumber \\
      &=& 3 N^2 \sum_\lambda \bar \Psi_S j_I^\mu \Psi_S  \nonumber \\
      & &+ 3 a N^2 \left\{\sum_\lambda \bar \Psi_{D3} j_I^\mu \Psi_S 
              +        \sum_\lambda \bar \Psi_{S} j_I^\mu \Psi_{D3}
                \right\}    \nonumber \\
      & &+ 3 b N^2 \left\{\sum_\lambda \bar \Psi_{D1} j_I^\mu \Psi_S 
              +        \sum_\lambda \bar \Psi_{S} j_I^\mu \Psi_{D1}
                \right\}  \nonumber \\
      & & 
+N^2\left[
{\cal O}(a^2)  +  {\cal O}(b^2) + {\cal O}(ab) \right],
\label{eqJ1}
\ea 
where the last three terms are corrections from the 
transitions between the same or different D-states. 
Those terms can be neglected 
if the admixture coefficients are small. 
We can write $J^\mu$ in a compact form
\be
J^\mu=
N^2 J^\mu_S + a N^2 J^\mu_{D3} +
 b N^2 J^\mu_{D1}.
\label{eqJ2}
\ee
with explicit forms for the currents 
$J_S^\mu$, $J_{D3}^\mu$ and $J_{D1}^\mu$ 
presented in Appendix~\ref{apFF}.

From the general form (\ref{eqJgen}) and for the wave functions defined by
Eqs.~(\ref{eqPsiS}) and (\ref{eqPsiD1})-(\ref{eqPsiD3}), 
we can write
\ba
G_{E0}(Q^2) &=&  N^2 \tilde g_\Delta {\cal I}_S  
\label{eqGE0a}  \\
G_{M1}(Q^2) &=&  
N^2 \tilde f_\Delta 
\left[
{\cal I}_S 
+\frac{4}{5} a  {\cal I}_{D3}
-\frac{2}{5} b {\cal I}_{D1} \right]
 \label{eqGM1a} \\
G_{E2}(Q^2) &=& 
3 (aN^2) \tilde g_\Delta  \frac{{\cal I}_{D3}}{\tau} 
\label{eqGE2a} \\
G_{M3}(Q^2) &=& 
\tilde f_\Delta N^2\left[ 
a \frac{{\cal I}_{D3}}{\tau} +
2  b \frac{{\cal I}_{D1}}{\tau} \right],
\label{eqGM3a}
\ea
where ${\cal I}_S$ is the overlap between the initial and final S-states,
\be
{\cal I}_S=
\int_k \psi_S(P_+,k) \psi_S(P_-,k),
\ee
and the other overlap integrals, between the initial 
S and each of the final D-states, are
\ba
& &
{\cal I}_{D1}=
\int_k b(\tilde k_+,\tilde q_+)\psi_{D1}(P_+,k) \psi_S(P_-,k) 
\label{eqID1}
\\
& &
{\cal I}_{D3}=
\int_k b(\tilde k_+,\tilde q_+)\psi_{D3}(P_+,k) \psi_S(P_-,k),
\label{eqID3}
\ea
 with
\be
b(\tilde k_+, \tilde q_+) =
\frac{3}{2} \frac{(\tilde k_+ \cdot \tilde q_+)^2}
{\tilde q_+^2}
-\frac{1}{2} \tilde k_+^2. 
\label{eqB}
\ee
The variable $\tilde k_+$ is defined by Eq.~(\ref{eqKtil});
the variable $\tilde q_+$ is defined the same way 
(just  by replacing $k$  by $q$). 
The derivation of the previous expressions 
is presented in Appendix~\ref{apFF}.

In Eqs.(\ref{eqGE0a})-(\ref{eqGM3a}),
we chose to have $N^2=1$ in order to 
reproduce the charge of the $\Delta$.
(the corrections due to the D-state 
to D-state transition are of the order of 
$a^2$, $b^2$ and $ab$, as indicated by Eq.~(\ref{eqJ1})). 

The first observation to be made on the results (\ref{eqGE0a})-(\ref{eqGM3a})  
is that, 
to first order in $a$ and $b$, the D-states do not contribute to 
$G_{E0}$. The second observation
is that the D-states provide the only 
nonvanishing contributions to 
the electric quadrupole and magnetic octupole form factors, 
$G_{E2}$ and $G_{M3}$
(opening
the possibility for a satisfactory selection 
of models for the D-states, otherwise specially difficult
due to the overshadowing of their effects in $G_{M1}$ by the dominant S-waves).

The definition of $b(\tilde k_+,\tilde q_+)$ implies 
that $b(\tilde k_+,\tilde q_+)=-{\bf k}^2 Y_{20}(\hat k)$
when $Q^2=0$ (see  Ref.~\cite{NDeltaD} for details). 
Since the wave functions are independent 
of $z=\cos \theta$ in that limit,
the angular integration 
of $b(\tilde k_+,\tilde q_+)$
vanishes in the limit $Q^2=0$, and therefore the D-state 
overlap integrals vanish also.
We stress that this result is model independent, since it 
is a consequence 
of the orthogonality between the S-state ($L=0$) 
and the D-states ($L=2$). 

Consider now the form factors $G_{E2}$ and $G_{M3}$.
They are proportional to $1/\tau$, which is infinite for $Q^2=0$. 
But, as ${\cal I}_{D3}$ or ${\cal I}_{D1}$
are themselves proportional to $\tau$, both form factors go to 
a finite value as $\tau$  and $ Q^2\to 0$, ensuring
no divergence. 
The proportionality ${\cal I}_{D3}, {\cal I}_{D1} \sim \tau \sim Q^2$ 
simply comes from 
the specific dependence of the 
scalar wave functions on the quark momentum.
The proof of that proportionality is done in Appendix~\ref{apIQ2}.

This is how a nonzero contribution 
from the D-state overlap integrals 
survives for each form factor,  in spite of the orthogonality 
between S and D-states.
This same behavior (overlap integral ${\cal I} \sim Q^2$ 
as $Q^2\to 0$) was already observed also in the
Coulomb quadrupole form factor
of the  $\gamma N \to \Delta$ reaction \cite{NDeltaD}.

To conclude, the form factors $G_{E2}$ and $G_{M3}$ are a consequence of the 
D-states alone, and they depend on both the admixture coefficients, $a$ and $b$, and the particular momentum dependence
of the D-wave components of the wave function.

A remark about the predictions for  the  $\Delta^0$ case can also be made:
In our approach  $G_{E2}(0)=0$ and \mbox{$Q_{\Delta^0} = 0$}, 
because the D-state to D-state transitions
were neglected. Once those currents are included 
we may expect a nonzero, although rather small, contribution to
$G_{E2}(0)$.


The formulas obtained for the form factors are considerably simplified 
in the case $Q^2=0$. 
Since in that limit the integrals 
associated to transitions to the $\Delta$ D-states vanish
due to angular momentum, via
$b(k_+,q_+) \approx Y_{20}(\hat k)$,
we are left with, 
setting  $N^2 \to 1$
\ba
G_{E0}(0)  &=& e_\Delta            \\
G_{M1}(0) &=&  f_\Delta            \\
G_{E2}(0) &=&  3 a  e_\Delta {\cal I}_{D3}^\prime
\\
G_{M3}(0) &=& 
f_\Delta \left[ a\, {\cal I}_{D3}^\prime + 
2\, b \, {\cal I}_{D1}^\prime \right].
\label{eqGM30}
\ea
(The factor
${\cal I}_S=1$ from the normalization of $\psi_S$ was suppressed.)
In these equations
\ba
& &
{\cal I}_{D3}^\prime =
\lim_{\tau \to 0}   \frac{{\cal I}_{D3}(\tau)}{\tau} = \frac{d {\cal I}_{D3}}{d\tau} (0)
\nonumber \\
& &
{\cal I}_{D1}^\prime =
\lim_{\tau \to 0} \frac{{\cal I}_{D1}(\tau)}{\tau} = \frac{d {\cal I}_{D1}}{d\tau} (0),
\nonumber 
\ea
where  ${\cal I}_{D3}$ and ${\cal I}_{D1}$ are defined in 
Eqs.~(\ref{eqID1})-(\ref{eqID3}).

The several moments, corresponding to each of the 
four multipole form factors, are calculated from Eqs.~(\ref{eqMoments}).
Note that since the magnetic moment is
defined by  $G_{M1}(0)$ and ${\cal I}_{D1}$ and ${\cal I}_{D3}$ 
vanish at the origin, it follows that the magnetic moment 
is fixed by the S-state alone.

\section{Results}
\label{secResults}

The algebraic results for the form factors given in the 
previous section were obtained within the
covariant spectator formalism, independently of the specific 
model ingredients. In this section we start to 
specify the scalar wave functions 
and the quark current, 
required for an actual numerical application.

For the quark current we took the parameterization inspired on
vector meson dominance (VMD)   \cite{Nucleon,NDelta,NDeltaD}:
\ba
f_{1\pm} (Q^2) &= &
\lambda + \frac{(1-\lambda)}{1+Q^2/m_v^2} +
\frac{c_{\pm} Q^2/M_h^2}{\left(1 + Q^2/M_h^2 \right)^2} \qquad
\nonumber\\ 
f_{2\pm} (Q^2) &= &
\kappa_{\pm} 
\left\{ 
\frac{d_\pm}{1+Q^2/m_v^2} 
+
\frac{(1-d_{\pm})}{1+Q^2/M_h^2} 
\right\}\, , \label{eqf1m}
\ea
where $m_v$ represents 
a light vector meson $m_v= m_\rho$ (or $m_\omega$),
and  $M_h = 2M_N$ is the mass of an heavy vector 
meson which simulates shorter range effects.
The parameter $\lambda$ was  
adjusted to give  the correct quark density number 
in deep inelastic scattering  \cite{Nucleon,NDelta},
$\lambda =1.22$.
The coefficients $c$ and $d$ were fixed 
in a previous work to describe the 
nucleon elastic form factor data.
We have $c_+=4.16$, $c_-=1.16$ 
and $d_+=d_-=-0.686$ \cite{Nucleon}.
The particular combination (\ref{eqf1m}) 
with $c_+\ne c_-$ gives $f_{1+} \ne f_{1-}$ 
and therefore breaks isospin symmetry 
for $Q^2 \ne 0$, essential for a good description of the nucleon data.
The anomalous magnetic moments $\kappa_+= 1.639$ 
and $\kappa_-= 1.823$ generate the nucleon magnetic moment 
exactly \cite{Nucleon}.
The parameterization (\ref{eqf1m}) was 
applied to the nucleon \cite{Nucleon,Lattice}, $\Delta$ 
electromagnetic form factors 
\cite{DeltaFF,Letter}
as well as to the three
$\gamma N \to \Delta$ 
transition form factors \cite{NDelta,NDeltaD,LatticeD,Lattice}.

The scalar wave functions 
depend on the 
kinematic and dimensionless variable $\chi$, as introduced in 
Refs.~\cite{Nucleon,Gross06,DeltaFF,NDelta,NDeltaD}:
\be
\chi= \frac{(M_\Delta-m_s)^2-
(P-k)^2}{M_\Delta m_s}.
\label{eqChi}
\ee
We then use \cite{NDeltaD,Lattice}:
\ba
& &\psi_S(P,k)= \frac{N_S}{m_s(\alpha_1+ \chi)^3} 
\label{eqPsiSs} \\
& &\psi_{D3}(P,k)= \frac{N_{D3}}{m_s^3(\alpha_2+ \chi)^4} \\
& &\psi_{D1}(P,k)= \frac{N_{D1}}{m_s^3}\left\{
\frac{1}{(\alpha_3+ \chi)^4}- \frac{\lambda_{D1}}{(\alpha_4+ \chi)^4} 
\right\}.
\label{eqPsiD1s}
\ea
The functions (\ref{eqPsiSs})-(\ref{eqPsiD1s}) 
were chosen for a good description of the $\gamma N \to \Delta$ 
data \cite{NDelta,NDeltaD,LatticeD}.
The coefficient $\lambda_{D1}$ in $\psi_{D1}$ was 
determined by imposing the condition that
the $\Delta$ D1-state and the nucleon S-state are orthogonal \cite{NDeltaD}.
 
In the applications we consider two different
parameterizations for the $\Delta$ wave function, 
both previously derived in the spectator formalism.
By comparing  the results of two models we can 
illustrate the sensitivity of the results to model building.
The two parameterizations are equivalent with respect 
to the dominant S-state, but differ substantially in 
the D-state contributions.
In both models the D-states give minor 
contributions for the quadrupole 
$\gamma N \to \Delta$ transition form factors, 
since they are basically dominated by 
the pion cloud contributions at low $Q^2$, which can be 
parameterized by a simple analytical structure \cite{NDeltaD}.
The first model applied here, labeled  
Spectator 1 (Sp 1),  was derived  
in Ref.~\cite{NDeltaD} (and named model 4 there)
to describe the $\gamma N \to \Delta$ 
transitions in the physical region.
The D-states were calibrated by fitting 
the experimental data.
The second parameterization, labeled
Spectator 2 (Sp 2), was given in Ref.~\cite{LatticeD}. 
Although it
has the same structure as Sp 1, it was adjusted 
to the 
lattice QCD data \cite{Alexandrou08b} for $\gamma N \to \Delta$, in a 
pion mass region 
where the pion cloud effects are negligible, 
and afterwards extended to the physical pion mass point.
The results for the physical region are therefore independent 
of the pion cloud mechanisms.   
This feature gives the latter model 
some robustness, making
the procedure, in our opinion,  more reliable, and indirectly, 
constrained by QCD as well. 
In the first model (Sp 1) there is an admixture of 0.88\% 
for the D3-state and a larger admixture of 4.36\%  for the D1-state.
In the Sp 2 model one obtains an admixture of 
0.72\% for both D3 and D1. These results supply
us with the first quantitative idea of the sensitivity of the
D-states to the fitting procedure. 
The D1-state, specially,  shows to be strongly sensitive to that procedure.

\begin{figure*}[t]
\vspace{.5cm}
\centerline{
\mbox{
\includegraphics[width=3.2in]{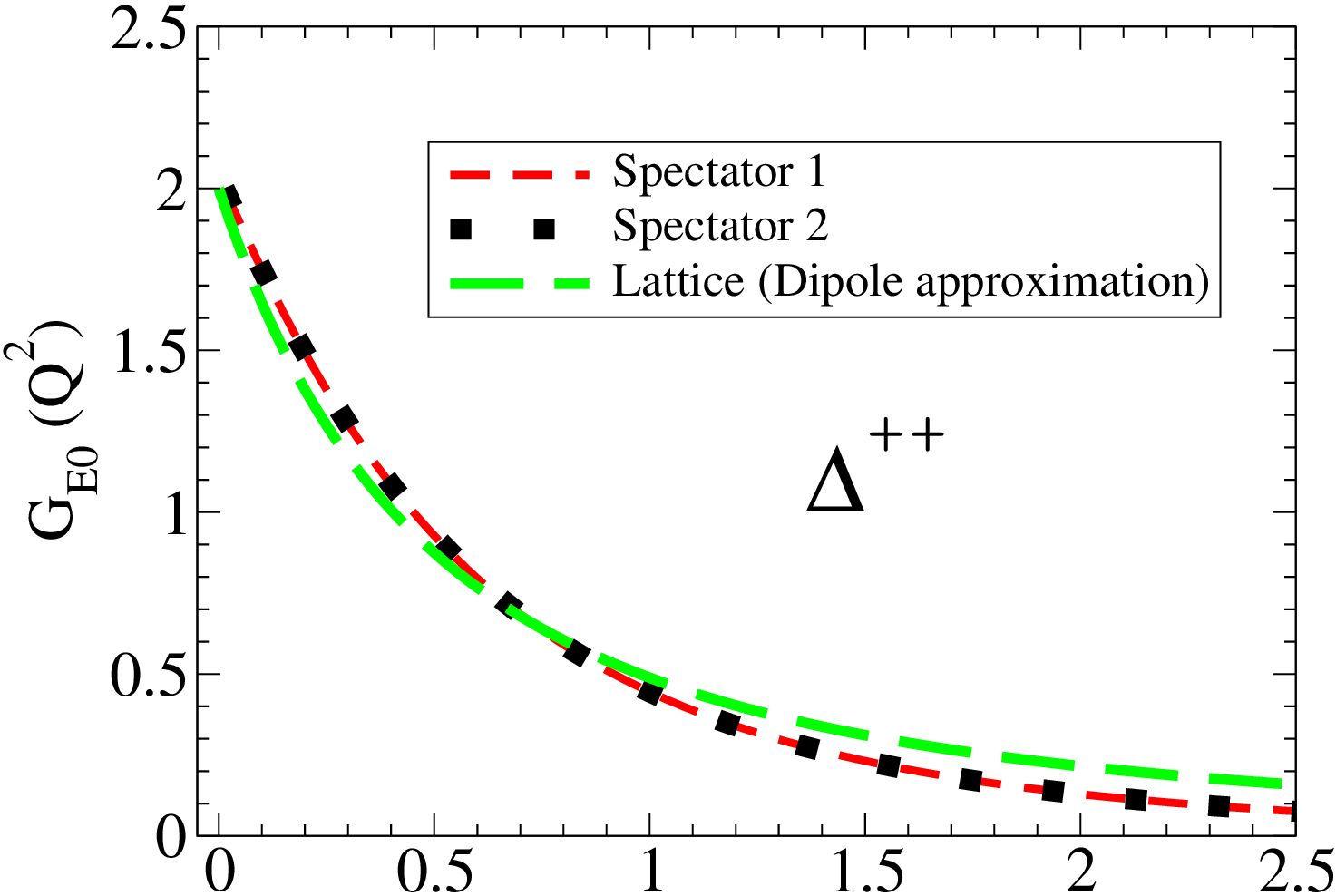} 
\includegraphics[width=3.2in]{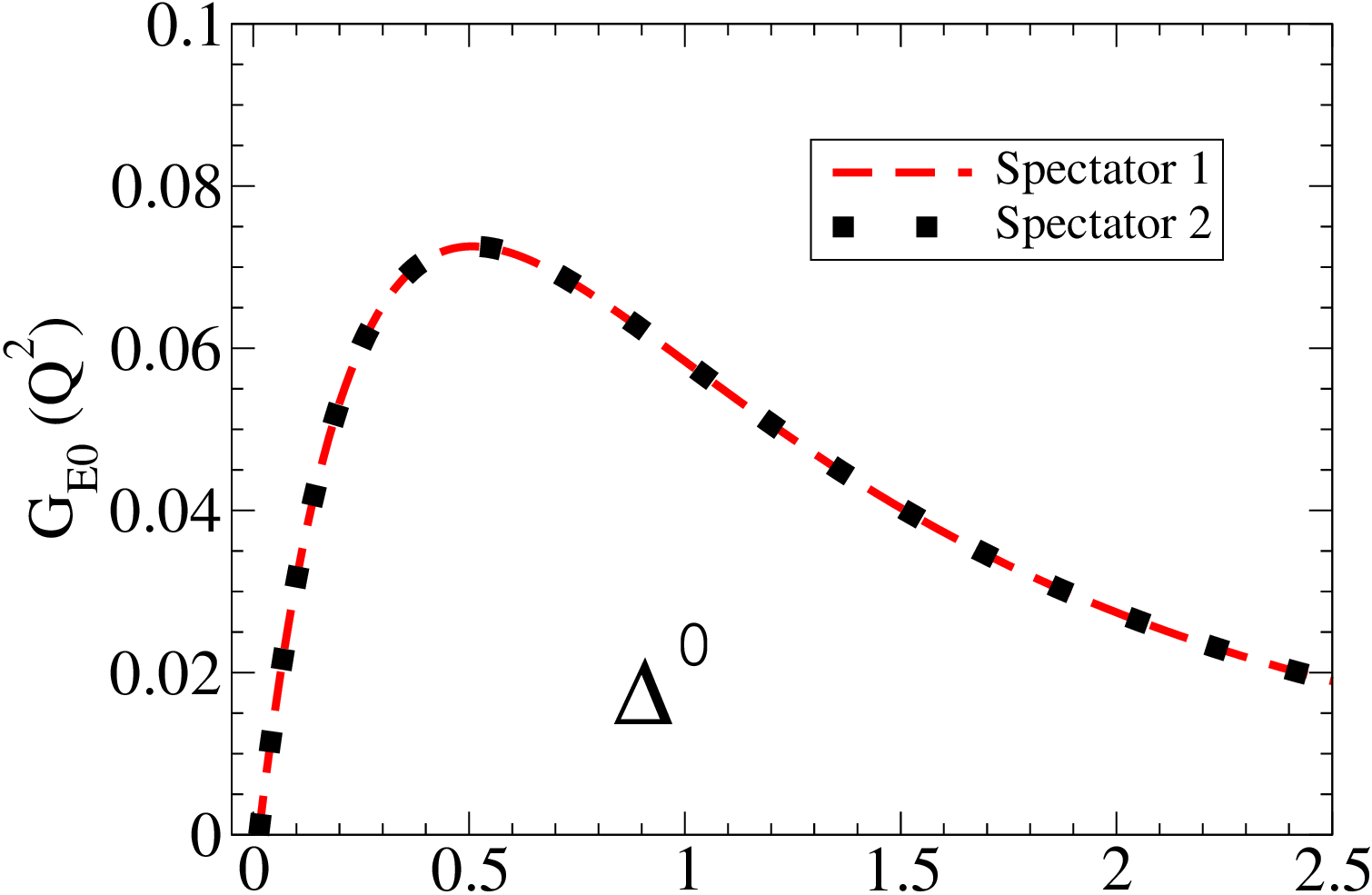}}}
\vspace{1.05cm}
\centerline{
\mbox{
\includegraphics[width=3.3in]{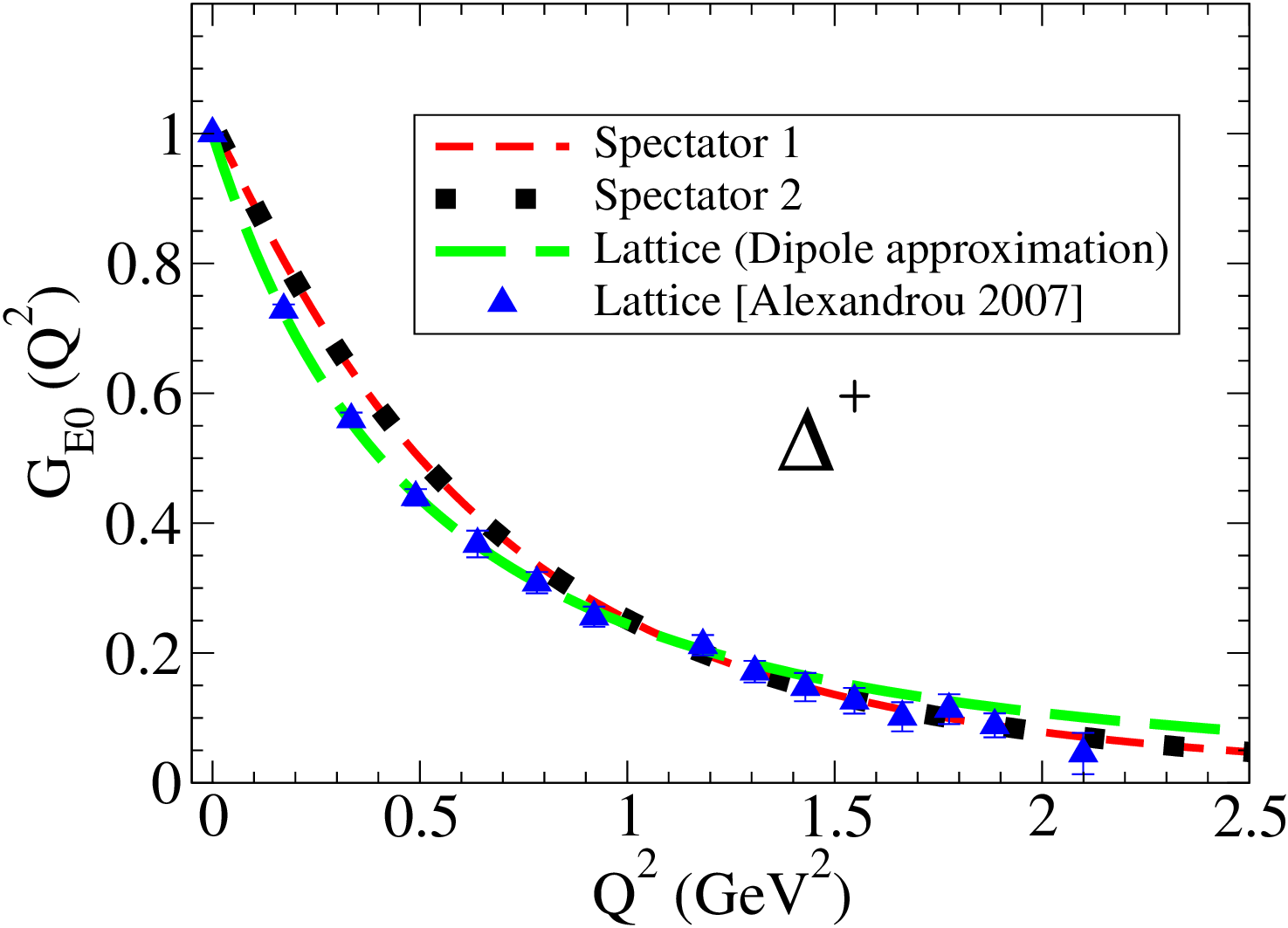}
\includegraphics[width=3.2in]{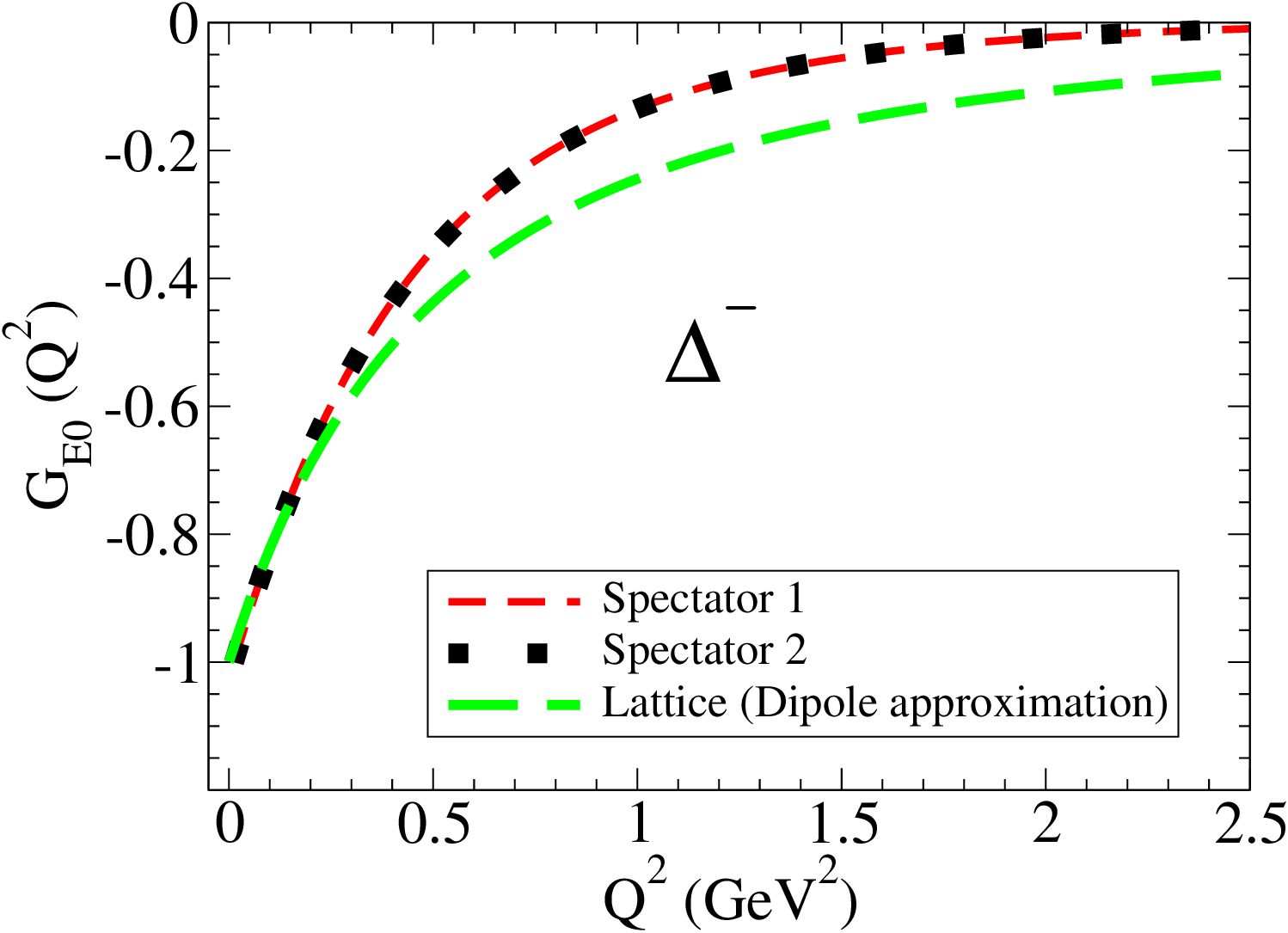}}}
\caption{\footnotesize{
$G_{E0}$ form factor. 
The Lattice data is the physical extrapolation 
from Refs.~\cite{Alexandrou07,Alexandrou07X}.}}
\label{figGE0}
\end{figure*} 

\begin{table*}
\begin{center}
\begin{tabular}{l c c c  c c c c  c c c c}
\hline
\hline
$<r_{E0}^2>$ & &  $p$  & & $\Delta^{++}$ & & $\Delta^+$ & & 
$\Delta^0$ & &  $\Delta^-$ \\
\hline
NRQM \cite{Giannini90}  & &  0.74  & &        & & 1.06     & &   & & \\
MIT bag model \cite{Gobbi92} & & 0.53
   & &   0.62     & &  &&  \\
Skyrme \cite{Schwesinger92} & & 0.52
                         & & 0.55 && 0.51  && -0.068   & & 0.65 \\
QSM \cite{Gobbi92}      & & 0.75 
& &  0.90 &&        &&          &&      \\
FT QM  \cite{Sahoo95}   & &  0.56
& & 0.70   & & 0.67     & &  0.03  & & 0.71 \\
RQM \cite{Barik95}      & &   
& &  0.520 && 0.523  && 0        && 0.523 \\
CQM (imp) \cite{Wagner00}    & & 
       & &    0.735     && 0.737  && 0
                                 & &   0.735 \\     
CQM \cite{Wagner00}     & &      
 & &    0.766     && 0.766  && 0
                                 & &   0.766 \\ 
GP/Large-$N_c$ \cite{Buchmann00b} & & 0.792$\pm$0.024
                          & & 0.85$\pm$0.09 & & 0.79$\pm$0.09 
                          & & -0.11$\pm$0.09 && 1.02$\pm$0.09 \\
 GBE \cite{He05}   & &  0.564 
        & &      & & 0.689    & &  0  & &  \\
$\chi$QSM  \cite{Ledwig08} & & 0.768
 & &        & & 0.794    & &   & &  \\
$\chi$QSM SU(3) \cite{Ledwig08} & & 0.770 
 & &        & & 0.815    & &   & &  \\  
$\chi$PT   ~\cite{Geng09}  &&   &&  && 0.328$\pm$0.016 &&  &&  \\
\hline 
Spectator 
\cite{DeltaFF}  & &  & &  & & 0.33  & &          & &      \\
Spectator 1 & &  &&  0.35       & &  0.29    & &  -0.104  & & 0.50 \\ 
Spectator 2 & & && 0.35      & &  0.29    & &  -0.104  & & 0.50 \\ 
\hline
Lattice: & & &&   &&  &&  && \\
Quenched \cite{Leinweber92} && & &  & &
0.397$\pm$0.088 & &     \\
Quenched \cite{Alexandrou07,Alexandrou07X}$^a$ && & &   & &
0.477$\pm$0.008 & &    & & \\
Quenched Wilson \cite{Alexandrou08,Alexandrou09} && & &  & &
  0.425$\pm$0.011 & &  & & \\
Dynamical Wilson \cite{Alexandrou08,Alexandrou09} && & & & &
  0.373$\pm$0.021 & &  & & \\
Hybrid  \cite{Alexandrou08,Alexandrou09} && & &   & &
  0.411$\pm$0.028 & &  & &  \\
Quenched \cite{Boinepalli09} && & & 0.410$\pm$0.057 && 0.410$\pm$0.057 && 0  && 
                                      0.410$\pm$0.057 \\               
\hline
\hline
\end{tabular}
\end{center}
\caption{Summary of existing theoretical and lattice results 
for $<r_{E0}^2>$ (fm$^2$). 
For the nucleon one has
$r_p^2=0.766$ fm$^2$ and $r_n^2=-0.116$ fm$^2$ \cite{Amsler08}.
The proton electric radius in 
the Spectator models is $r_{p}^2=0.79$ fm$^2$ 
(see model II from  Ref.~\cite{Nucleon}). \\
$^a$ Extrapolation to the physical point.}
\label{tableRE0}
\end{table*}

\begin{figure}
\vspace{.5cm}
\centerline{
\mbox{
\includegraphics[width=3.2in]{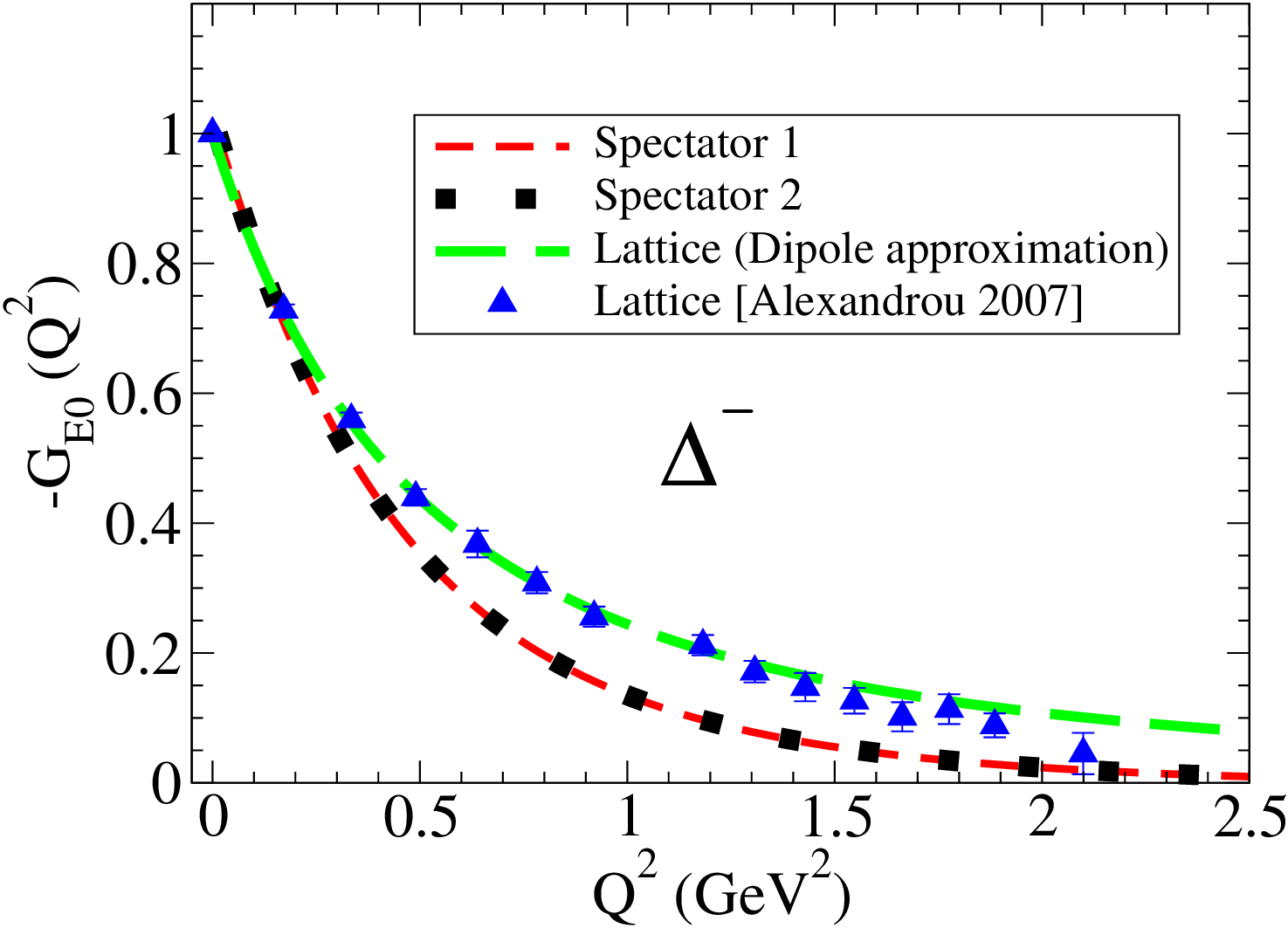}}}
\caption{\footnotesize{
Comparing our model results for $-G_{E0}^{\Delta^-}$ 
with lattice data and the dipole 
form for  $G_{E0}^{\Delta^+}$.}}
\label{figGE0x}
\end{figure}

\begin{figure}
\vspace{.5cm}
\centerline{
\mbox{
\includegraphics[width=3.2in]{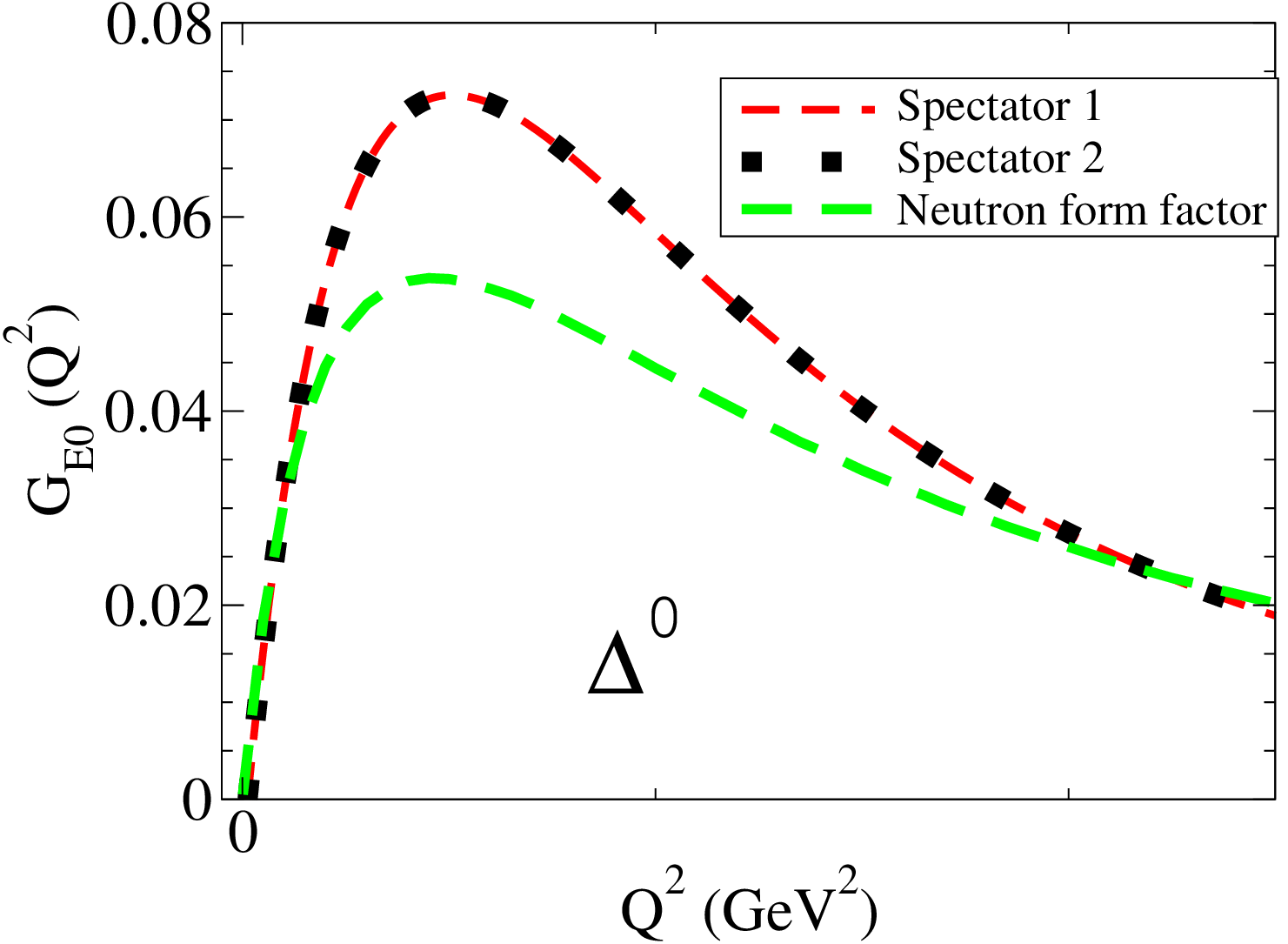} }}
\caption{\footnotesize{
Comparing
$G_{E0}^{\Delta^0}$ with the neutron form factor.}}
\label{figGE0y}
\end{figure}

\begin{figure*}[t]
\vspace{.5cm}
\centerline{
\mbox{
\includegraphics[width=3.2in]{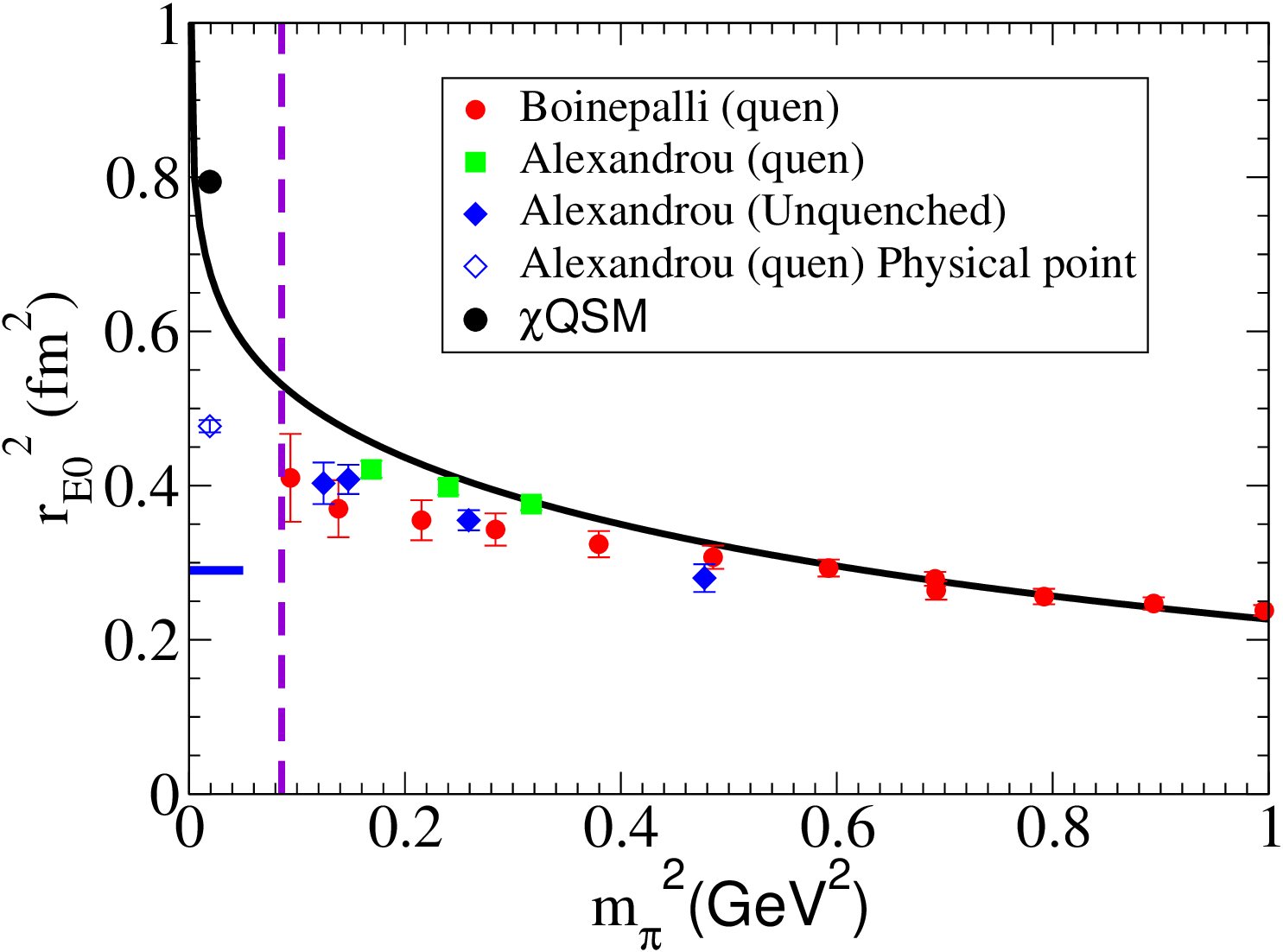} 
\includegraphics[width=3.2in]{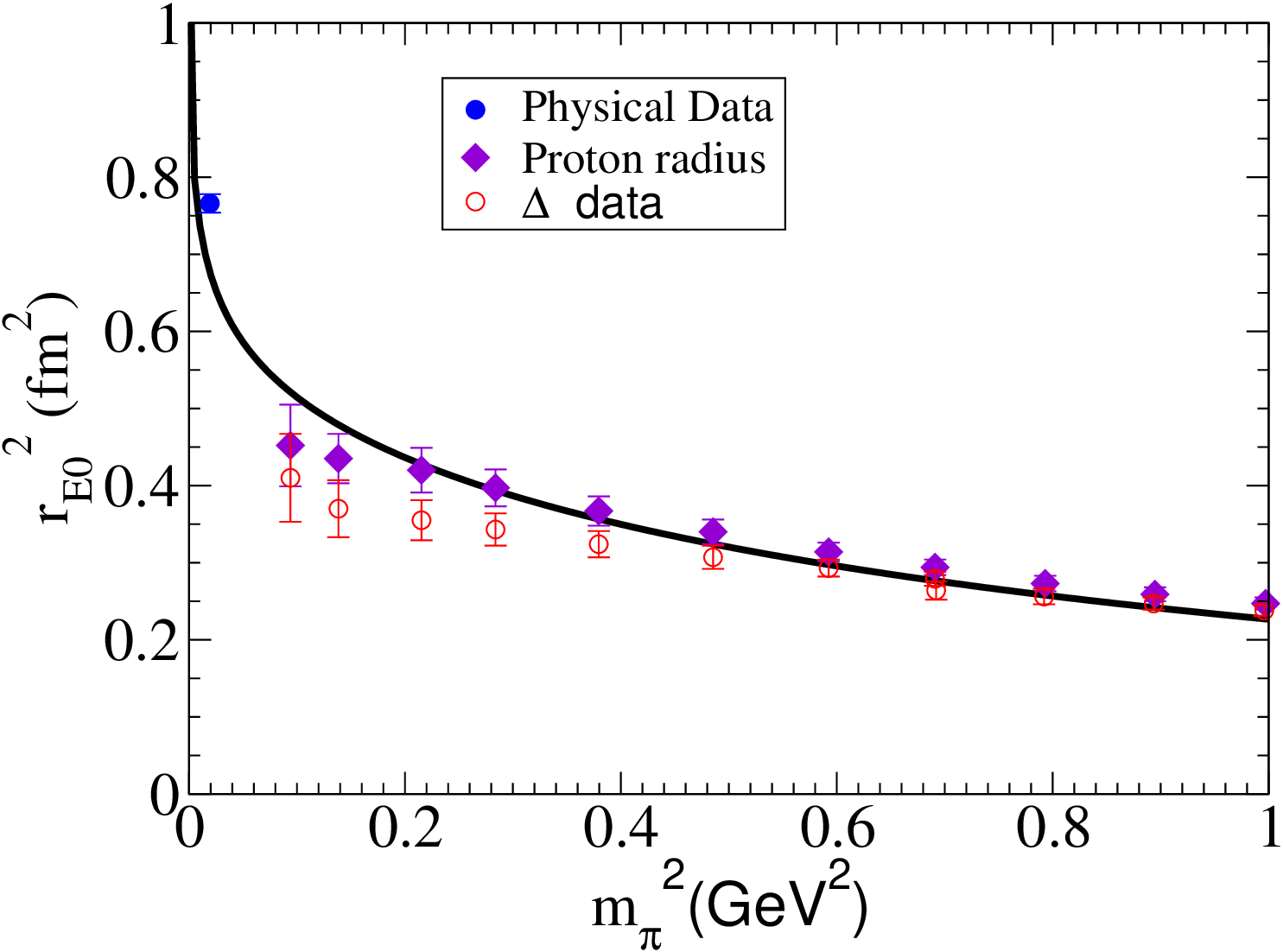}}}
\caption{\footnotesize{
Left panel: 
$\Delta^+$ electric charge radius squared in
lattice QCD \cite{Boinepalli09} as a function of the pion mass, compared 
with the chiral extrapolation (solid line)
of the proton charge radius \cite{HackettJones00}.
We show also the result for $\chi$QSM \cite{Ledwig08}
and our prediction (horizontal line).
The vertical line indicates $m_\pi=M_\Delta-M_N$.
Right panel: $\Delta^+$ \cite{Boinepalli09} and proton
\cite{Boinepalli06} charge radius 
in lattice QCD as a function of the pion mass, 
compared with the chiral extrapolation 
of the proton 
charge radius \cite{HackettJones00} (solid line).
The experimental result for $r_p^2$ 
\cite{Amsler08} is also included.}}
\label{figRE0mpi2}
\end{figure*}

\subsection{Electric charge form factor}

The results for $G_{E0}(Q^2)$ are shown in Fig.~\ref{figGE0}. 
The results for Sp 1 and Sp 2 are exactly 
the same because both models have the same S-state 
parameterization, and we chose $N=1$.
In the figure, for the $\Delta^+$ case we compare our results 
with the physical extrapolation of the  $\Delta^+$ 
form factor from 
quenched lattice QCD  data \cite{Alexandrou07,Alexandrou07X}. 
For the other charge cases, in the
absence of lattice data 
one can use exact isospin symmetry,
which amounts to take
\be
G_{E0}^{\Delta^{++}} (Q^2)= 2 G_{E0}^{\Delta^+} (Q^2)= 
- 2 G_{E0}^{\Delta^-} (Q^2),
\label{eqGEsim}
\ee
where for $G_{E0}^{\Delta^+}$ we used the dipole
approximation of the lattice data \cite{Alexandrou07,Alexandrou07X}.  
Although in quenched approximation this symmetry holds, there is 
no rigorous theoretical justification for it,
since breaking of the isospin symmetry can be expected
for $Q^2 \ne 0$, and was seen to be consistent with the nucleon data.
In fact, the spectator models accommodate 
isospin asymmetry 
(through $f_{1+} \ne f_{1-}$ and  $f_{2+} \ne f_{2-}$).
This is why,
in 
Fig.~\ref{figGE0x}, where $-G_{E0}^{\Delta^-}$ 
is compared directly with $G_{E0}^{\Delta^+}$, the differences are more
significant than what can be observed in Fig.~\ref{figGE0} for ${\Delta^+}$.
As expected, 
the disagreement between $-G_{E0}^{\Delta^-}$ and $G_{E0}^{\Delta^+}$
increases with $Q^2$.
Finally, in Fig.~\ref{figGE0y} we compare
the results for $\Delta^0$ with 
the neutron electric form factor. 
Note that the slope near $Q^2=0$
is very similar, for both models and  $G_{En}$.

Comparing our calculations with the 
lattice data (extrapolated to the physical region) 
from Ref.~\cite{Alexandrou07,Alexandrou07X}
we note, as observed previously in Ref.~\cite{DeltaFF}
where only S-states were taken,
that both spectator models differ from 
the lattice data for low $Q^2$,  but are significantly close to 
the lattice data at larger $Q^2$.
Nevertheless, our models cannot be simulated by a pure dipole dependence 
as the lattice QCD data can, and one notices 
that they have a slightly slower falloff 
with $Q^2$, implying  a smaller charge radius
or, equivalently, a stronger charge concentration, 
than suggested by the lattice simulations.
In our calculations the squared charge radius of the $\Delta^+$
is 0.29 fm$^2$ in both models,
to be compared with  
0.48 fm$^2$ from the lattice simulation \cite{Alexandrou07,Alexandrou07X}.

\subsubsection{Electric charge radius: comparison with other models}

The analysis of the charge distribution is naturally done first
in terms of the charge radius.
For the charged $\Delta$ states, the expansion of the charge form factor 
in powers of $Q^2$,
($e_\Delta= G_{E0}(0)\ne 0$) 
\be
G_{E0}(Q^2) = G_{E0}(0) \left[ 1 -\frac{Q^2}{6} <r_{E0}^2> 
+ {\cal O}(Q^4)\right],
\ee
defines the 
charge squared radius as
\be
<r_{E0}^2> = - \frac{6}{G_{E0}(0)} \left. 
\frac{d G_{E0}}{d Q^2}\right|_{Q^2=0}.
\label{eqRE02}
\ee 
For the case of neutral states 
($\Delta^0$),  $G_{E0}(0)=0$,
and we use 
\be
<r_{E0}^2> = - 6 \left. 
\frac{d G_{E0}}{d Q^2}\right|_{Q^2=0}.
\label{eqRE0b}
\ee 
In some works Eq.~(\ref{eqRE0b}) [with no normalization to $G_{E0}(0)$ 
included] instead of Eq.~(\ref{eqRE02}), defines the charge radius also
for charged particles.
The definition (\ref{eqRE02})
has the advantage of being suitable 
to higher order form factors, namely $G_{M1}$,  
without loss of generality.  For instance, 
for the proton, the electric charge 
and the magnetic dipole form factors 
have the same dipole dependence on $Q^2$ at
low $Q^2$,  however with no normalization at $Q^2=0$ as in Eq.~(\ref{eqRE0b}),
one is led to very different 
electric charge and magnetic dipole radii.
In addition, with Eq.~(\ref{eqRE02}) the $\Delta^-$ and the $\Delta^{++}$ 
radii can be directly compared with $\Delta^+$.
In particular, in a model with exact  
isospin symmetry  [see Eq.~(\ref{eqGEsim}] 
the charge radius is equal for all 
the charged $\Delta$ states if Eq.~(\ref{eqRE02}) is used.

The results from our two models, together with a 
summary of the literature, is presented in Table
\ref{tableRE0}. 
For a better interpretation of the results we write  
in the first column the prediction of each calculation
for the squared charge radius of the proton, 
when it is available.  
Our prediction for the $\Delta^+$ radius  is 0.29 fm$^2$.
This number is below the value predicted by a variety of models.

The first estimate of the $\Delta$ charge radius was 
based on NRQM and
suggested $r_{\Delta^+}^2 \simeq 1.06$ fm$^2$ \cite{Giannini90},
implying a larger spatial distribution for the $\Delta$ than for the proton 
($r_p^2 \simeq 0.77 $ fm$^2$).
That effect was traditionally explained as a result of 
the repulsive hyperfine interaction of the 
quarks in spin-triplet state,
in contrast with the attractive effect in the singlet state 
\cite{Giannini90}. However,
there was, at the time of this explanation, no direct experimental 
evidence of that fact.

The $\Delta$ charge radius was also computed 
within the MIT bag model \cite{Gobbi92,Thomas84}.
One should mention for completeness
other approaches, as a Skyrme model \cite{Schwesinger92}, 
a quark-soliton Model 
(QSM) \cite{Gobbi92},
a field theory quark model (FT QM) \cite{Sahoo95},
a relativistic quark model (RQM) \cite{Barik95},
a constituent quark model (CQM) \cite{Buchmann00},
a general 
parameterization of QCD combined with large-$N_c$ 
(GP/Large $N_c$) \cite{Buchmann00b}
a Goldstone boson exchange (GBE) model \cite{He05},
a chiral soliton model ($\chi$QSM) \cite{Ledwig08}
and a chiral perturbation theory model ($\chi$PT)
\cite{Geng09}. 
Although the results differ 
from model to model,  they all have in common the feature 
that  
the $\Delta$ has a larger charge radius than the proton.
The exception is the $\chi$PT approach from Ref.~\cite{Geng09}.

The same trend is seen 
in CQMs.
By taking a nonrelativistic CQM 
with two-body exchanges, 
Buchmann \etal~\cite{Buchmann97} 
obtains for the $\Delta^+$ electric radius $r_{\Delta^+}^2$ squared,
\be
r_{\Delta^+}^2=r_p^2-r_n^2,
\label{eqRdel1}
\ee
where $r_p^2$ and $r_n^2$ 
are the proton and neutron 
squared radius.
In the same model $r_{\Delta^0}^2=0$.
Using recent
values for the nucleon radii ($r_p^2 = 0.77$ fm$^2$; 
$r_n^2 =-0.116$ fm$^2$) \cite{Amsler08}
one obtains $r_{\Delta^+}^2=0.88$ fm$^2$, which, in any case, is larger than
$r_p^2$.

The relation (\ref{eqRdel1}) was improved 
by Dillon and Morpurgo using a 
general parameterization (GP) of QCD \cite{Dillon99}
with corrections of the order of 10\%-20\%.
Ameliorating upon Ref.~\cite{Buchmann97},
Ref.~\cite{Wagner00} establishes that
the contributions of the impulse approximation 
are dominant to the charge radius, and that 
the two-body currents associated with 
the quark-antiquark pairs are only a small
correction ($\approx 0.03$ fm$^2$).
Using an operator method based on 
a GP \cite{Dillon99} and the large-$N_c$ limit, 
Buchamnn and Lebed \cite{Buchmann00b}
related the four $\Delta$ charge radii
with the neutron and proton radii.
Those results and their uncertainty 
are also shown in Table \ref{tableRE0}.
In this case $r_{\Delta^+}^2 \approx r_p^2$.
%
A similar model, as the GBE model \cite{He05}, 
predicts however a smaller $\Delta$ charge radius 
($r_{\Delta^+}^2 \simeq 0.7$ fm$^2$). 
Finally, a
calculation within
$\chi$QSM suggests 
$r_{\Delta^+}^2 \simeq 0.8$ fm$^2$ \cite{Ledwig08}.

Interestingly, the most recent estimates of the
$\Delta$ electric squared radius 
\cite{Buchmann97,Wagner00,Buchmann00b,He05,Ledwig08}
suggest
$r_{\Delta}^2  \approx 0.8$ fm$^2$, 
slightly larger, but not significantly larger than the experimental 
result for the proton $r_p^2 \simeq 0.77$ fm$^2$.
In general we can say that the several calculations 
in the literature predict $r_{\Delta^+}^2 > r_p^2$.

Our prediction for the $\Delta^+$ radius (0.29 fm$^2$)
is below all the model calculations
identified above.
In general these last ones also overpredict the 
lattice calculations values ($\sim 0.40$ fm$^2$ 
and $\sim 0.48$ fm$^2$ from \cite{Alexandrou07,Alexandrou07X}).
Only the estimate of $\chi$PT \cite{Geng09}
is close to these lattice results. 

As for the $\Delta^0$ charge radius,
it is interesting to note that it is very close to 
the neutron radius ($r_n^2= -0.116$ fm$^2$) 
in our two models.
This comes as no 
surprise, since, as noticed in  
Fig.~\ref{figGE0y}, $G_{E0}$ and 
the neutron electric form factor. 
have almost the same shape for low $Q^2$.
The result is also consistent  with
partial quenched $\chi$PT (PQ$\chi$PT)
\cite{Arndt}.

\subsubsection{Electric charge radius: comparison with the 
lattice QCD results}

Interestingly, our results are comparable in magnitude
with the lattice QCD results in
Refs.~\cite{Alexandrou08,Alexandrou09,Boinepalli09}. 
The lattice results shown in Table \ref{tableRE0} 
correspond to  
the lowest pion mass $m_\pi$ taken by each method
($m_\pi \sim$ 300, 400 MeV).

It is expected that the lattice results
for the charge radius increase
as $m_\pi$ approaches the physical pion mass value, since 
it is what happens for the
proton, where
$r_p^2$ diverges as $m_\pi \to 0$.
In fact, the $\Delta^+$ lattice data 
\cite{Boinepalli09,Alexandrou07,Alexandrou08,Alexandrou09}
follows approximately 
the behavior of the proton charge radius 
as parameterized by $\chi$PT in Ref.~\cite{HackettJones00}.
This is seen in the left panel of 
Fig.~\ref{figRE0mpi2}.
The vertical line  shown
indicates the 
inelastic threshold for pion production, 
defined by $m_\pi = M_\Delta-M_N$.
When this inelastic channel is crossed from 
above we can conjecture the suppression of  
$\Delta^+$ charge radius below that point, 
as suggested by the
$\Delta$ magnetic moment 
studies within chiral effective field theory 
\cite{Cloet03,Pascalutsa05,Alexandrou08}.  
In the graph we include also the result
of $\chi$QSM in Ref.~\cite{Ledwig08}
at the physical point, $r_{\Delta^+}^2=0.794$ fm$^2$, 
which is close to the 
proton charge radius. 

In the right panel 
of the Fig.~\ref{figRE0mpi2} we illustrate that the $\Delta^+$ lattice charge 
radius results \cite{Boinepalli09}, although comparable with the 
proton charge radius lattice data \cite{Boinepalli06}, 
are slightly larger, for the 
same lattice QCD conditions (lattice space $a=0.128$ fm) 
and approximation (quenched). 

Some care must be taken in
comparing  model results like ours, and lattice QCD results.
First, there are several methods to 
simulate the quark field in lattice, 
corresponding to different choices for the actions, as the Wilson (quenched or dynamical) 
\cite{Alexandrou07,Alexandrou08,Alexandrou09,Gockeler05,Alexandrou06},
Clover \cite{Boinepalli09,Boinepalli06} or 
hybrid \cite{Alexandrou08,Alexandrou09} actions.
The different choices lead to different results.
Besides, lattice methods can differ 
in the calibration needed to make the connection with the physical limit.
That can be done by adjusting
the lattice spacing using the Sommer method 
\cite{Boinepalli06}, or the 
physical nucleon mass 
\cite{Gockeler05}.
The two methods lead to very different 
predictions for the proton charge radius
\cite{Boinepalli06,Gockeler05}.

To finish the discussion on the charge radius, 
we mention that the method from
Ref.~\cite{Arndt}, based on quenched and partial 
quenched chiral perturbation theory, is useful to extrapolate 
lattice QCD calculations to the physical limit, and
predicts results similar 
to the ones from our models Sp 1 and Sp 2.

\subsubsection{Overview of the electric charge form factor}

Apart from some deviation at low $Q^2$, our models 
give a reasonable description of the 
lattice data in all their range, when 
lattice uncertainties are considered.

Our two models  (Sp 1 and Sp 2) gives the 
same result, because there are no 
D-state contributions for $G_{E0}$ 
and we have used $N^2=1$.

In comparison to other quark models, 
ours underestimate the charge radius, while it is very close 
to the lattice data from 
Refs.~\cite{Alexandrou07,Alexandrou07X,Alexandrou08,Alexandrou09}.
We predict, as the lattice data does, a stronger charge concentration 
at the origin for the $\Delta$
than for the proton. Still, we obtain a slightly weaker charge concentration 
than lattice QCD.
The quality of the agreement with the 
quenched lattice data increases 
for high $Q^2$, where the meson cloud contribution should be smaller.
This supports the idea that at least
the valence sector is well described 
by our covariant spectator model.
However, some corrections are expected from the
pion cloud at low $Q^2$, and
they
are not yet included 
in our valence quark model. In principle, they will
increase the charge radius.
Another effect to be added in the future is 
the inclusion of the D-state to D-state transition. 
But this effect is expected to be 
small since it is 
proportional to the D-state to D-state contributions 
of the order of $a^2$, $b^2$  and $ab$,
with $a$ and $b$ already very small.

\begin{table*}
\begin{center}
\begin{tabular}{l c  c c c c  c c c}
\hline
\hline
$G_{M1}(0)$ & & $\Delta^{++}$ & & $\Delta^+$ & & 
$\Delta^0$ & &  $\Delta^-$ \\
\hline
Exp.  \cite{Yao06,Kotulla02}
& & 7.34$\pm$2.49      & & 3.54$^{+4.59}_{-4.72}$    & &   & &    \\
SU(6)  & & 7.31               & & 3.65  & & 0  & & -3.65 \\
$\chi$QSM  \cite{Ledwig08}   & & 6.36      & &  3.09    & & -0.18  & & -3.45 \\
Large $N_c$ \cite{Ledwig08}  & &           & &  3.24    & &   & &  \\
$\chi$PT  \cite{Geng09}   & & 7.92$\pm$0.17   & &  3.73$\pm$0.03    
& & -0.47$\pm$0.12  & & -4.67$\pm$0.26 \\
Large $N_c$-$\chi$PT~\cite{Mendieta09}  && 7.07  && 3.13 && -0.82 &&   -4.77  \\
PQ$\chi$PT \cite{Arndt}      & & 7.58      & &  3.79    & &   0  && -3.79 \\
 SU(2) 
$\chi$PT \cite{Jiang09}   & &    & &      
& & -0.97  & & -5.51 \\
GBE \cite{He05}              & & 7.34      & &  3.67    & &  0  & & -3.67 \\
QCDSR \cite{Azizi08}         & & 6.34$\pm$1.50    & &  3.17$\pm$0.75    
                             & & 0         & & -3.17$\pm$0.75 \\ 
DM \cite{Machavariani08}     & & 7.16      & &  4.80    & & 
                                -3.29      & & -4.93    \\
Gauge/String Duality
\cite{Hashimoto08}           & & 5.81      & &  3.04    & &  0.27 & & -2.51\\
$(qqq)q\bar{q}$ model 
\cite{An06}    
              & & 7.66      & &          & &       & &     \\
HCQM  ($\nu=0.7$)     \cite{Thakkar10}   & & 5.93   & &  3.00   
& & 0.066  & & -2.95 \\
U-spin \cite{Slaughter09}   & &   & &    & &       & & -1.76$\pm$0.08 \\
\hline
Spectator \cite{DeltaFF}     & & 6.71      & &  3.29    & & -0.12 & & -3.54 \\
Spectator 1                  & & 6.66      & &  3.27    & & -0.12 & & -3.51 \\ 
Spectator 2                  & & 6.66      & &  3.27    & & -0.12 & & -3.51 \\
\hline
Lattice: &&   &&  &&  && \\
Background \cite{Lee05} & & 6.86$\pm$0.24 & & 1.27$\pm$0.10 & & 
-0.046$\pm$0.003 & & -3.90$\pm$0.25 \\
$\chi$-extrapolation
\cite{Cloet03}$^a$  & &    6.54$\pm$0.73   & &   3.26$\pm$0.35   
& &  0.079  & &  -3.22$\pm$0.35   \\
Quenched (ext) \cite{Alexandrou07,Alexandrou07X}$^a$ &&  
&& 3.04$\pm$0.21 && && \\
Quenched Wilson \cite{Alexandrou08,Alexandrou09} & &  & &
   2.635$\pm$0.094 & &  & & \\
Dynamical Wilson \cite{Alexandrou08,Alexandrou09} & & & &
   2.35$\pm$0.16 & &  & & \\
Hybrid  \cite{Alexandrou08,Alexandrou09} & &   & &
   3.05$\pm$0.24 & &  & &  \\
Quenched \cite{Boinepalli09}  
 & & 5.28$\pm$0.92  & & 2.64$\pm$0.46 && 
                             0              & & -2.64$\pm$0.46 \\
Background \cite{Aubin}$^b$  
& & 4.85$\pm$0.16 & & 3.15$\pm$0.08 & & 
0.0013$\pm$0.0210  & & 
-2.42$\pm$0.08 \\
\hline
\hline
\end{tabular}
\end{center}
\caption{Summary of recent results for $G_{M1}(0)$
compared with the experimental data and the SU(6) result.
A compilation of earlier results can be found in 
Ref.~\cite{DeltaFF}. 
To obtain $\mu_\Delta$ in $\mu_N$ units (nucleon magneton)
use $\mu_\Delta = G_{M1}(0) \sfrac{M_N}{M_\Delta} \mu_N$. \\ 
$^a$ Extrapolation for the physical point;  
$^b$ Ref.~\cite{Aubin} uses 
$\mu_{\Delta^-}=-\sfrac{1}{2}\mu_{\Delta^{++}}$.  \\
The lattice results 
from Refs.~\cite{Alexandrou08,Alexandrou09,Boinepalli09,Aubin}
include no extrapolation to the physical point. 
The lattice results correspond respectively to 
$m_\pi=411$ MeV for quenched Wilson, 
$m_\pi=384$ MeV for the dynamical Wilson and 
$m_\pi=353$ MeV for the Hybrid action \cite{Alexandrou08,Alexandrou09};
$m_\pi=306$ MeV for quenched from Ref.~\cite{Boinepalli09}
and $m_\pi=366$ MeV from Ref.~\cite{Aubin}.}
\label{tableGM1}
\end{table*}

\begin{figure*}[t]
\vspace{.15cm}
\centerline{
\mbox{
\includegraphics[width=3.3in]{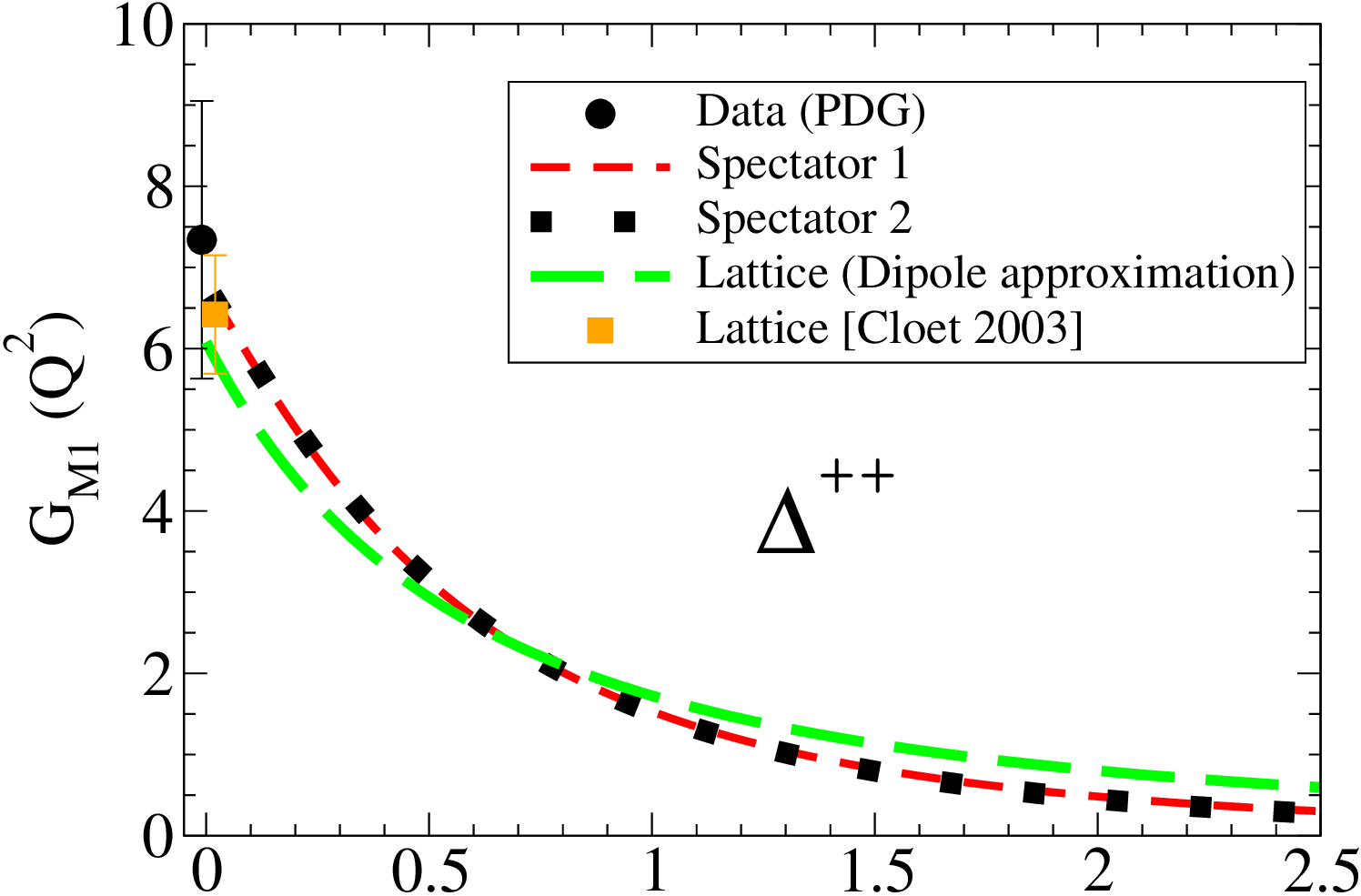} \hspace{.40cm}
\includegraphics[width=3.2in]{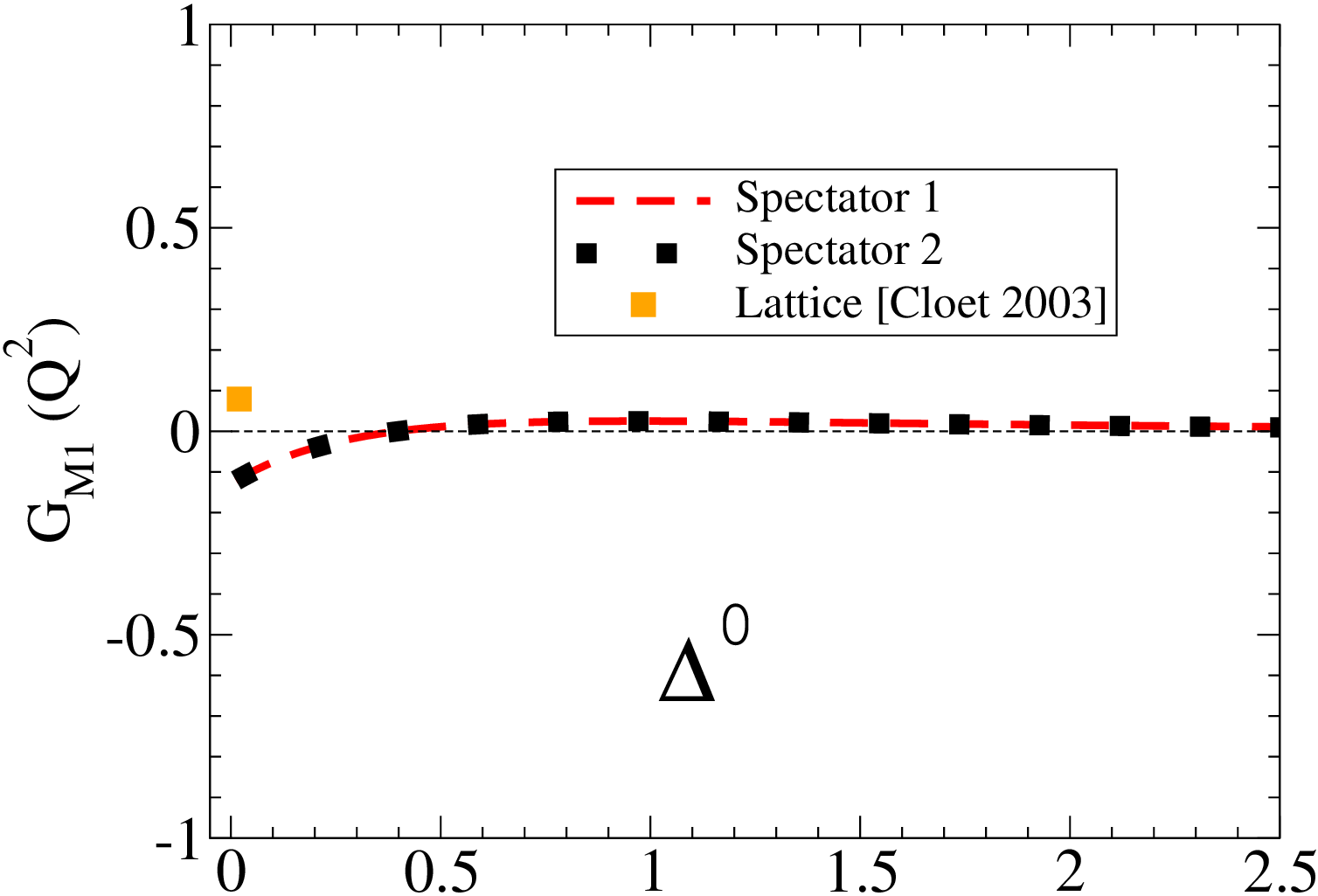}}}
\vspace{0.9cm}
\centerline{
\mbox{
\includegraphics[width=3.35in]{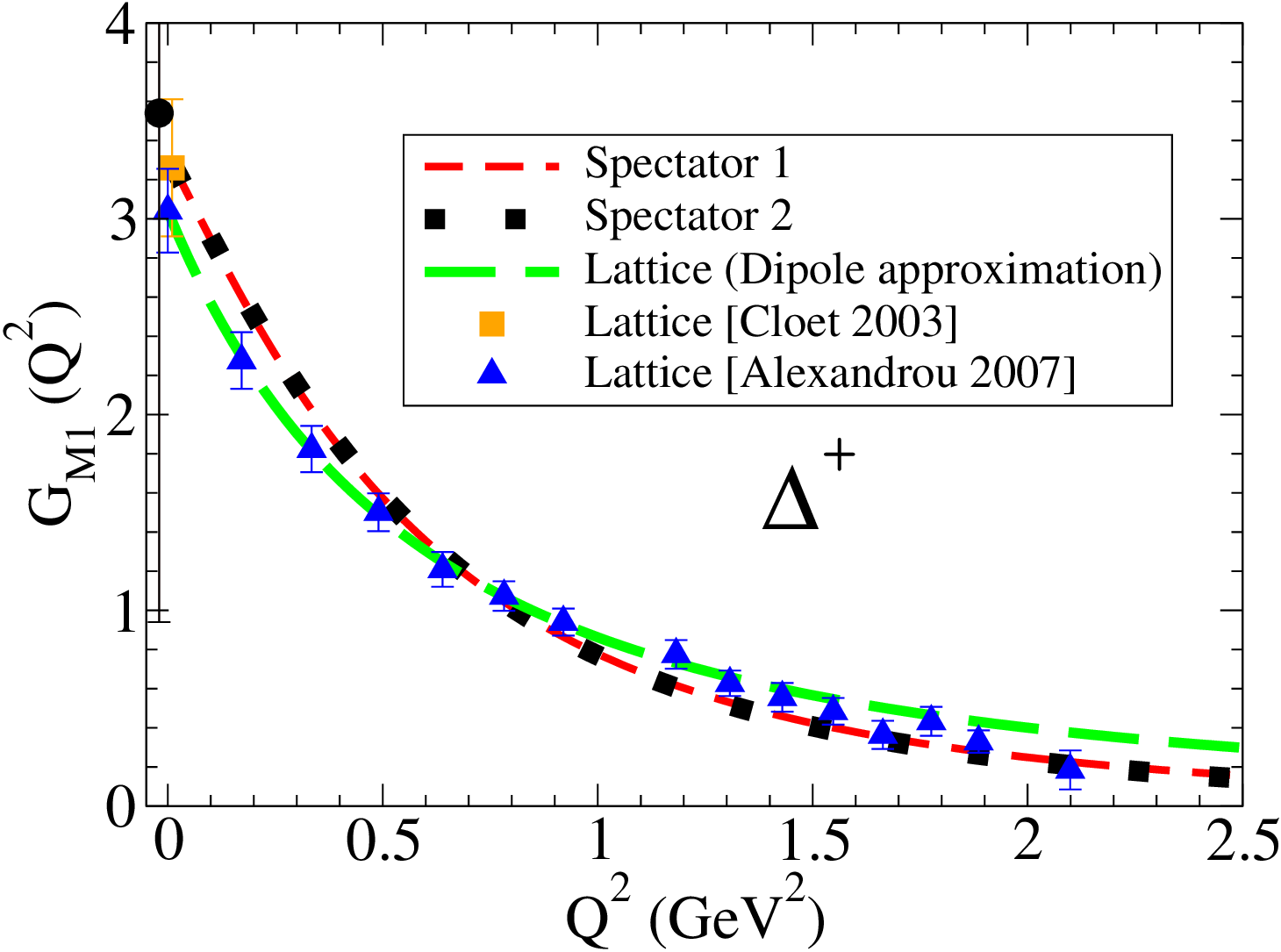} \hspace{.40cm}
\includegraphics[width=3.2in]{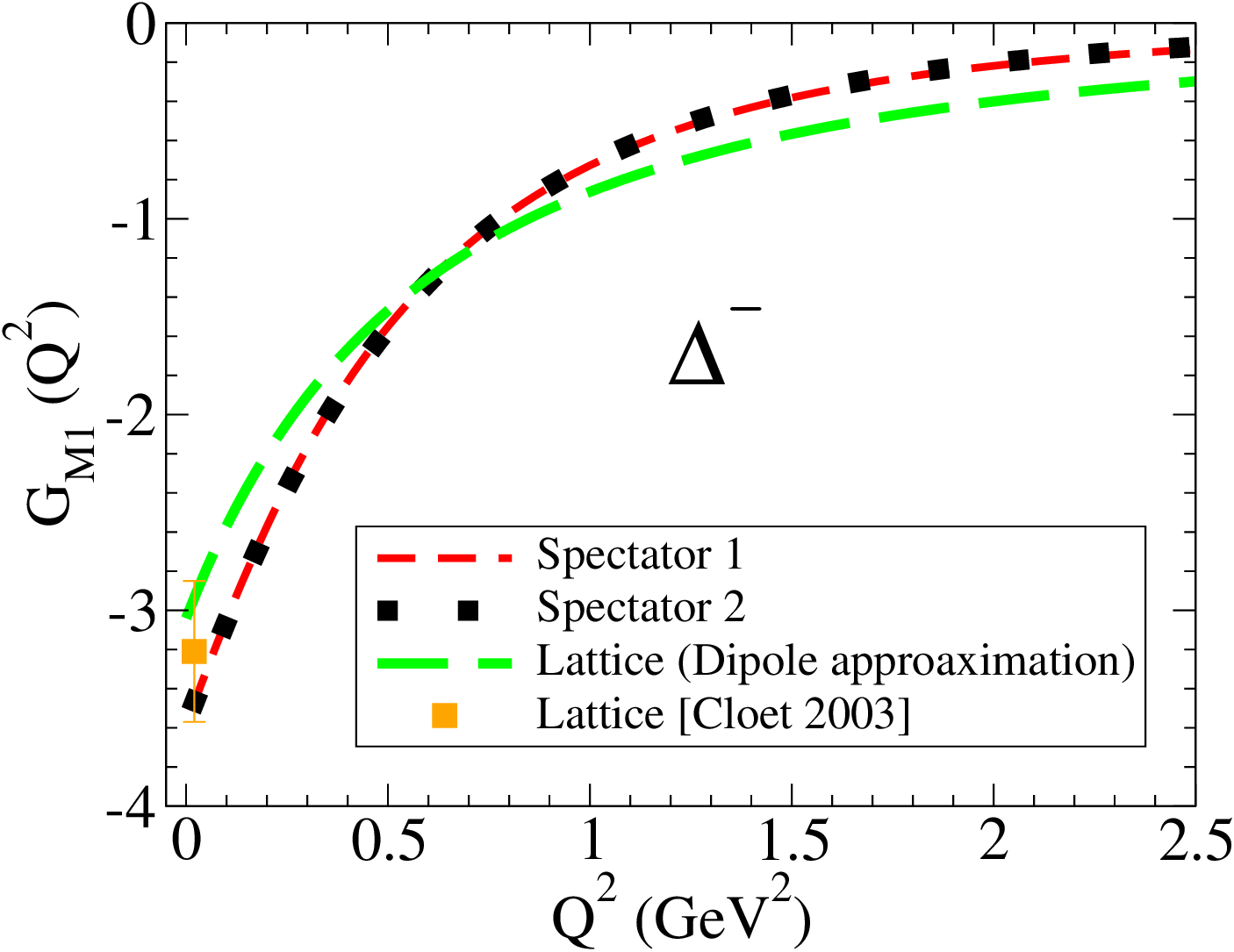}}}
\caption{\footnotesize{
$G_{M1}$ form factor. 
Quenched lattice data from Ref.~\cite{Alexandrou07,Alexandrou07X}.
For $\Delta^+$ we include the experimental 
result from Ref.~\cite{Kotulla02}:
$G_{M1}(0)= 3.54^{+2.37}_{-2.60}(\mbox{stat+sys})
\pm 3.94(\mbox{theor})$ \cite{Kotulla02}.
The theoretical uncertainty for the $\Delta^+$ case is not presented in the figure.}}
\label{figGM1}
\end{figure*}

\begin{figure}
\centerline{
\mbox{
\includegraphics[width=3.2in]{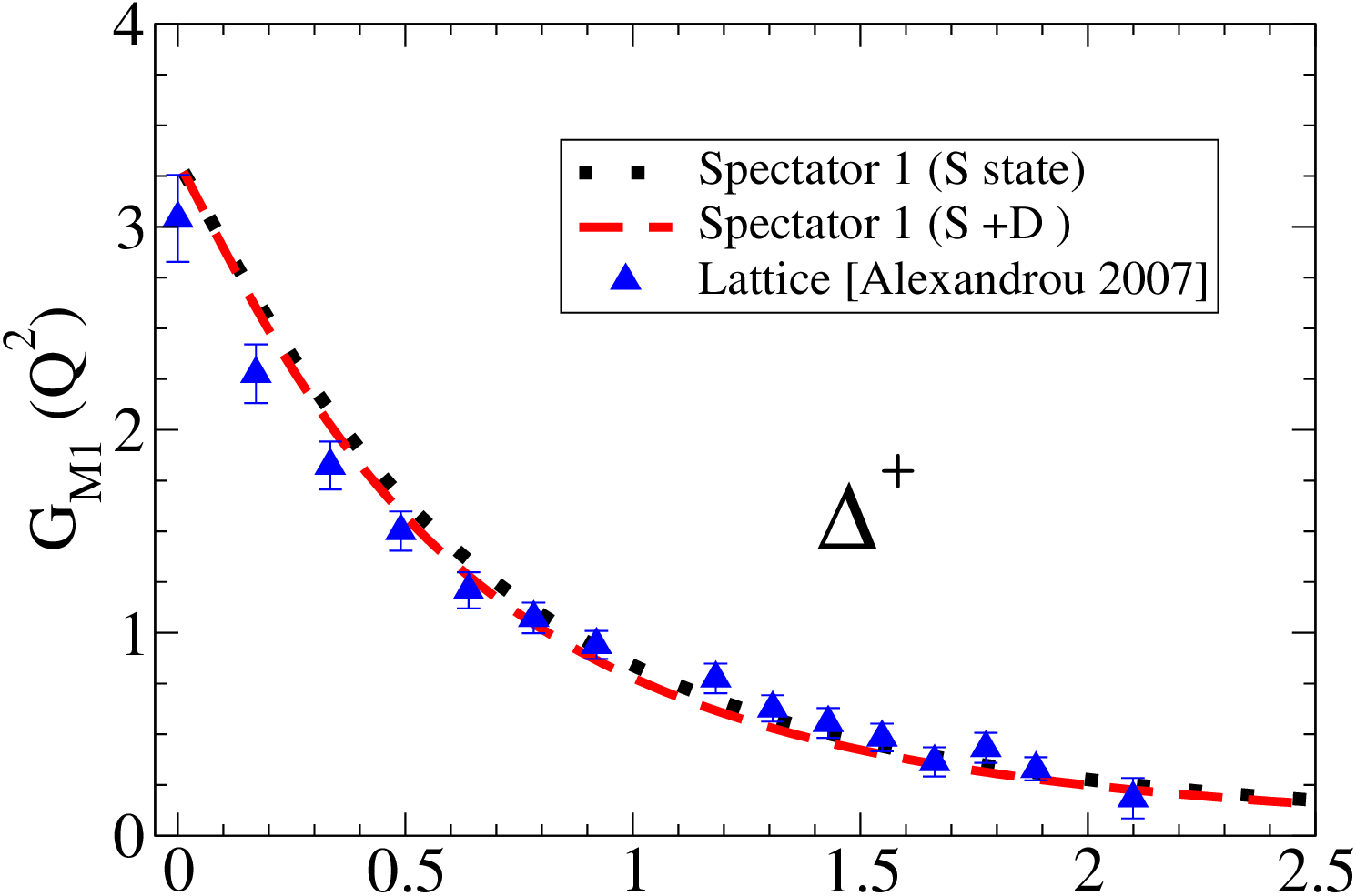}}}
\vspace{1.0cm}
\centerline{
\mbox{
\includegraphics[width=3.2in]{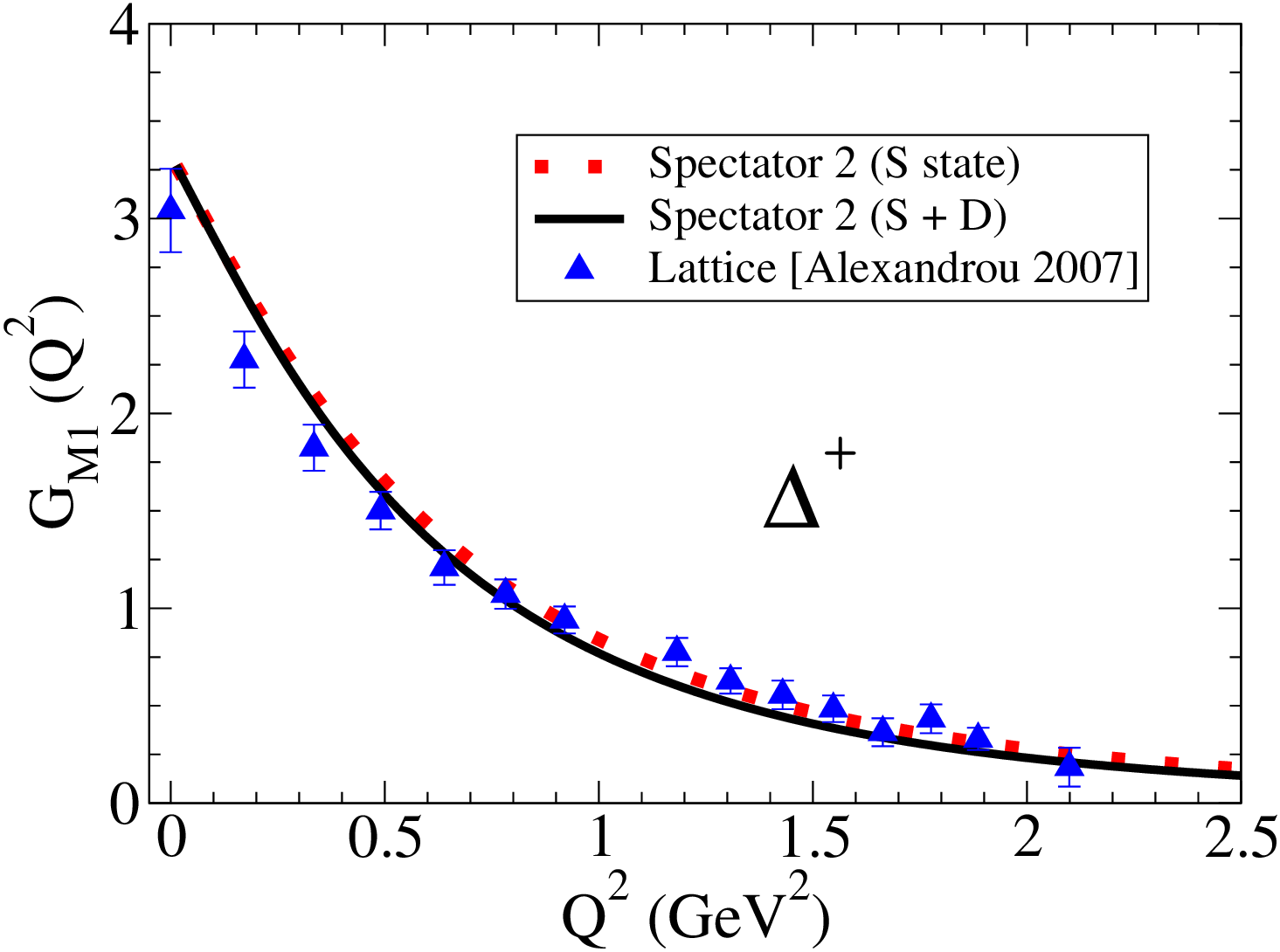} }}
\caption{\footnotesize{
Comparing the S-state approach with the effects of the D-states 
in model Sp 1 and Sp 2. Lattice data from 
Ref.~\cite{Alexandrou07}.}}
\label{figGM1SD}
\end{figure}

\begin{table*}
\begin{center}
\begin{tabular}{l c l  c  c c c c c  c c c}
\hline
\hline
$<r_{M1}^2>$ & & $r_{Mp}^2$ && $\Delta^{++}$ && $\Delta^+$ && 
$\Delta^0$ &&  $\Delta^-$ \\
\hline
MIT bag model \cite{Gobbi92} &&0.38  &&   0.46   &&       &&         && \\
Skyrme \cite{Schwesinger92}  &&       &&   0.47   && 0.39  && -0.16   && 0.71 \\
QSM \cite{Gobbi92}      
&& 0.61 &&  0.60 &&        &&          &&      \\
$\chi$QSM  \cite{Ledwig08}  && 0.656  &&         & & 0.634    & &   & &  \\
$\chi$QSM SU(3) \cite{Ledwig08} && 0.665 &&     && 0.658    & &   & &  \\
CQM (imp) \cite{Wagner00}        & & &&    0.656     && 0.656  && 0
                                 & &   0.656 \\     
CQM \cite{Wagner00}            &&  && 0.623     && 0.623  && 0
                               &&   0.623 \\  
GBE \cite{He05}       &&   0.752     &&     & & 0.689    & &  0  & &  \\
\hline
Spectator \cite{DeltaFF}  &&   && && 0.37 && && \\
Spectator 1 && 
&&  0.32  &&    0.30      & &  1.16   & &  0.36 \\
Spectator 2 && 
&&  0.31   &&    0.30      & &  1.15  & &  0.35 \\
\hline
Lattice: &&  &&  &&  &&  && \\
Quenched \cite{Alexandrou07,Alexandrou07X}$^a$ && && & & 0.412$\pm$0.049 && && \\
Quenched Wilson \cite{Alexandrou08,Alexandrou09} &&  && &&
  0.240$\pm$0.023 & &  & & \\
Dynamical Wilson \cite{Alexandrou08,Alexandrou09} && &&  &&
  0.230$\pm$0.017 & &  & & \\
Hybrid  \cite{Alexandrou08,Alexandrou09} &&   &&  &&
  0.250$\pm$0.033 & &  & &  \\
Quenched \cite{Boinepalli09} &&  && 0.410$\pm$0.057
 && 0.410$\pm$0.057 &&  0 && 
                                      0.410$\pm$0.057 \\      
\hline
\hline
\end{tabular}
\end{center}
\caption{Summary of existing theoretical and lattice results 
for $<r_{M1}^2>$ (fm$^2$).
Ref.~\cite{Boinepalli09} the electric and magnetic 
radius are assumed to be the same. 
For the nucleon we can consider the 
result associated with the dipole form $\Lambda^2=$ 0.71 GeV$^2$.
The proton magnetic radius in 
the Spectator models is $r_{Mp}^2=0.74$ fm$^2$ as 
extracted from Ref.~\cite{Nucleon} (model II). \\
$^a$ Extrapolation for the physical point.}
\label{tableRM1}
\end{table*}

\subsection{Magnetic dipole form factor}
\label{secGM1}

The magnetic dipole form factor results are
shown in Fig.~\ref{figGM1}, 
for the four charged $\Delta$ states.
Only the magnetic moments of $\Delta^{++}$ and $\Delta^+$ 
are experimentally known, and their data points are represented 
in the graphs for $Q^2=0$, according 
to  $\mu_\Delta = G_{M1}(0) \sfrac{e}{2 M_\Delta}$  
\cite{SIN,UCLA,Exp,Yao06,Castro01,Kotulla02}. 
The extrapolation of the quenched lattice QCD data to 
the physical point exists only for the $\Delta^+$ case.
In that case the dependence on $Q^2$ 
of our results is directly compared 
to that extrapolation.
For the other charge cases, in order to compare with the lattice QCD 
data we use the relation 
based on exact isospin symmetry  
\be
G_{M1}^{\Delta^{++}} (Q^2)= 2 G_{M1}^{\Delta^+} 
(Q^2)= - 2 G_{M1}^{\Delta^-} (Q^2),
\label{eqGMiso}
\ee
together with the dipole approximation  for the  
$G_{M1}^{\Delta^+}(Q^2)$ lattice data. 
The result extracted from the lattice 
data \cite{Leinweber92} using $\chi$PT \cite{Cloet03}
for $Q^2=0$ is also shown in the graphs.

Our predictions for $\Delta^+$ 
are very close to the lattice data. 
In detail, our results are 
7\% above the lattice data for low $Q^2$.  
However, the lattice 
errorbars are larger than the ones for the $G_{E0}$  
calculation, and our results
are close to the limit of the errorbars.
For high $Q^2$ the agreement is very good.  
In the limit $Q^2=0$ our results are 
consistent with both the experimental points 
and with the lattice data extrapolation from 
Refs.~\cite{Cloet03,Alexandrou07,Alexandrou07X}.
As found before in the calculation with S-waves only  \cite{DeltaFF}, 
the 
exception to this good agreement occurs in the 
$\Delta^0$ case, where 
\cite{Cloet03} predicts a sign different from ours.
This disagreement is not problematic, 
since in that work the theoretical 
uncertainty was not explicitly evaluated,
and $\mu_{\Delta^0}$ 
is expected to be smaller relatively to the
charged cases.

In Fig.~\ref{figGM1SD} we  compare 
the effect of the D-states to $G_{M1}$,
for both models Sp 1 and Sp 2.
We conclude that
the effects of the D-states are small
although they increase with $Q^2$.
The small magnitude of these effects
is easily understood from Eq.~(\ref{eqGM1a}) 
with small admixture coefficients.

\subsubsection{Magnetic moment: comparison with other models}


Our predictions for the magnetic moments are 
$\mu_{\Delta^{++}}=5.08 \mu_N$ and  $\mu_{\Delta^+}=2.49 \mu_N$ 
for both Sp 1 and Sp 2.
The reason why both models lead to exactly the same 
result is that $G_{M1}(0)$ is determined 
only by the S-state parameterization,
which is the same for the two models.
There are some discrepancies between 
the different experiments, and the Particle Data Group \cite{Yao06} 
reports the interval of  3.7-7.5 for 
the $\Delta^{++}$ magnetic moment.
Our results are in good agreement  
with that result. 
Also the result for ${\Delta^+}$ is 
consistent if the isospin symmetry is used to estimate 
the experimental value of $\mu_{\Delta^+}$. 

The magnetic moment of $\Delta^+$ is traditionally 
compared with the proton magnetic moment $\mu_p$.
In the heavy quark limit (static) approximation they coincide.
The SU(6)  
prediction $G_{M1}(0)=3.65$ \cite{Beg64}
implies that, 
when expressed in the nucleon magneton units ($\mu_N= \sfrac{e}{2M_N}$),
the $\Delta^+$ and the proton have the same magnetic moment
[$\mu_{\Delta^+}= 3.65 \, \sfrac{e}{2M_\Delta}= 
3.65 \sfrac{M_N}{M_\Delta} \sfrac{e}{2M_N} =2.79 \mu_N$].
Our results for the $\Delta^+$ are $G_{M1}(0)=3.29$ in both models.

In a previous work \cite{DeltaFF},
we compared our results for $G_{M1}(0)$ without D-waves
with several formalisms,
namely relativistic quark models \cite{Schlumpf93},
QCD sum rules \cite{Lee98,Aliev00},
chiral and soliton models \cite{Kim,Butler94}
and dynamical reaction models (DM) based 
on hadron degrees of freedom 
\cite{Pascalutsa05,Chiang05}. 
Now we compare our calculations, with 
D-waves included, with a comprehensive update of the more 
recent results based on  
$\chi$QSM \cite{Ledwig08},
large-$N_c$ \cite{Ledwig08},
SU(2) $\chi$PT  \cite{Jiang09},
$\chi$PT \cite{Geng09},
large $N_c$-$\chi$PT~\cite{Mendieta09}, 
PQ$\chi$PT \cite{Arndt}, 
GBE \cite{He05}, QCD sum rules \cite{Azizi08}, 
a quark models with sea quark contributions \cite{An06},
a hypercentral quark model (HCQM) \cite{Thakkar10},
a Gauge/String Duality model \cite{Hashimoto08},
a recent DM \cite{Machavariani08} 
and  the U-spin symmetry \cite{Slaughter09}. For completeness.
to this list we add the CQM with one
and two-body currents \cite{Buchmann97}.  
The predictions for $G_{M1}(0)$ from all these works, 
together with our own results, 
are displayed in
Table \ref{tableGM1}.
The experimental and the SU(6) results are also included.
Finally, we present the lattice 
QCD results
\cite{Leinweber92,Cloet03,Lee05,Alexandrou07,Alexandrou08,Alexandrou09,Boinepalli09}.

In general, all the different 
models are consistent with the available experimental information. 

The PDG interval for $G_{M1}(0)$ corresponds to 
$\Delta^{++}$ and is $4.8-9.8$.
One has $2.4-4.9$ for $\Delta^{+}$ assuming the isospin symmetry.
The symmetric interval hold for $\Delta^{-}$, in the same approximation.
Only Refs.~\cite{Slaughter09,Jiang09} for $\Delta^-$ are 
clearly out of the interval.

As for the $\Delta^0$ case, SU(6) 
and exact isospin symmetry predicts $G_{M1}(0)=0$, 
and our models, and in general all models
predict contributions 
an order of magnitude smaller than the magnetic 
moment of the charged $\Delta$ or even zero
(excluding the results from DM~\cite{Machavariani08},
SU(2) $\chi$PT~\cite{Jiang09}
$\chi$PT~\cite{Geng09} and 
large $N_c$-$\chi$PT~\cite{Mendieta09}).
When $\mu_{\Delta^0}$ is not zero
a negative value is frequently obtained, with 
a few exceptions. 
Note that the measurement 
of $\mu_{\Delta^0}$
is technically extremely difficult, 
since on top of the very short $\Delta$ lifetime,
there is in addition the problem of tracking neutral particles.
For all these reasons our knowledge of $\mu_{\Delta^0}$ 
will probably be restricted in the near future 
to values from theoretical models and lattice simulations.

\subsubsection{Magnetic moment: comparison with the lattice QCD results}

Since the lattice simulations of the $\Delta$
magnetic moment are performed 
for heavy quark masses,
to compare with 
phenomenological models the extrapolation
to the physical point is needed.

Fortunately, there are nowadays 
methods based on $\chi$PT that 
can be used for such extrapolations 
\cite{Pascalutsa05,Cloet03,Alexandrou08}. 
The magnetic moment from Ref.~\cite{Cloet03} is the result
of a chiral extrapolation.
References \cite{Lee05,Alexandrou07} use a simple 
functional dependence on the pion mass to 
extrapolate to the physical point.
All the other lattice results
in Table \ref{tableGM1} refer to pions masses larger 
than the physical one.

\begin{figure}[t]
\centerline{
\mbox{
\includegraphics[width=3.2in]{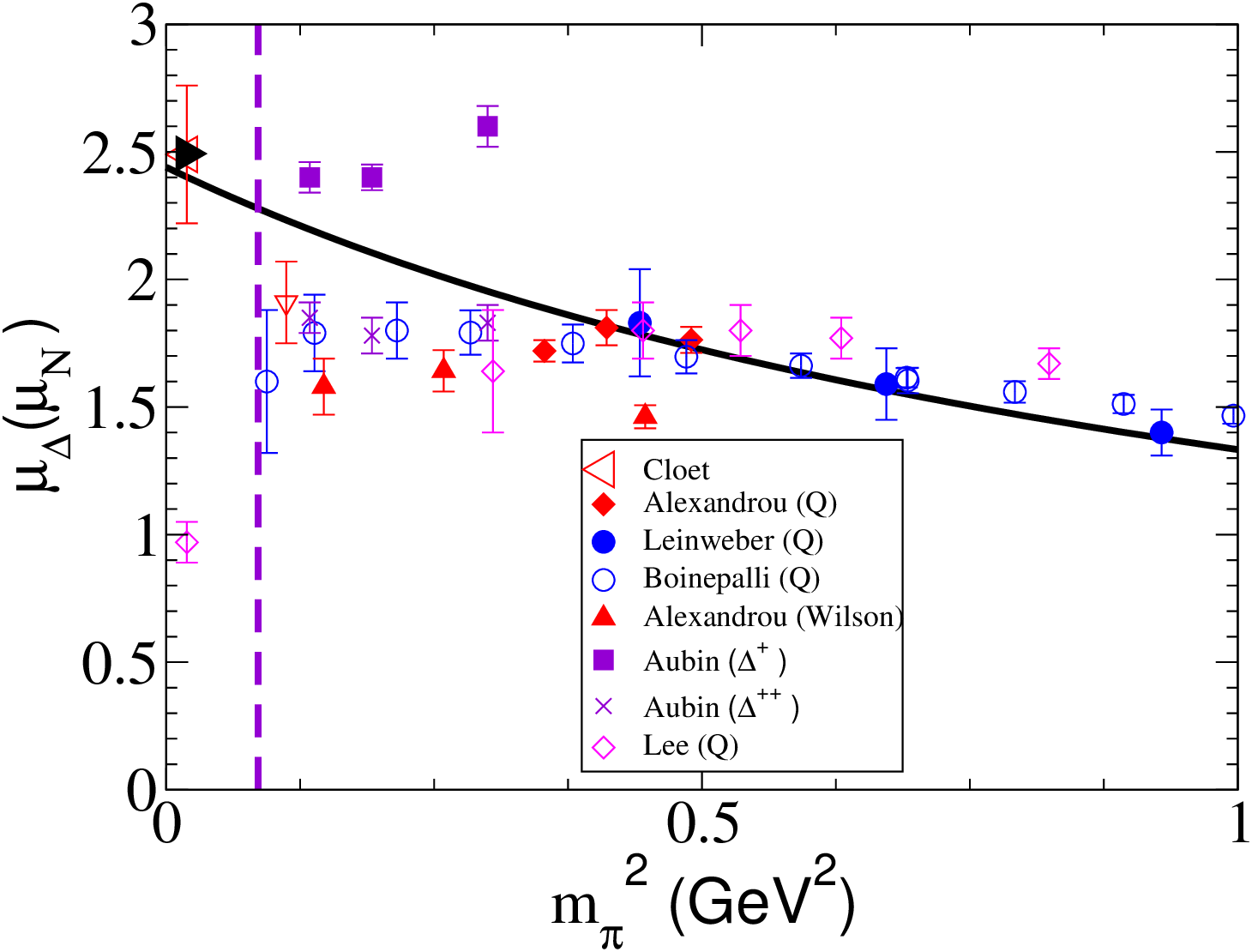}}}
\caption{\footnotesize{
$\Delta^+$ 
magnetic moment from lattice calculations.
Quenched lattice data (labeled Q) from 
\cite{Leinweber92,Lee05,Alexandrou07,Alexandrou07X,Alexandrou08,Alexandrou09}.
Unquenched lattice data from 
\cite{Alexandrou08,Alexandrou09,Aubin}.
The solid line is the analytical contribution 
for $\mu_{\Delta^+}$ as derived in Ref.~\cite{Cloet03}. 
The result of Sp 1 and Sp 2 
(filled $\triangleright$) is also represented.
}}
\label{figMuLat}
\end{figure}

The dependence of the $\Delta^+$ magnetic moment
on the pion mass is presented in Fig.~\ref{figMuLat} 
in addition to the result from models 
Sp 1 and Sp 2 at the physical point.
The lattice data in that figure comprises 
the quenched results from 
\cite{Leinweber92,Lee05,Alexandrou08,Alexandrou09,Boinepalli09},
and the unquenched results from 
\cite{Alexandrou08,Alexandrou09} obtained with the Wilson and hybrid actions 
\cite{Aubin} using Clover fermions  
and also 
the background field method \cite{Bernard82,Lee05,Aubin}. 
The chiral extrapolations of Cloet~\cite{Cloet03} 
using the pion loop corrections, 
and Lee~\cite{Lee05} using a simple analytical 
form, is also on the graph.
The inelastic point ($m_\pi < M_\Delta-M_N$) is
represented by the vertical line.
The solid line in the figure 
corresponds only to the 
analytical part of the  chiral extrapolation from 
Cloet \etal~\cite{Cloet03}, but the analytical line  
and the full result
lead to similar results for $\mu_{\Delta^+}$ 
at the physical point \cite{Cloet03}
with an accuracy better that 10\%.
The extrapolation given by  Cloet \etal~compares well
with our results.

The main conclusion from the figure is that there are 
large discrepancies between several lattice
calculations. 
This makes harder to draw conclusions from the comparison of our results with
the available lattice data.
The unquenched results from Aubin \etal~\cite{Aubin} for the $\Delta^+$ 
exceed all the other calculations, 
including the analytical contribution to 
the chiral extrapolation \cite{Cloet03}.
Similarly to Ref.~\cite{Lee05},
in the Aubin results
a significant deviation of the 
isospin symmetry (see Eq.~(\ref{eqGMiso})) is manifest
in the $\Delta^{++}$ and $\Delta^+$  predictions.
Note however that the Aubin results 
for $\sfrac{1}{2} \mu_{\Delta^{++}}$,
while different from the $\Delta^+$ results, 
are similar to the results from other calculations.
The significant violation of the condition
$\sfrac{1}{2} \mu_{\Delta^{++}}= \mu_{\Delta^+}$,
shows that the Aubin calculation violates strongly 
isospin symmetry, given by  Eq.~(\ref{eqGMiso}).

To understand the discrepancies between the different lattice calculations,
it is important to realize that besides the extrapolation to the physical point,
certain lattice calculation of $G_{M1}$ demand
an additional extrapolation of $Q^2$ down to the $Q^2=0$ point, given that
the minimum lattice momentum $Q^2$ is non-zero.
The most common method uses 
a given analytical form for the
extrapolation.
A dipole \cite{Alexandrou07}, or an exponential 
falloff form \cite{Alexandrou08,Alexandrou09} have been used.
Another possibility, used by the Adelaide group
\cite{Leinweber92,Boinepalli09,Boinepalli06}, 
applies the scaling between the electric charge form factor 
and the magnetic dipole form factor,
$\sfrac{G_{M1}(Q^2)}{G_{M1}(0)}= \sfrac{G_{E0}(Q^2)}{G_{E0}(0)}$.
This assumption implies that the magnetic dipole radius squared 
is the same as the electric charge radius.
Note, however that, 
although the scaling approximation 
is valid for the proton form factors \cite{Nucleon},
there is no reason to believe that it holds 
also for the $\Delta$ system.

As a complete different alternative, the background field method 
\cite{Bernard82,Lee05,Aubin} takes the three-quark system interacting  
at rest ($Q^2=0$)  with a static magnetic field.
Considering the energy shift in the system,
which is proportional to the magnetic 
dipole and also with the magnetic field, 
the magnetic moment is determined 
avoiding the uncertainty present in the extrapolation
of the form factor method. The different results of \cite{Aubin} 
may be a consequence
of the combination of a strong magnetic field with
a not so large lattice volume used in the numeric lattice simulations.

The other problem to consider when comparing between 
and to different lattice calculations lies in estimating how 
the quenched approximation deviates from the results
from the full QCD calculation, with the same pion mass. 

In the SU(6) limit (exact SU(3) flavor symmetry) gives
$\mu_{\Delta^+}= \mu_p$.
This limit is obtained in quenched QCD  
when and light and strange quarks
have the same mass,  i.e.  $m_K=m_\pi$, with $m_K$ the kaon mass.
In a recent quenched simulation \cite{Boinepalli09}
the SU(3) symmetry point corresponds to 
$m_\pi \simeq 700$ MeV ($m_\pi^2 = 0.485$ GeV$^2$) \cite{Boinepalli09}.
Below this point ($m_\pi < 700$ MeV) 
$\mu_p > \mu_{\Delta^+}$. 
This result can be understood analyzing 
the pion cloud contributions in 
quenched and in full QCD.
As explained in Ref.~\cite{Boinepalli09}, 
considering again $\Delta^+$, for $m_\pi < 700$ MeV, 
those contributions are negative in quenched approximation
while positive in full QCD.

Therefore, for $\Delta^+$ in the region $m_\pi < 700$ MeV,
the quenched result underestimates 
not only full QCD, but even the 
the core contribution, which  
is already below the full QCD result.
This is an artifact of the quenched approximation also observed in Refs.~\cite{Lee05,Young}.

As for $\mu_{\Delta^0}$, the extrapolations 
of Cloet \cite{Cloet03} 
and Lee \cite{Lee05} give different signs.
The Aubin result for $m_\pi = 366$ MeV
\cite{Aubin}  is  positive but the 
errorbars are so large that it can also be consistent
with zero and negative values. 
Note that the results of Lee 
\cite{Lee05} are quenched 
and  the results of Aubin \cite{Aubin} are unquenched.

\begin{figure}[t]
\vspace{.4cm}
\centerline{
\mbox{
\includegraphics[width=3.3in]{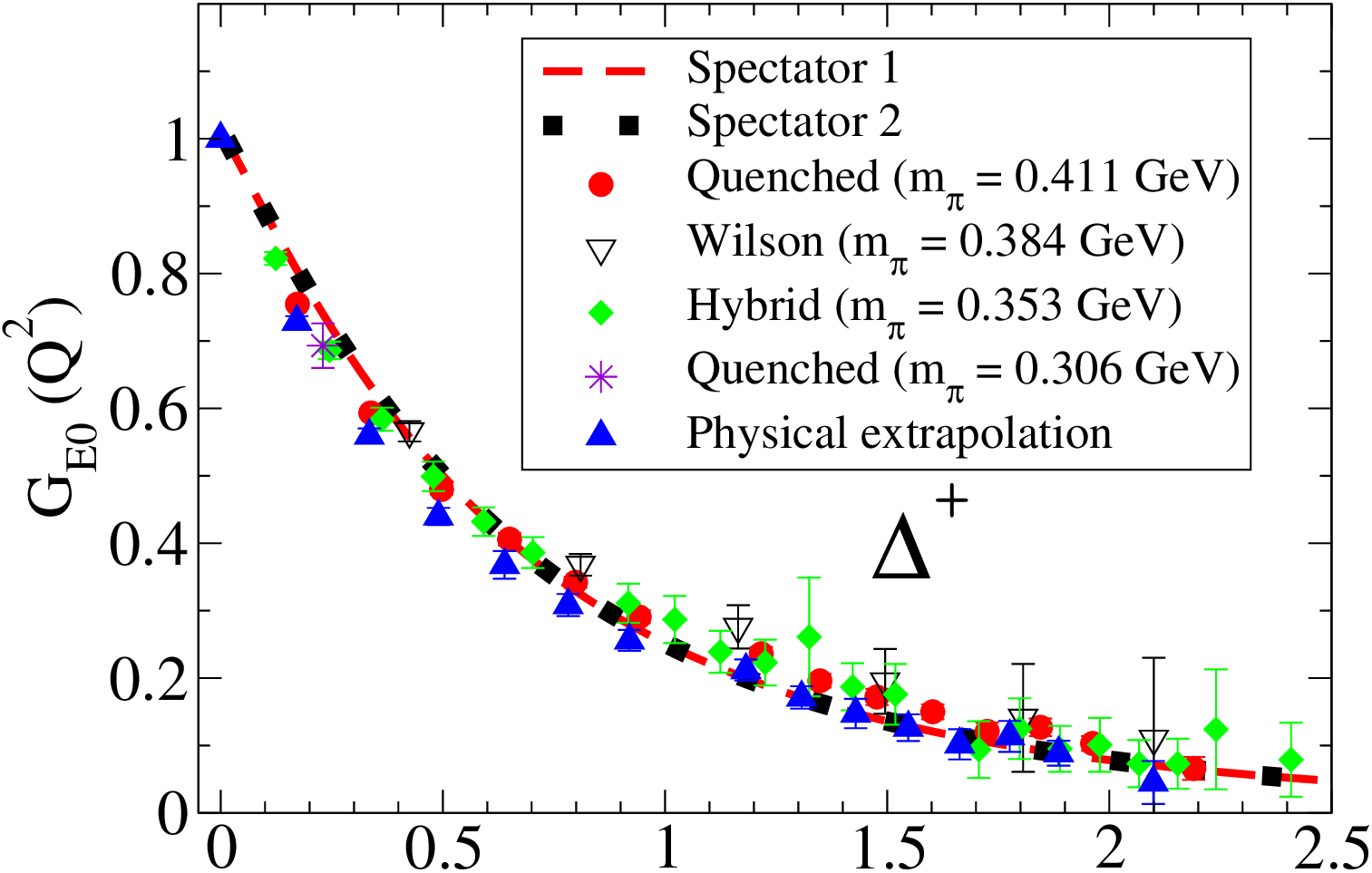}}}
\vspace{1.00cm}
\centerline{
\mbox{
\includegraphics[width=3.2in]{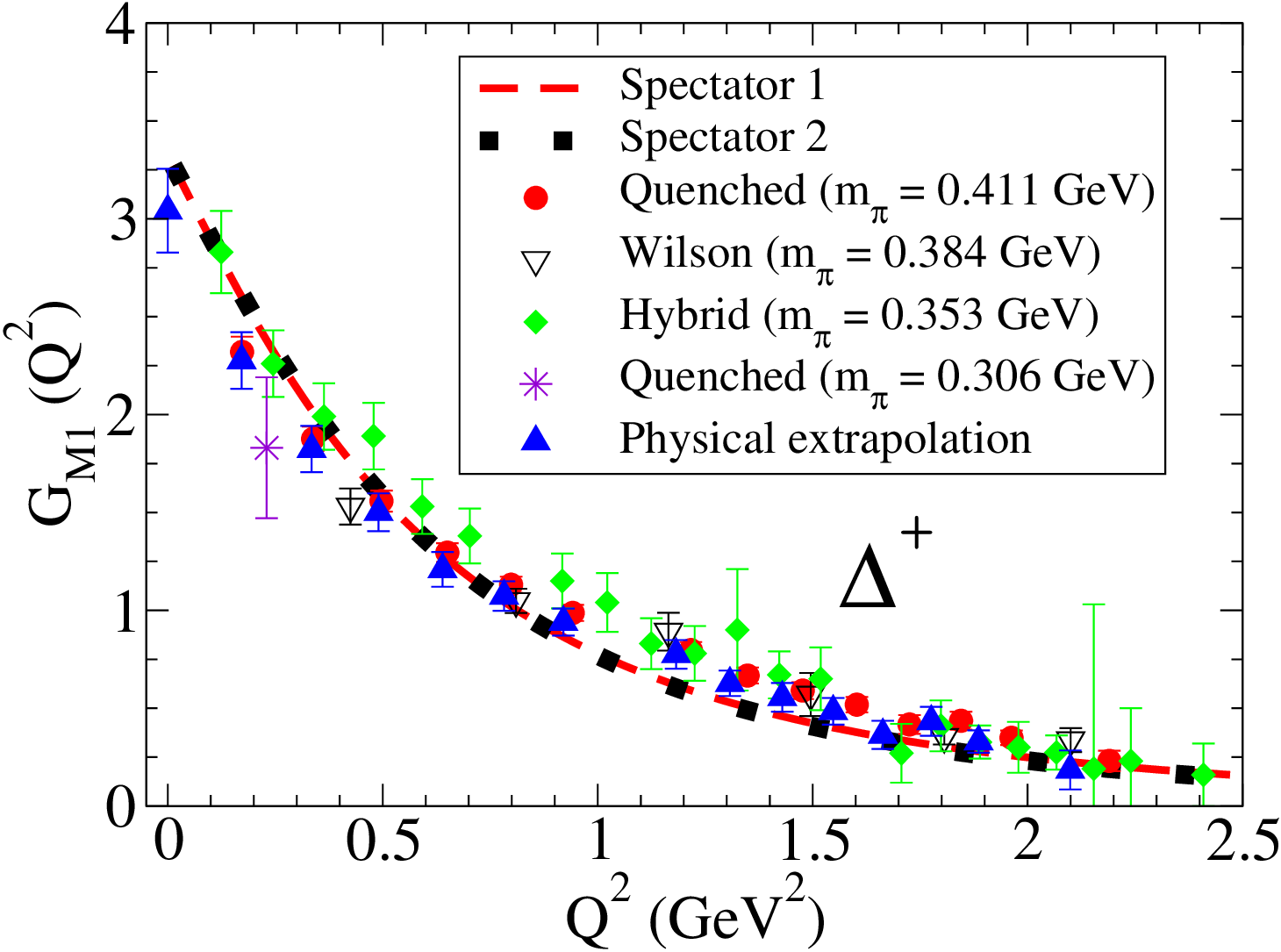} }}
\caption{\footnotesize{
Lattice data from Refs.~\cite{Alexandrou07,Alexandrou09,Boinepalli09}.
underestimates the other results. 
}}
\label{figGLat}
\end{figure}

\subsubsection{Magnetic dipole radius}

We define the magnetic dipole radius squared $<r_{M1}^2>$
by means of the expansion of $G_{M1}$ specified analogously 
to Eq.~(\ref{eqRE02}).
The results of the 
magnetic dipole radius squared are listed on Table \ref{tableRM1} for  
different models.
At low $Q^2$ they
are similar to the ones obtained 
for the electric charge radius, 
suggesting a scaling between the
electric charge and magnetic dipole 
form factors, as 
it was already observed for the proton form factors.

In the comparison between our results and others, the main difference is seen
in the $\Delta^0$ magnetic 
dipole radius, where we obtain a value larger 
than the radii predicted by all models discussed till this point.
This can be just a consequence 
of the small value of $G_{M1}$ for $Q^2=0$.
As it happened for the electric radius, 
the results for the magnetic dipole radius of our two models, 
as well as the lattice QCD data, 
underestimate the results 
obtained with other quark models
(see results of Refs.~\cite{Ledwig08,Wagner00,He05}).
This can be a consequence of not 
including the pion cloud,
expected to be important at low $Q^2$.

\subsubsection{Overview of the magnetic dipole form factor}

Our results
indicate that the S-state contributions are 
dominant for the $\Delta$ 
magnetic dipole 
form factors. In the previous section we saw that the same holds for the
electric charge form factor.
The D-states contribute only with small corrections. 
In spite of pion cloud effects  
not being included explicitly here, we predict magnetic dipole 
distributions similar 
to the lattice results.
When compared with other quark models 
we predict 
a larger concentration 
of the magnetic dipole 
distribution at the origin.

\begin{table*}
\begin{center}
\begin{tabular}{l c  c c c c  c c c}
\hline
\hline
$G_{E2}(0)$ & & $\Delta^{++}$ & & $\Delta^+$ & & 
$\Delta^0$ & &  $\Delta^-$ \\
\hline
NRQM (Isgur)  \cite{Isgur82,Krivoruchenko91}  
&& -3.82 &&  -1.91 && 0 && 1.91 \\
NRQM 
\cite{Krivoruchenko91}  && -3.63    && -1.79   && 0 && 1.79 \\
Skyrme \cite{Kroll94}  &&  -3.39  &&  -1.21    && 0.94     && 3.12     \\    
Buchmann (imp) \cite{Buchmann97}  & & -2.49 && -1.25 && 
  0  &&  1.25 \\
Buchmann  (exc) \cite{Buchmann97}  & & -9.28 && -4.64 && 
  0  &&  4.64 \\
$\chi$PT \cite{Butler94}    & &  -3.12$\pm$1.95 && -1.17$\pm$0.78 &&
                                 0.47$\pm$0.20  && 2.34$\pm$1.17 \\ 
$\chi$PT  \cite{Geng09}   & & -1.05$\pm$1.29   & &  -0.94$\pm$0.58    
& & -0.86$\pm$0.94  & & 0.78$\pm$0.78 \\
 QMCM \cite{Leonard90} & & -2.34  && -0.81 && 0.70 && 2.22 \\
$\chi$QSM  \cite{Ledwig08}  & &        & & -2.15    & &   & &  \\
QCDSR \cite{Azizi08} & & -0.0452$\pm$0.0113 && 
 -0.0226$\pm$0.0057 & & 0  && 0.0226$\pm$0.0057 \\
GP(QCD)
\cite{Blanpied01} && && -7.02$\pm$4.05  &&    && \\
\hline
Spectator 1 & &  -4.08   & &  -2.04   & & 0  & &  2.04 \\ 
Spectator 2 & &  -3.41   & &  -1.71   & & 0  & &  1.71 \\ 
\hline
Lattice: &&   &&  &&  && \\
Quenched \cite{Leinweber92} & & -0.7$\pm$2.8 & & -0.4$\pm$1.4 & &
                                0            & & 0.4$\pm$1.4 \\
Quenched Wilson  
\cite{Alexandrou09} & &  & &
  -0.81$\pm$0.29 & &  & & \\
Dynamical Wilson  \cite{Alexandrou09} & & & &
  -0.87$\pm$0.67 & &  & & \\
Hybrid  \cite{Alexandrou09} & &   & &
  -2.06$^{+1.27}_{-2.35}$ & &  & &  \\
\hline
\hline
\end{tabular}
\end{center}
\caption{Summary of existing theoretical and lattice results 
for $G_{E2}(0)$.  
To obtain the electric quadrupole moment 
use $Q_\Delta= G_{E2}(0) \sfrac{e}{M_\Delta^2}$ 
with $M_\Delta= 6.24$ fm$^{-1}$.}
\label{tableGE2}
\end{table*}

\begin{table*}
\begin{center}
\begin{tabular}{l c  c c c c  c c c}
\hline
\hline
$<r_{E2}^2>$ & & $\Delta^{++}$ & & $\Delta^+$ & & 
$\Delta^0$ & &  $\Delta^-$ \\
\hline
Spectator 1 & &  0.58   & &   0.53     & &  0.21  & &  0.74 \\ 
Spectator 2 & &  0.40   & &   0.34     & &  0.18  & &  0.55 \\ 
\hline
Lattice: &&  && && \\
Quenched Wilson  
\cite{Alexandrou09} & &  & &
  0.336$\pm$0.096 & &  & & \\
Dynamical Wilson  \cite{Alexandrou09} & & & &
  0.134$\pm$0.106 & &  & & \\
Hybrid  \cite{Alexandrou09} & &   & &
  0.43$^{+1.35}_{-0.20}$ & &  & &  \\
\hline
\hline
\end{tabular}
\end{center}
\caption{Summary of results  
for $<r_{E2}^2>$ (fm$^2$).}
\label{tableRE2}
\end{table*}

\begin{figure*}[t]
\centerline{
\mbox{
\includegraphics[width=3.2in]{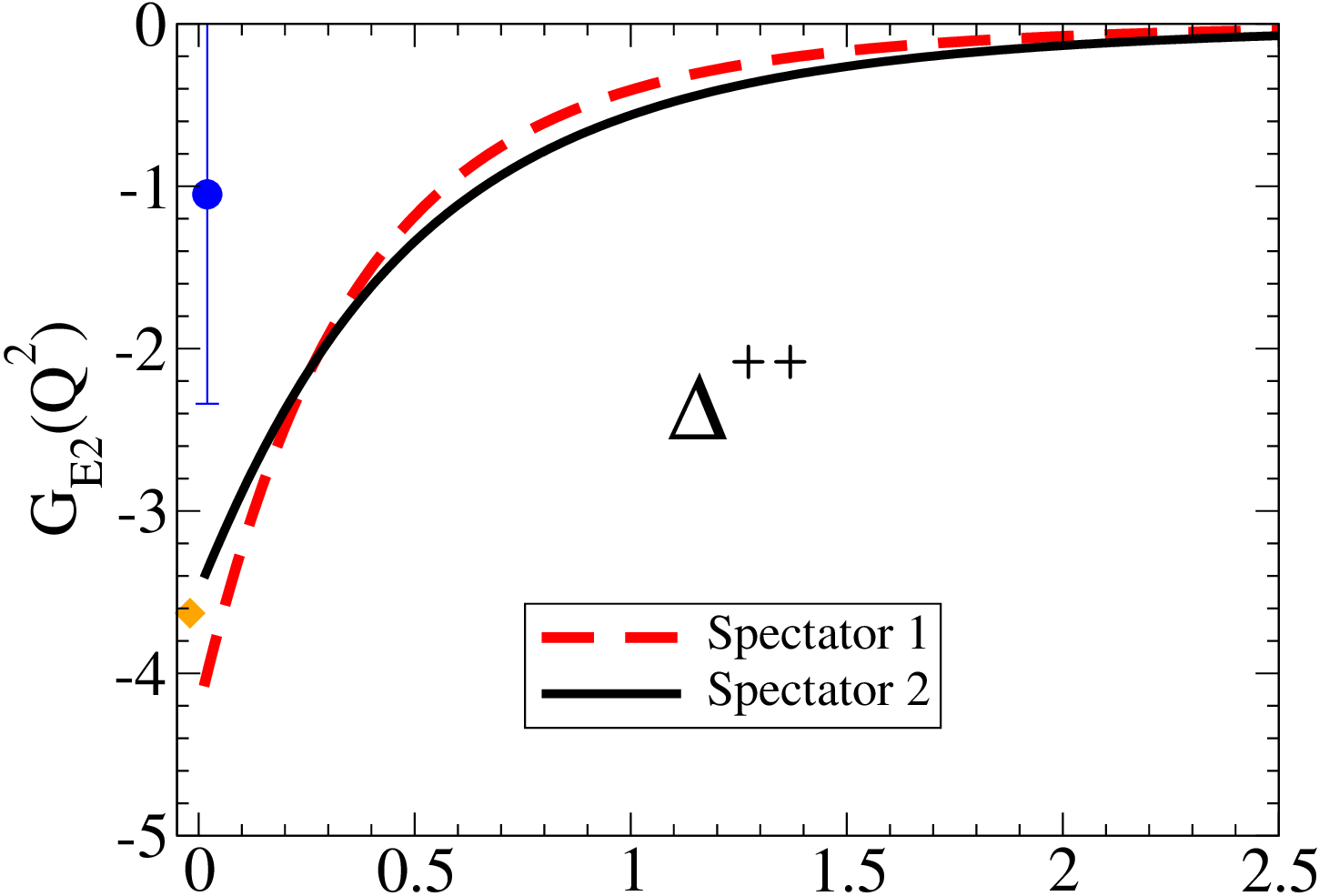} 
\hspace{.5cm} 
\includegraphics[width=3.3in]{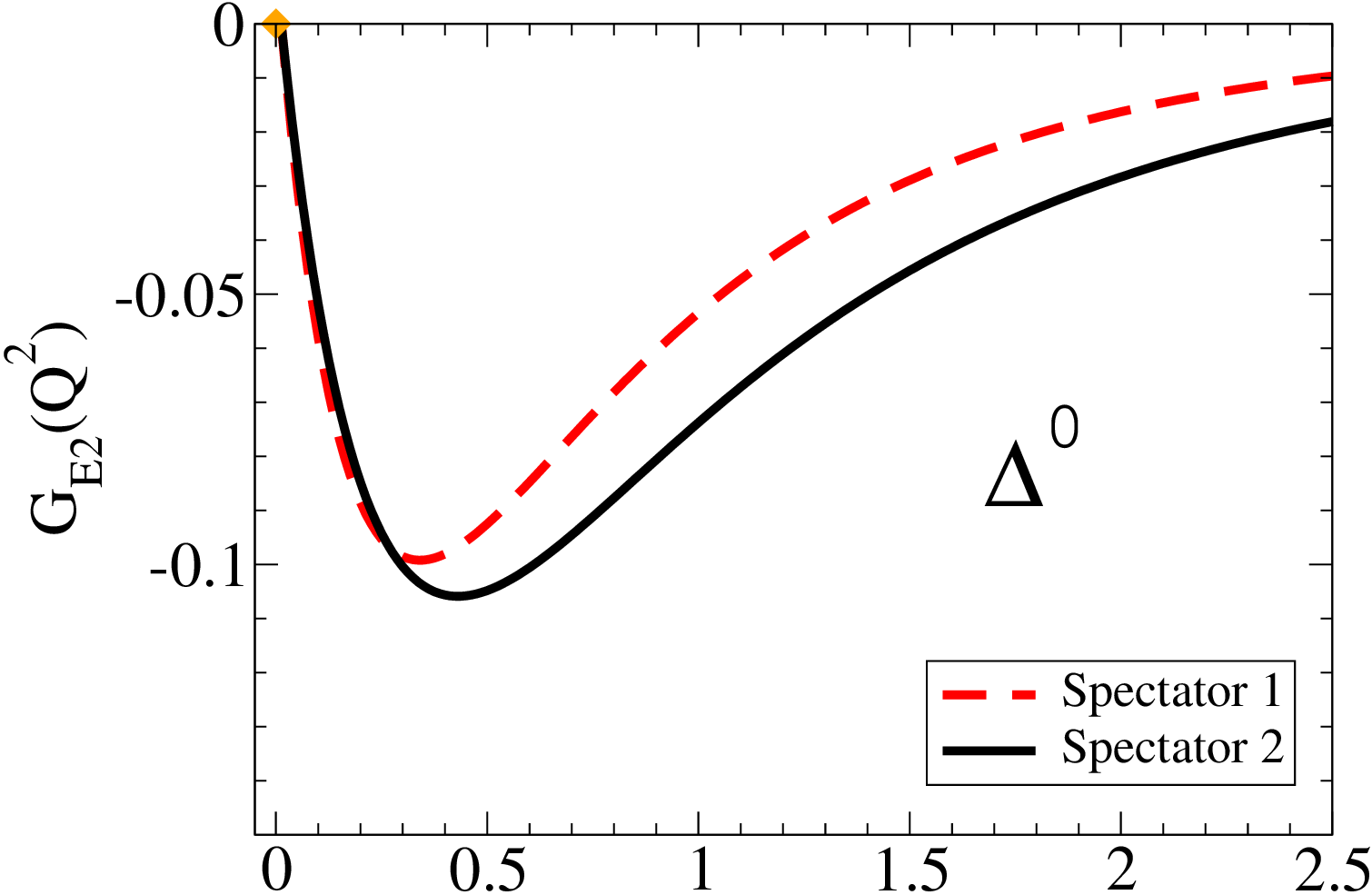}}}
\vspace{1cm}
\centerline{
\mbox{
\includegraphics[width=3.2in]{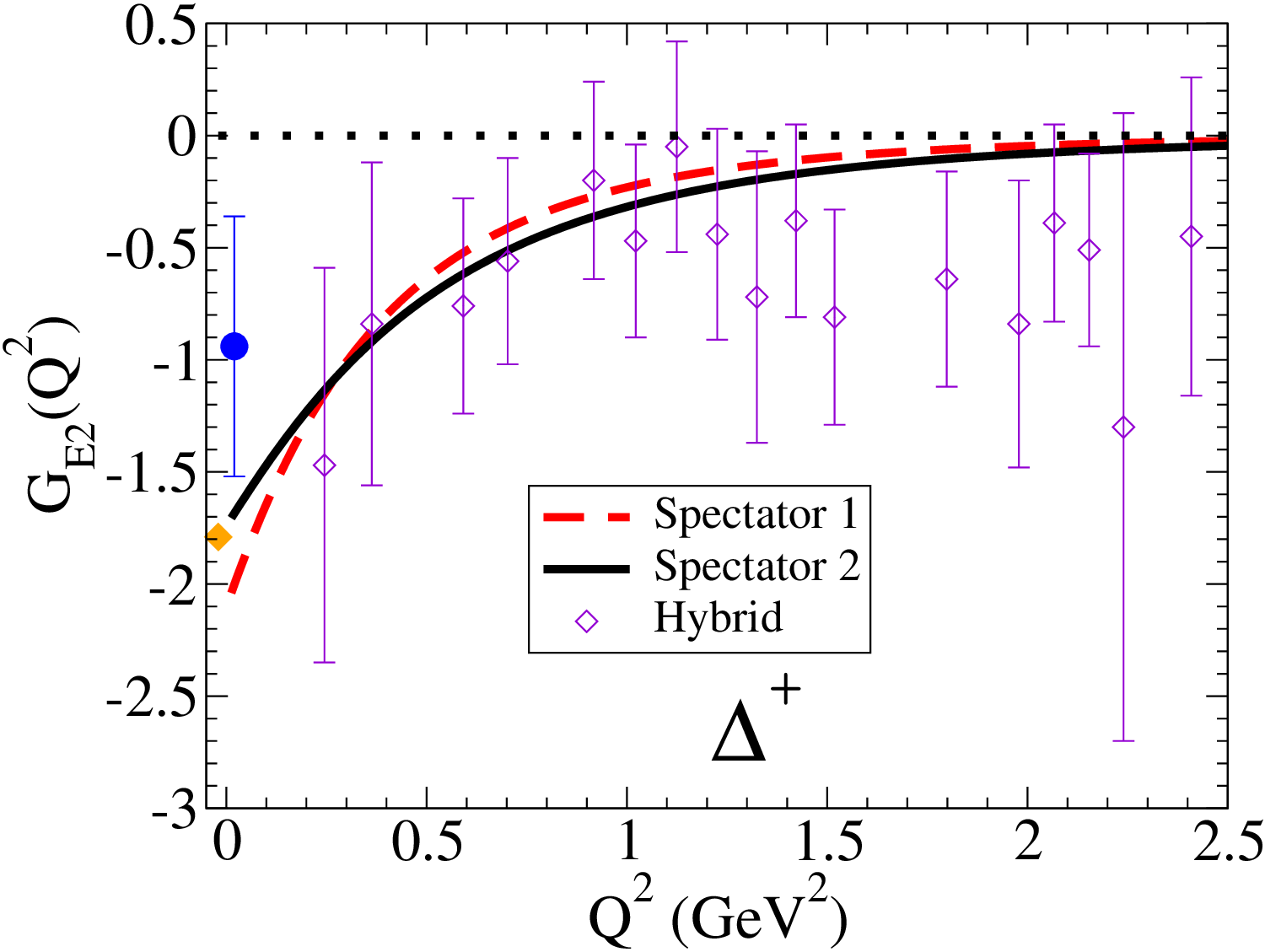}
\hspace{.6cm} 
\includegraphics[width=3.35in]{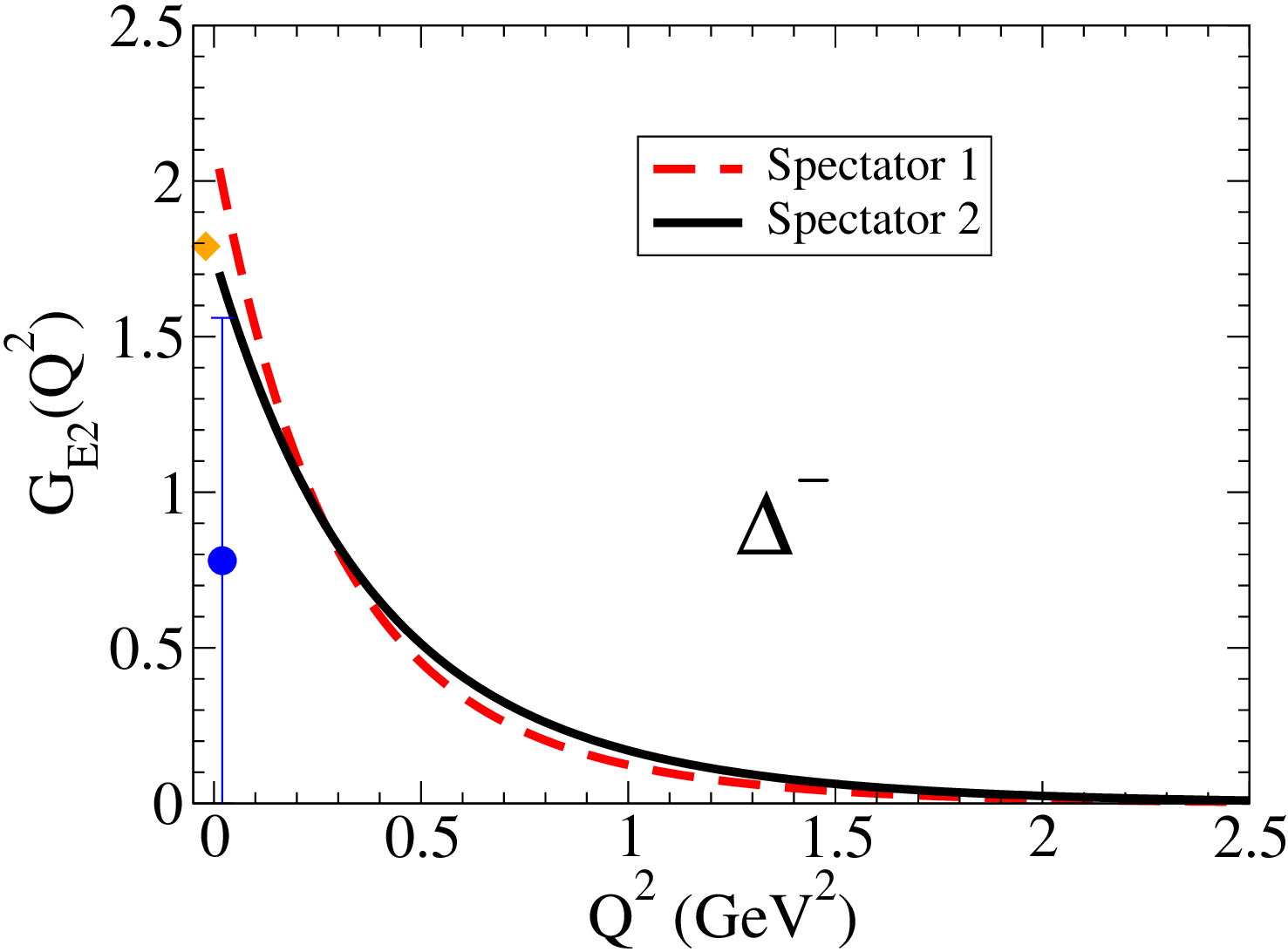}}}
\caption{\footnotesize
Form factor $G_{E2}$.  
The $\Delta^+$ lattice QCD points 
correspond to $m_\pi=353$ MeV from \cite{Alexandrou09}.
The NRQM ($\diamond$) is the result of Eq.~(\ref{eqQimp})
and \cite{Giannini90}; 
$\chi$PT ($\circ$) is from \cite{Geng09}.
The $\chi$PT result for $\Delta^0$ is out of the scale
(see Table \ref{tableGE2}).}
\label{figGE2}
\end{figure*}

\subsection{Direct comparison with lattice QCD}

We have compared our results for the 
magnetic dipole form factor with the
extrapolation to the physical pion mass point
of the quenched lattice QCD data.
But quenched data include pion cloud 
effects only partially, 
and lattice results 
based on quenched approximation are known to 
underestimate the full QCD result, as
discussed.
Fortunately, all  $\Delta$ form factors 
were recently also evaluated with
unquenched  methods. Those include 
sea quark effects from up and down quarks 
(Wilson action), and also from the strange quark
(hybrid action) \cite{Alexandrou08,Alexandrou09}.  
For those calculations 
there are no extrapolations to the physical
pion mass limit available.
In these cases we could only compare our models directly 
to the published lattice data.
Figure \ref{figGLat} shows the results
for $G_{E0}$ and $G_{M1}$
corresponding to the lowest 
pion mass in each case: 411 MeV (quenched),
384 MeV (Wilson action) and 353 MeV (hybrid action).
In addition to the results of 
Refs.~\cite{Alexandrou09} we include
the point corresponding to $m_\pi=306$ MeV from Ref.~\cite{Boinepalli09}.

Surprisingly, our two models also reproduce the magnitude
of the lattice QCD data even in the nonphysical pion mass region.
The agreement between our predictions for $G_{M1}$ 
and the lattice data at low $Q^2$
explains why the $<r_{M1}^2>$ values obtained 
with our models are close to the lattice values, 
as seen in Table~\ref{tableRM1}. 
Note that although
different methods 
give very similar results for $G_{M1}$
they exceed slightly 
the physical point (one data point from Ref.~\cite{Boinepalli09} 
is the exception, probably as a consequence of 
the small pion mass considered).

\subsection{Electric quadrupole form factor}
\label{secGE2}

Because of
Eqs.~(\ref{eqGE0a})-(\ref{eqGM1a}),
the charge and magnetic dipole form 
factors are essentially determined by
the S-state component of the $\Delta$ 
wave function. But without D-states in the $\Delta$ wave function
the electric quadrupole form factor
is identically zero. As such, $G_{E2}$ measures
a transition between the S-state and the D3-state 
(${\cal S}=3/2$, $L=2$)
of the $\Delta$ 
[see Eq.~(\ref{eqGE2a})]. 
Differently said, the D3-state induces
a deformation in the $\Delta$ system 
that builds up a nonzero
electric quadrupole proportional to
the D3-state admixture parameter $a$.

\subsubsection{Electric quadrupole moment}

The electrical quadrupole moment $Q_\Delta$
is defined as $Q_\Delta=G_{E2}(0)\sfrac{e}{M_\Delta^2}$.
The shape of the $\Delta$ can be interpreted  
according to the sign of $Q_\Delta$, if one relates in the Breit frame the charge distribution to
the form factors 
in the nonrelativistic limit.
For a positive charge, a
deformation extended along the equatorial region ($Q_\Delta < 0$) 
corresponds to an oblate distribution (pancakelike) and
a deformation along the polar axis 
($Q_\Delta > 0$) to a
prolate distribution (cigarlike). 
Alternative interpretations of deformation
can be introduced, 
and give different insight on the structure
\cite{Alexandrou08,Alexandrou09}. 
More details about this issue of
deformation are 
presented in a separate work  \cite{Deform}.

\begin{figure*}[t]
\vspace{.2cm}
\centerline{
\mbox{
\includegraphics[width=3.2in]{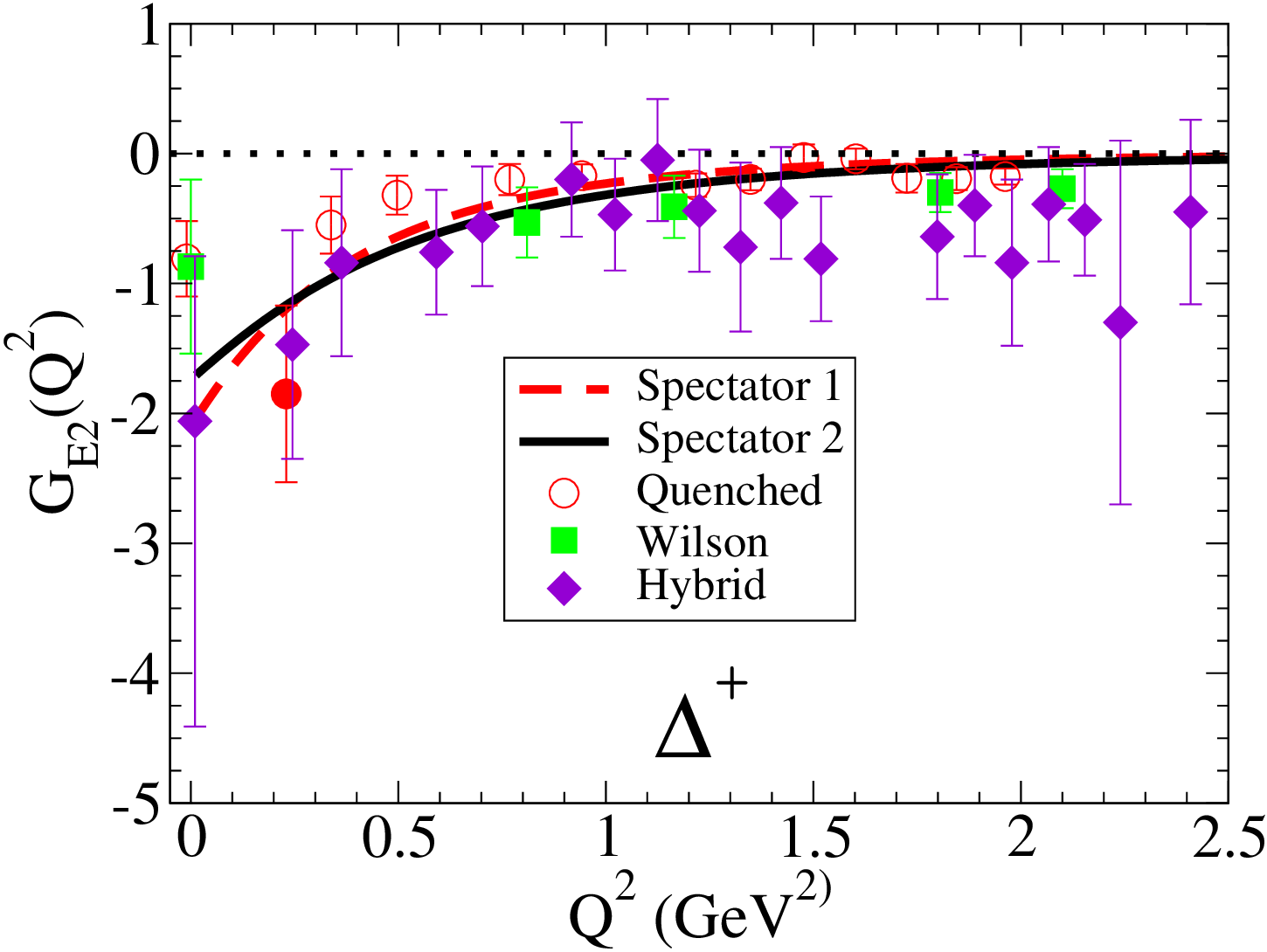}
\hspace{.5cm} 
\includegraphics[width=3.2in]{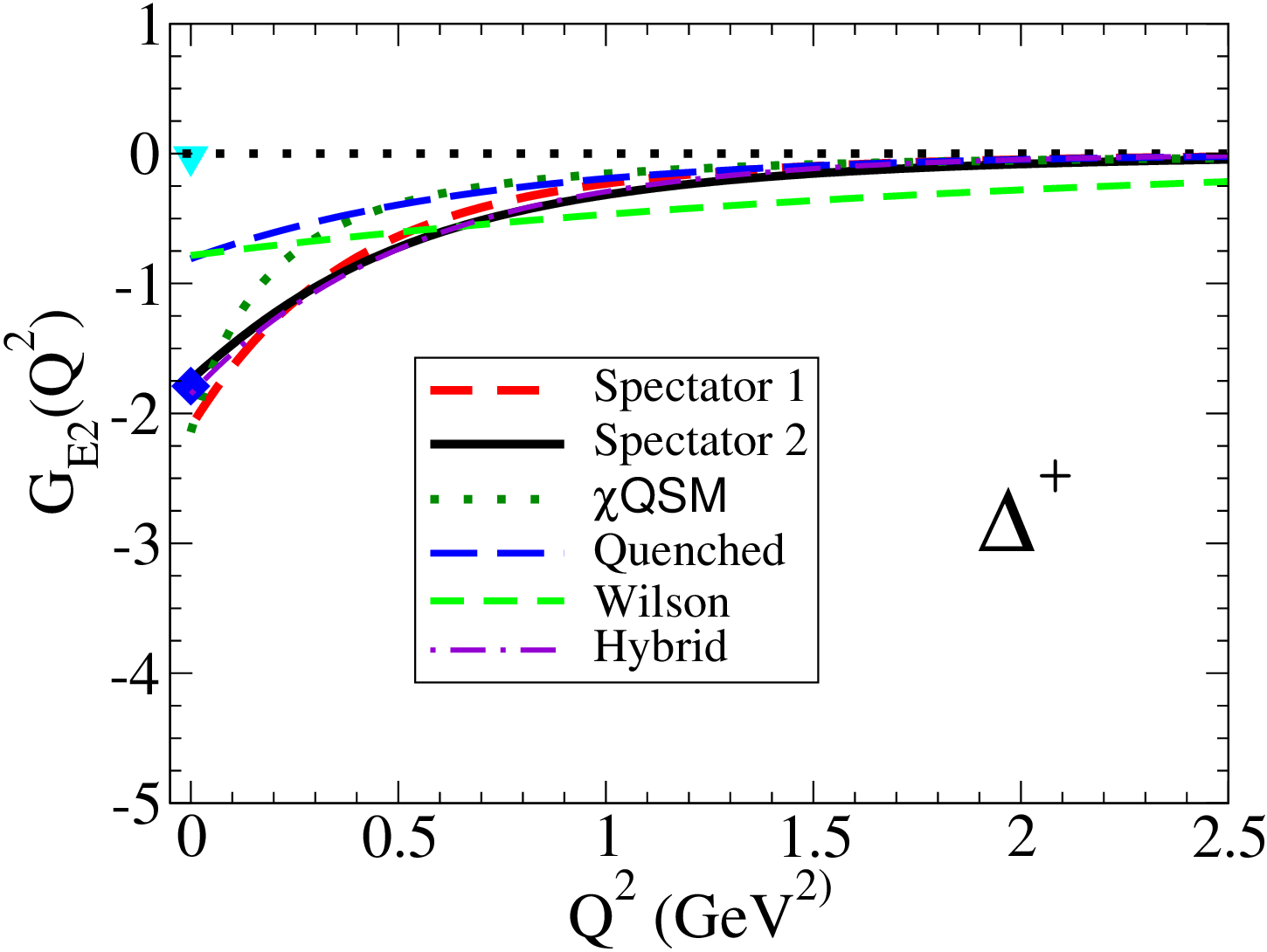}}}
\caption{\footnotesize{
Comparing  
$G_{E2}$ with lattice data from Refs~\cite{Alexandrou08,Alexandrou09}.
Left panel: quenched data (open $\circ$); 
Wilson action (filled $\square$) and hybrid action
(filled $\diamond$).
The quenched result from \cite{Boinepalli09} 
for $m_\pi=306$ MeV is also included (filled $\circ$).
Right panel: 
interpolation of 
the lattice data 
\cite{Alexandrou08,Alexandrou09} and 
$\chi$QSM  \cite{Ledwig08}; 
NRQM ($\diamond$) \cite{Giannini90}
and QCDSR  ($\bigtriangledown$) \cite{Azizi08} results
are also shown.
}}
\label{figGE2lattice}
\end{figure*} 

Our results for  $G_{E2}$ at $Q^2=0$ are shown in 
Table \ref{tableGE2}.
Since $G_{E2}$ depends only on the state D3,
$G_{E2}(0)$ is determined by ${\cal I}_{D3}^\prime$ 
which is $-7.00$ (Sp 1) or $-6.65$ (Sp 2). Besides,
the D3-state admixture coefficents are not very different
(0.88\% versus 0.72\%) and therefore both models give
similar values for $G_{E2}(0)$.
In our case, $Q_{\Delta^0} =0$
because the D-state to D-state contributions 
were neglected. But nonzero 
although small contribution for $Q_{\Delta^0}$ 
may appear otherwise.

In Table \ref{tableGE2} we make comparisons to other works, 
starting with nonrelativistic quark models (NRQM) 
e.g.~the classic Isgur model \cite{Isgur82}. In those models
the quark D-states are a consequence 
of the hyperfine interaction in the presence 
of a confinement mechanism. 
Then $Q_\Delta$ was
estimated in terms of the 
admixture coefficients for the symmetric, 
antisymmetric and with mixed symmetry components in the 
S and D-state wave functions,
together with a confinement parameter 
associated to the spherical harmonic oscillator
(interpreted as the quark core radius) \cite{Buchmann97,Giannini90}, 
and assuming, as we do here,
that the transition 
between D-states is negligible.

In this type of NRQM description, 
where the valence quark degrees of freedom and the  
electromagnetic interaction are reduced to 
the impulse approximation (which includes only the one-body current),  
it was possible to relate
the $\Delta$ charge distribution 
with the neutron charge distribution 
\cite{Krivoruchenko91,Richard84} according to 
\be
Q_\Delta^{(imp)} = \sfrac{2}{5}e_\Delta r_n^2,
\label{eqQimp}
\ee
where $r_n^2$ is the neutron 
squared radius, and
the index ({\it imp}) holds for impulse 
approximation (one-body current).
Equation (\ref{eqQimp}) is a parameter free relation, since it is independent 
of the admixture coefficients and of
the confinement parameter.  
Considering a recent estimate
$r_n^2= -0.116$ fm$^2$ \cite{Amsler08},
one has  $Q_{\Delta^+} \simeq -0.0464$ fm$^2$, or 
$G_{E2}^{\Delta^+}(0) \simeq -1.81$, in 
close agreement to  
the result of Ref.~\cite{Krivoruchenko91}.
Similar  numerical results 
were obtained by Buchmann \etal~\cite{Buchmann97} 
with a 
different
combination of values for
the admixture coefficients together with the confining parameter.

To estimate the nonvalence contribution for $Q_\Delta$ 
absent in the formulas above,
Buchmann \etal~\cite{Buchmann97}
included two-body currents 
for the quark-antiquark production mechanisms, and obtained
\be
Q_\Delta^{(exc)} = e_\Delta r_n^2.
\label{eqQexc}
\ee
In this case no D-state admixture is considered.
Although it is a consequence of the constituent quark 
formalism,  this result has again the feature of being parameter independent 
\cite{Buchmann97}, and could even be obtained in the large $N_c$ limit 
\cite{Buchmann02}.
Later, Eq.~(\ref{eqQexc}) was 
improved by means of a  
GP of QCD \cite{Blanpied01,Buchmann02b,Dillon99},
for the inclusion of higher order terms.
From the 
$\gamma N \to \Delta$ electric quadrupole data
\cite{Blanpied01} the value
$G_{E2}(0)=-7.02\pm4.05$ was obtained.
Note that this result, and already the result from Eq.~(\ref{eqQexc})
\cite{Buchmann97}, are large when compared 
to the other calculations compiled
in Table \ref{tableGE2}.

The other results in the literature are based on  $\chi$PT 
\cite{Butler94,Tiburzi09,Geng09},
on a quark-meson coupling model (QMCM) \cite{Leonard90},
on a Skyrme model \cite{Kroll94}, and more recently  
$\chi$QSM \cite{Ledwig08}
and QCD sum rules \cite{Azizi08}.
For completeness we mention as well
calculations
of PQ$\chi$PT in the heavy 
quark mass limit \cite{Arndt,Tiburzi05},
and relations between the electric quadrupole moments for
different charge $\Delta$ states \cite{Buchmann03,Lebed95}.

From lattice QCD there are earlier calculations
from Leinweber \cite{Leinweber92},
and recent simulations 
\cite{Alexandrou08,Alexandrou09,Ledwig08,Boinepalli09} 
based on different approaches: 
quenched, dynamical Wilson ($u$ and $d$ sea quarks)
and hybrid action (also $s$ quarks included).
The lattice results are also presented in 
Table \ref{tableGE2}.

Our model compares well with the NRQM 
from Refs.~\cite{Buchmann97,Isgur82,Krivoruchenko91}
based only on the valence quark degrees 
of freedom (see Table \ref{tableGE2}),
although it overshoots  the 
QCD sum rules \cite{Azizi08}.
Mixed descriptions, involving quarks and mesons, 
in particular pions, as degrees of freedom 
as $\chi$PT \cite{Butler94,Geng09}, QMCM \cite{Leonard90} 
and $\chi$QSM \cite{Ledwig08} 
suggest that the meson cloud can be important, 
although not as significant as Eq.~(\ref{eqQexc}) 
seems to imply.

\subsubsection{Dependence of $G_{E2}$ with $Q^2$}

Before we compare our results with lattice QCD data
we investigate the dependence of 
$G_{E2}$ on $Q^2$.
The results for $G_{E0}(Q^2)$ are 
presented in Fig.~\ref{figGE2}, for 
the models Sp 1 and Sp 2, for all the $\Delta$ charge cases.
For $\Delta^+$ the results are also compared 
with the more complete unquenched lattice QCD simulation 
(hybrid action) for the lightest pion mass available 
($m_\pi=353$ MeV) \cite{Alexandrou09}.
As a reference result we include also the 
$\chi$PT calculation from \cite{Geng09} 
and the NRQM from Eq.~(\ref{eqQimp}).
[The $\chi$PT prediction for $\Delta^0$ is out 
of the scale although the errorbars are consistent 
with our result (zero)]. 
The hybrid action considers $u$ and $d$ quarks 
as valence quarks, 
and $u$, $d$ and $s$ as sea quarks.
Therefore pion cloud contributions are included, 
although the results are not extrapolated 
to the physical pion mass.

In the left panel of Fig.~\ref{figGE2lattice} our models 
for $\Delta^+$ electric quadrupole are 
compared with other lattice QCD methods,
namely the quenched calculation with $m_\pi=411$ MeV
and the dynamical Wilson action calculation with $m_\pi=384$ MeV,
and again the hybrid action results for $m_\pi=353$ MeV.
All the cases shown correspond to the smallest
pion mass values of each method.
In the same graph,
also the lattice analytical extrapolations for $Q^2=0$  
assuming an exponential dependence in $Q^2$ are included
[Note the significant errorbars].

In the
right panel of Fig.~\ref{figGE2lattice}
we compare Sp 1 and Sp 2 
with the dipole approximation of the model  $\chi$QSM
from \cite{Ledwig08}, 
that provides results for $G_{E2}$ 
as function of $Q^2$. Additionally, the figure depicts the results from the 
analytical approximation of lattice QCD of 
Refs.~\cite{Alexandrou08,Alexandrou09},
as well as the
NRQM and QCDSR \cite{Azizi08} results for $Q^2=0$.

From Fig.~\ref{figGE2lattice} we conclude 
that our predictions are close to 
the hybrid action
and the dynamical Wilson action results. Still,
our results are larger, in absolute value, than
the quenched data, particularly in the low $Q^2$ region
(note that the dynamical hybrid data are consistent 
with the Wilson and quenched data, within
their statistical errors).
The difference between our results and  different lattice approaches 
seem to suggest that contrarily to the 
leading form factors $G_{E0}$ and $G_{M1}$ \cite{DeltaFF},
the meson cloud effect may be important for $G_{E2}$.
Also, the model $\chi$QSM, where the contribution 
of the meson cloud (sea quark contributions) near 
$Q^2=0$ is around 70\% 
\cite{Ledwig08} 
falls faster that those models as $Q^2$ increases. 
With the exception of 
Ref.~\cite{Alexandrou07} where an extrapolation 
to the physical point was published,
there are no extrapolations 
of $G_{E2}$ results for the physical point.

In the left panel of Fig.~\ref{figGE2lattice}  we depict also
the lattice data point of Ref.~\cite{Boinepalli09}
estimated in quenched approximation 
for $m_\pi=306$ MeV at $Q^2=0.230$ GeV$^2$. 
In the limit of the errorbars,
this point is also consistent with both our models (Sp 1 and Sp 2),

To finish the discussion about $G_{E2}(Q^2)$ 
we present in Table \ref{tableRE2} 
the result of our models and the ones 
from lattice QCD for $<r_{E2}^2>$.
Also for this observable Sp 1 and Sp 2 are similar,
as we would expect from the graphs from Fig.~\ref{figGE2}.
In both models the radius is in agreement in magnitude with 
the results from lattice.

\subsubsection{Overview of the electric quadrupole form factor}

The $\Delta$ electric quadrupole form factor is a measure 
of the deformation in the charge distribution, 
and is proportional to the 
transition coefficient between S and D3-states.
Except for the lattice QCD calculations
\cite{Alexandrou07,Alexandrou08,Alexandrou09},
and for the model $\chi$QSM \cite{Ledwig08},
previous studies of the dependence of the 
$\Delta$ electric quadrupole form factors on $Q^2$
are almost nonexistent.
There are however predictions of several models 
at the photon point. 
Our model is consistent with most 
of these calculations based on 
valence quark degrees of freedom.  

Our results agree also with the 
dynamical lattice QCD simulations 
with (Wilson and hybrid actions)
for the lightest pion cases.
%
Lattice QCD data seems to suggest 
that sea quark effects play a more important role 
for $G_{E2}$ than for $G_{E0}$ and $G_{M1}$, 
but the statistics must still be increased 
to allow more definitive conclusions.

The similarity of the D3-state parameterization in both our models
Sp 1 and Sp 2 
leads to very close results for the 
electric quadrupole form factor.
We conclude that the quadrupole electric form factor 
is not suitable to discriminate between our models.

\begin{figure*}[t]
\centerline{
\mbox{
\includegraphics[width=3.2in]{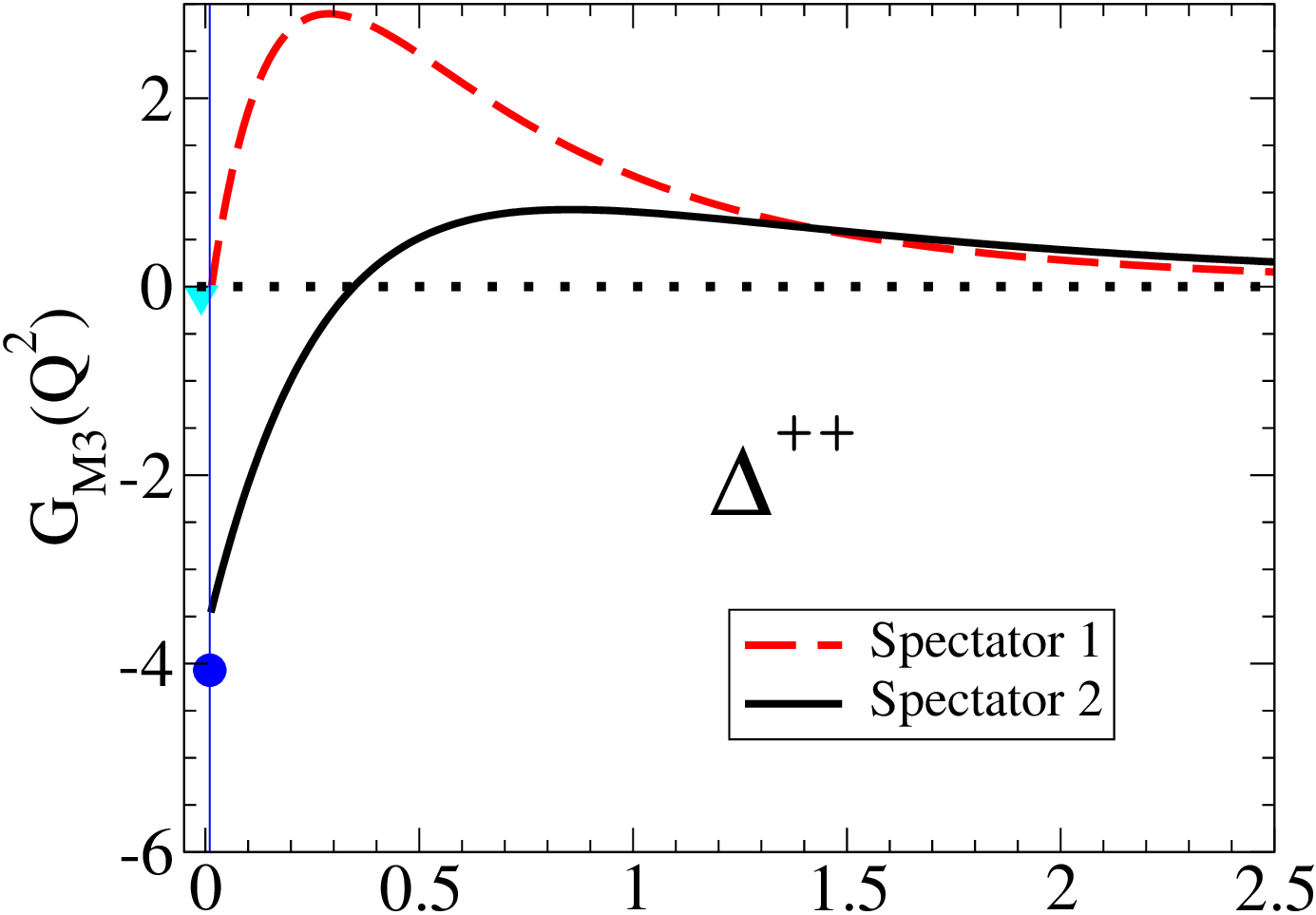}
\hspace{.5cm} 
\includegraphics[width=3.3in]{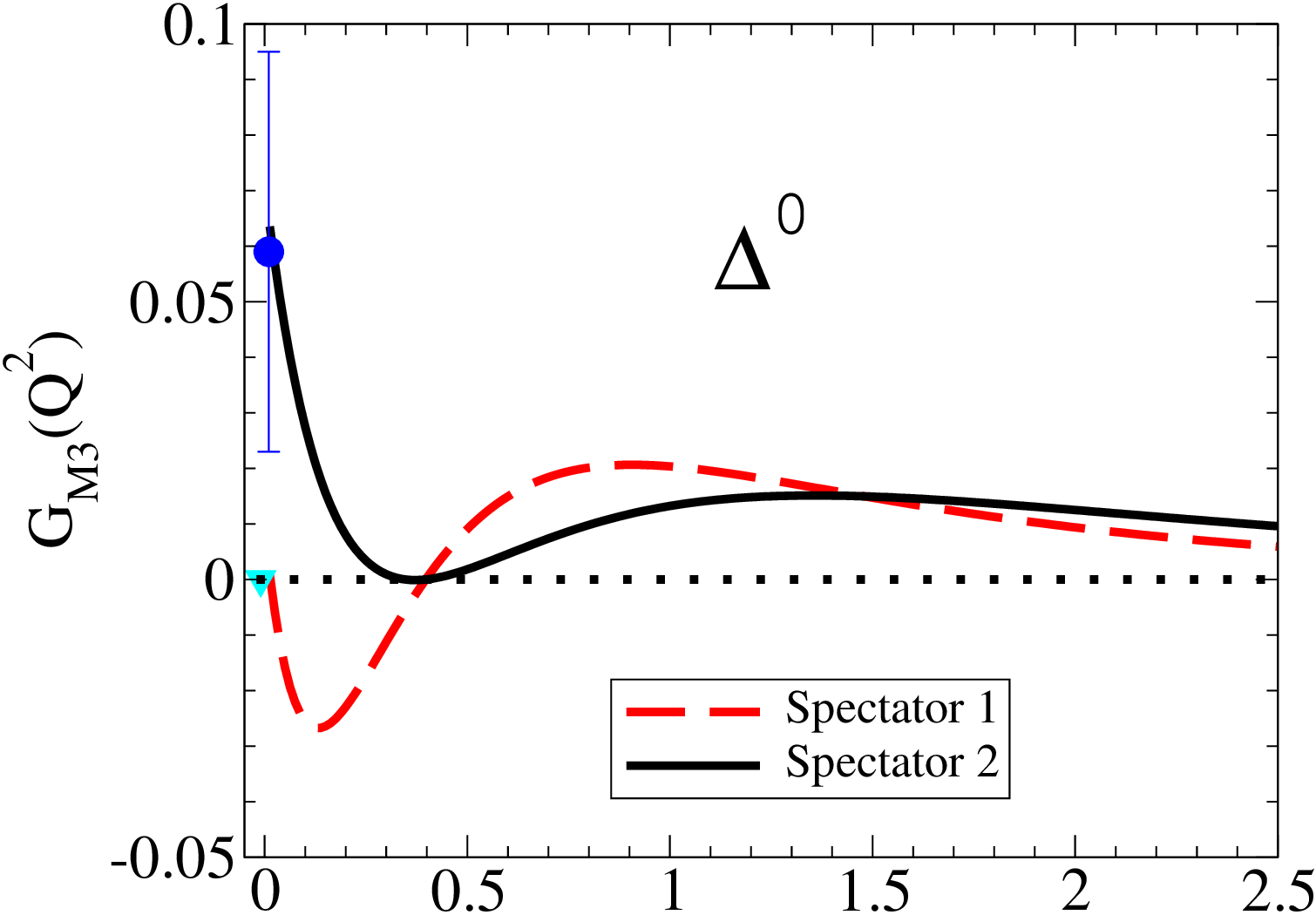}}}
\vspace{1cm}
\centerline{
\mbox{
\includegraphics[width=3.3in]{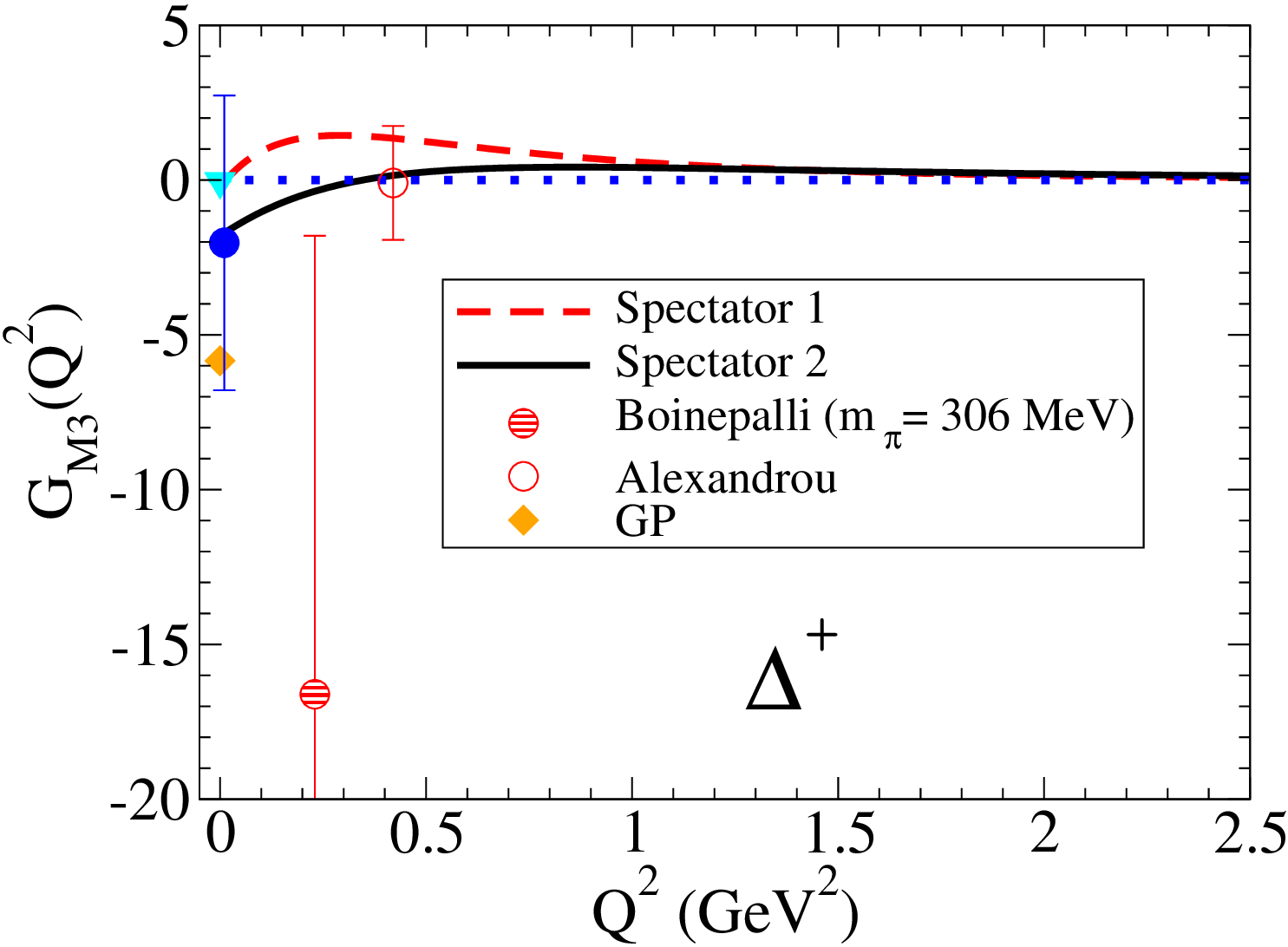}
\hspace{.8cm} 
\includegraphics[width=3.2in]{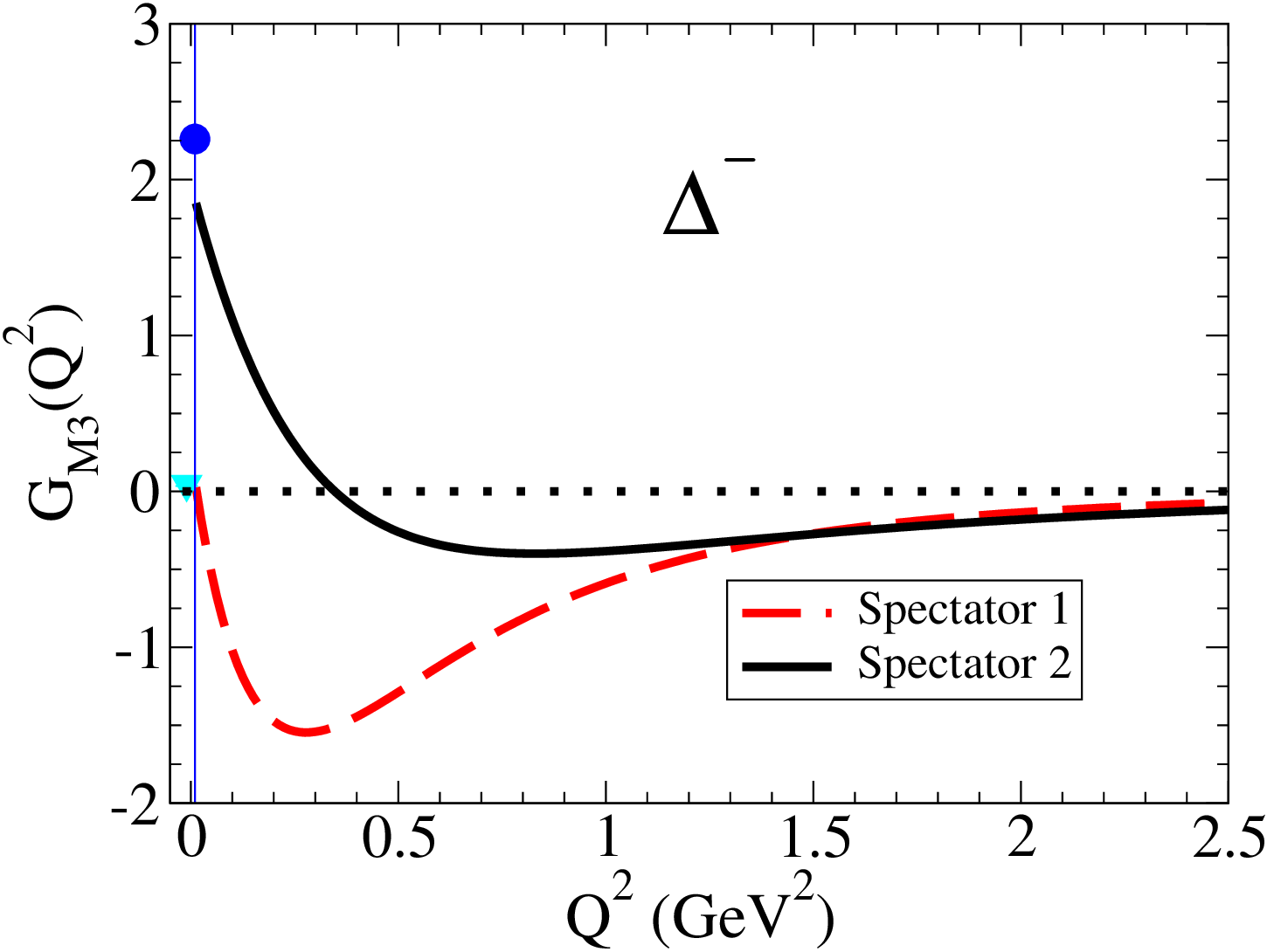}}}
\caption{\footnotesize{
$G_{M3}$ form factor. 
The results for QCDSR ($\triangledown$) \cite{Azizi08}
and  $\chi$PT (filled $\circ$) \cite{Geng09} for $Q^2=0$ are shown.
Lattice data for $\Delta^+$ from Refs.~\cite{Alexandrou07,Boinepalli09}
($m_\pi=306$ MeV) for $Q^2=0.230$ GeV$^2$, 
and Alexandrou \etal~\cite{Alexandrou07} for $Q^2=0.42$ GeV$^2$.
The error bar for the data from Alexandrou 
\etal~\cite{Alexandrou07} represents the interval of values 
associated with several pion masses. 
For $\Delta^+$ the result of GP \cite{Buchmann08} is also included.
}}
\label{figGM3}
\end{figure*}

\begin{table*}
\begin{center}
\begin{tabular}{l c  c c c c  c c c}
\hline
\hline
$G_{M3}(0)$ & & $\Delta^{++}$ & & $\Delta^+$ & & 
$\Delta^0$ & &  $\Delta^-$ \\
\hline
QCDSR \cite{Azizi08} & & -0.0925 $\pm$ 0.0234 & & -0.0462 $\pm$ 0.0117 & & 0 
                     & &  0.0462 $\pm$ 0.0117 \\ 
GP  \cite{Buchmann08}  & & -11.68    & &  -5.84 && 0 && 5.84 \\ 
$\chi$PT  \cite{Geng09}   & & -4.07$\pm$9.49   & &  -2.03$\pm$4.76    
& & 0.059$\pm$ 0.036  & & 2.26$\pm$4.52 \\
\hline
Spectator 1 & &   -0.049  & & -0.024   & &  0.00084  & &  0.026 \\ 
Spectator 2 & &   -3.51    & &  -1.72    & & 0.064  & &   1.85 \\
\hline
\hline
\end{tabular}
\end{center}
\caption{Summary of existing theoretical results 
for $G_{M3}(0)$.  
To obtain the magnetic octupole moment 
use ${\cal O}_\Delta= G_{M3}(0) \frac{e}{2 M_\Delta^3}$, 
with $M_\Delta= 6.24$ fm$^{-1}$.}
\label{tableGM3}
\end{table*}

\begin{table*}
\begin{center}
\begin{tabular}{l c  c c c c  c c c}
\hline
\hline
$<r_{M3}^2>$ & & $\Delta^{++}$ & & $\Delta^+$ & & 
$\Delta^0$ & &  $\Delta^-$ \\
\hline
Spectator 1 & &  $\sim$150   & &  $\sim$150     & &  $\sim$150  & &  $\sim$150 \\ 
Spectator 2 & &  1.2   & &   1.1    & &  2.0  & &  1.2 \\ 
\hline
\hline
\end{tabular}
\end{center}
\caption{Summary of results $<r_{M3}^2>$ (fm$^2$).}
\label{tableRM3}
\end{table*}

\subsection{Magnetic octupole form factor}

The third order form factor,
$G_{M3}$, 
did not attract as much attention 
as the first order ($G_{E0}$, $G_{M1}$) 
and the second order ($G_{E2}$) form factors.
Even for  the magnetic octupole moment, ${\cal O}_\Delta$,
there is no experimental information whatsoever, 
although ${\cal O}_\Delta$ 
was estimated considering the 
earlier NRQM \cite{Giannini90}. Interestingly, though, 
the lattice QCD simulations 
in Ref.~\cite{Alexandrou07} 
raised the question whether it is possible 
to determine the sign of ${\cal O}_\Delta$ 
accurately, or equivalently, with respect to the magnetic 
dipole distribution, whether the $\Delta$ is prolate 
(${\cal O}_\Delta$ 
positive) or oblate 
(${\cal O}_\Delta$  negative).
The first lattice QCD simulation 
\cite{Leinweber92} was not accurate enough 
to answer this question.
The recent work of the MIT-Nicosia group
\cite{Alexandrou07} based on the 
quenched approximation increased the hope 
that $G_{M3}$ or at least 
its sign can determined 
unequivocally by
lattice QCD.
This was a challenge 
for both MIT-Nicosia \cite{Alexandrou09}
and Adelaide groups \cite{Boinepalli09}.
The results from Ref.~\cite{Boinepalli09} 
(which are quenched) suggest a negative
sign (i.e.~a prolate distribution) for $G_{M3}$.
Nevertheless, one has to check whether the results 
for $G_{M3}$ are 
not significantly affected by the
sign of the pion loop contributions, 
as it happens for $G_{M1}$
(see discussion in Sec.~\ref{secGM1}).
One should note that those results correspond to large mass
pions, and no physical extrapolation was performed yet.

From the theoretical point of view, 
the publication of lattice QCD results 
for $G_{M3}$ triggered the application 
of quark models to compute
${\cal O}_\Delta$ and $G_{M3}$.
Buchmann and Henley \cite{Buchmann08} using 
a CQM with pion cloud, two-body currents 
and the GP formalism \cite{Dillon99}
predicted  ${\cal O}_{\Delta^+}= -0.012$ $e$fm$^3$.
The magnetic octupole was also recently 
predicted by QCD sum rules \cite{Azizi08} and 
$\chi$PT \cite{Geng09}.
Predictions for $G_{M3}(Q^2)$, based on the 
$\chi$SQM are expected for a near future \cite{Ledwig08}.
$\chi$PT predicts a nonzero 
octupole moment only at two meson loops level \cite{Butler94}.
Within PQ$\chi$PT in the heavy quark mass limit 
at one meson loop level, one has $G_{M3}(0)=0$ \cite{Arndt,Tiburzi05}, 
consistently with the quenched QCD data from 
Ref.~\cite{Boinepalli09} for large pion mass values.

A summary of the predictions for  $G_{M3}(0)$,  
including our results from models Sp 1 and Sp 2,
is presented in Table  \ref{tableGM3}.

In the covariant spectator quark model, 
according to Eq.~(\ref{eqGM30}), $G_{M3}(0)$
depends only 
on the integrals ${\cal I}_{D3}^\prime$, ${\cal I}_{D1}^\prime$ 
and the admixture coefficients $a$ and $b$.
For model Sp 1 one was
${\cal I}_{D3}^\prime=-7.00$, 
${\cal I}_{D1}^\prime= 1.59$.
As for model Sp 2,
${\cal I}_{D3}^\prime=-6.65$, 
${\cal I}_{D1}^\prime= 0.24$. 
For the model Sp 1 the large contribution 
of the D1-state (4\%) cancels the D3 contribution
leading to an almost zero value to $G_{M3}(0)$.
This entails a dramatic difference between 
the results of models Sp 1 and Sp 2,
as we can see from Table  \ref{tableGM3}.
Model Sp 1 gives negligible contributions for $G_{M3}(0)$.
The result from Sp 2 lies between the 
negligible predictions of QCD sum rules,  
and the large value of Buchmann \cite{Buchmann08}
based on a pion cloud model and the GP formalism
\cite{Dillon99,Buchmann02,Buchmann08}.
For completeness, and following the previous sections,   
the results for $<r_{M3}^2>$ are also
presented in Table \ref{tableRM3}.

The differences between models Sp 1 
and Sp 2 for $G_{M3}$ are clear from
Fig.~\ref{figGM3}. 
For $Q^2\ne 0$ the
cancellation between D1 and D3 contributions in model Sp 1 still exists,
although it is not as spectacular as in the $Q^2=0$ case.
We note also the change of sign in Sp 2 
for $Q^2 \approx 0.4$ GeV$^2$.
It would be interesting in the future to 
compare the $Q^2$ dependence of model Sp 2 
with alternative models or with lattice data.
The estimates of 
QCD sum rules \cite{Azizi08}, $\chi$PT \cite{Geng09} 
and GP \cite{Buchmann08} for $Q^2=0$ 
are also included in the graphs.
For the $\Delta^+$ we show  
the lattice data of Ref.~\cite{Boinepalli09}
for the pion mass $m_\pi=306$ MeV  at $Q^2=0.230$ GeV$^2$. 
In the graph we represent also the 
range of variation obtained in Ref.~\cite{Alexandrou07}
from calculations with several pion masses.

The results for $\Delta^+$ from model Sp 2 underestimate
in absolute value the lattice QCD data point
from Ref.~\cite{Boinepalli09} for $m_\pi=306$ MeV, but it is 
consistent with the central value of $\chi$PT \cite{Geng09}. 
We note, although, that the errorbands are significant.
The predictions of Sp 2 are also consistent 
the data from  Ref.~\cite{Alexandrou07},
within an interval
associated with several pion masses.
It will be interesting in the future 
to verify if the quenched approximation affects
crucially $G_{M3}$ 
and if the extrapolation 
to the physical point introduces significant corrections.
To achieve that goal the errorbands must be reduced.
At that point, Sp 2 can then be better tested.

We can conclude that, 
contrarily to the form factors analyzed before
(electric charge, magnetic dipole 
and electric quadrupole) the 
magnetic octupole form factor is 
extremely sensitive to the D-state parameterization.
As for our calculation enabling a decision  
in favor of Sp 1 or Sp 2,
we consider  model Sp 2 our best model,
because it
is better constrained, i.e.~it is constrained by not only the 
experimental but also the lattice data 
of the reaction $\gamma N \to \Delta$
\cite{LatticeD} 
(see discussion at the beginning of the section). 
Model Sp 2  predicts a much
stronger deformation from the symmetric 
magnetic dipole distribution.
It is also peculiar
that the Sp 2 predictions for $G_{M3}(0)$ are 
similar to the predictions of $G_{E2}(0)$.
This implies an almost identical 
deformation for the electric charge 
and for the magnetic dipole distribution, if
one uses 
the classical perspective of deformation.

A look at the magnetic octupole charge radius 
shows also a dramatic difference between the two models.
Model Sp 2 predicts  typical (although slightly
large) values, but model Sp 1 breaks all the scales.
That result might however be due to the 
equation used for $<r_{M3}^2>$, which is the analogue of Eq.~(\ref{eqRE02}), and
breaks down in the limit when $G_{M3}(0)$ gets close to zero.
But it might also be a consequence 
of the pion cloud effects not being
explicit in this work.

\subsubsection{Overview of the magnetic octupole form factor}

Model Sp 2, which is our best model, gives a result 
for the magnetic octupole 
closer to the available lattice QCD data.
We admit that model Sp 2 is not the last word 
in the $\Delta$ wave function parameterization.
Future lattice QCD simulations with increasingly 
smaller systematic errors may still
discard the parameterization of Sp 2 \cite{LatticeD}. 
Note that the quadrupole transition
form factors  
can nowadays be experimentally measured for that reaction 
(the interaction time in $\gamma N \to \Delta$
is larger than in the reaction $\gamma \Delta \to \Delta$).
A direct test of the parameterization 
will be an even more accurate study of 
the $\gamma N \to \Delta$ reaction, 
once more precise data is provided. 
By using
quadrupole form factors (electric and Coulomb) 
at the physical point \cite{NDelta,NDeltaD} 
or, in alternative,
their data
from lattice QCD \cite{LatticeD,Alexandrou08},
the valence contributions of the 
$\Delta$ wave function, and in particular 
the small D1 and D3-state contributions, can eventually be better constrained,
to further refine our the predictions for the $G_{E2}$ and  $G_{M3}$
$\Delta$ form factors.

\section{Conclusions}
\label{secConclusions}

In this work we extended the spectator quark model 
to include the $\Delta$ wave function D-states into the matrix elements of the electromagnetic current, 
in first order
in the admixture coefficients of those states. This allowed us to
evaluate all the four $\Delta$ elastic form factors.
We took two models previously calibrated by the two
nucleon form factor data \cite{Nucleon} and by the form factors for
the $\gamma N \to \Delta$ reaction \cite{NDeltaD,LatticeD}.
The two models differ in the way they 
describe the $\gamma N \to \Delta$ quadrupole data.
The first model (Sp 1) was derived 
by fitting the experimental data 
to the sum of the valence 
quark contribution with the pion cloud 
contribution \cite{NDeltaD}. 
The second model (Sp 2) was obtained
by adjusting only the valence quark contributions to 
that reaction to the lattice QCD \cite{LatticeD}, data 
while still describing the physical data.
Because in the last case we fixed the $\Delta$ D-state 
parameters independently of the 
pion cloud mechanisms, 
in our view model Sp 2 
is more reliable and robust than Sp 1.

Our predictions, based on the valence quark degrees 
of freedom alone, reproduce well the main features of the lattice results for
$G_{E0}$ and $G_{M1}$ when
extrapolated to the physical point,
except for a 
some deviation at low $Q^2$. 
In particular, the agreement for high $Q^2$ (say $Q^2 > 1$ GeV$^2$)
is excellent, indicating that 
pion cloud effects (not included explicitly in our model) 
are small for high $Q^2$ values. This confirms that 
the valence quarks are the dominant effect 
in the $G_{E0}$ and $G_{M1}$ form factors,
as anticipated in a previous work \cite{DeltaFF}
without the explicit contribution of the D-states.
In the low  $Q^2$ region, however, our results for the $\Delta$ suggest 
a smaller charge and magnetic charge radii, 
when compared with the proton.
This can be a limitation caused by the absence 
of the chiral behavior (or pion cloud effects).  Alternatively,
it can also represent the real physical case
if, as for the $\Delta$ magnetic moment,
there is a suppression of the charge radius 
due to the nonanalytical contribution 
of the  inelastic region  ($m_\pi < M_\Delta-M_N$).
Further lattice QCD studies, 
preferably unquenched studies, are needed to clarify this point further.

The inclusion of the D-states in the 
$\Delta$ wave function is required to explain 
the electrical quadrupole and magnetic octupole form factors.
The D3-state (spin 3/2 state with an orbital D-wave) 
state is needed for a 
nonvanishing electric quadrupole moment for the $\Delta$.
The D1-state (spin 1/2 with an orbital D-wave), as well as
the D3-state (spin 3/2 with an orbital D-wave), 
generates nonvanishing 
contributions for the magnetic octupole form factor.
By comparing our results 
with the available variety of lattice results, we 
conclude that valence quark contributions are 
of the same magnitude as the total result. 
This shows that to establish the magnitude 
of the contributions of the pion cloud
to these observables requires
more progress, with more accurate 
quenched and unquenched
lattice data.
In its absence for the near future \cite{DeltaMM}, we may estimate 
the pion cloud effects similarly to 
what was done for the nucleon and  the octet in Ref.~\cite{Octet}.
That work concluded that, although significant 
for the nucleon magnetic moments,
pion cloud effects give corrections 
of less than 8\% for the quark anomalous magnetic moments.

Another interesting result to be further explored in the future is that
contrary to $G_{E2}$, $G_{M3}$ is very sensitive to 
model parameterization.
This oversensitivity can be observed in 
the $G_{M3}$ radius. The value of
$G_{M3}(0)$ depends 
strongly on the D-states admixture and 
on the overlap between S and D1-states.
We obtained an almost 
complete cancellation of $G_{M3}$ in a particular model (Sp 1). 
If $G_{M3}(0)$ is in fact very small, then  
the study of D-state to D-state transitions 
may became important 
and 
should also be explicitly evaluated.
Since the direct measurement of a third order multipole 
is technically difficult,
a way out is, in our opinion, the use of the 
$\gamma N \to \Delta$ (experimental and lattice) data for an even more accurate determination 
of the D-state admixture
following Refs.~\cite{NDeltaD,LatticeD}, 
to predict ${\cal O}_\Delta$ as done here.

Our best model (Sp 2) gives for 
$\Delta^+$
\ba
 & &G_{M1}(0)= 3.27, \hspace{.5cm} G_{E2}(0)= -1.71 \nonumber \\
 & & G_{M3}(0)=-1.72.
\label{eqFinalFF}
\ea
These values correspond to the multipoles:
\ba
& &
\mu_{\Delta^+}= 2.49 
\mu_N, \hspace{.5cm}
Q_{\Delta^+}= -0.044 \;e \mbox{fm}^2 \nonumber \\
& &{\cal O}_{\Delta^+}= -0.0035\; e \mbox{fm}^3.
\label{eqFinalMM}
\ea 
As the lattice QCD data for $G_{M3}$ is 
still affected by a large uncertainty,
our results for ${\cal O}_{\Delta^+}$ are
independent predictions to be tested in the future.

The results (\ref{eqFinalFF})-(\ref{eqFinalMM}),
give an 
oblate deformation for the 
electric charge and magnetic dipole distributions, 
considering  the  {\it classical} interpretation of deformation.
In Ref.~\cite{Deform} we 
compare this analysis with the 
one based on transverse density distributions.

The dependence of the $G_{M3}$ form factor 
on $Q^2$ that we obtain is very atypical, and it will be
interesting if future lattice QCD calculations or 
alternative hadronic models may confirm our results. 
It can also happen that forthcoming lattice calculations 
may require a reparametrization 
of our model, which would affect in turn 
our results for the $\gamma N \to \Delta$
(where the quadrupole form factors, electric and Coulomb, 
are small and 
sensitive to the D-states.)

In the future, we can extend the model to the lattice QCD regime
as suggested in Refs.~\cite{Lattice,LatticeD,Omega}.
That possibility was already explored 
in previous applications to 
the nucleon \cite{Lattice}
and $\gamma N \to \Delta$ transition \cite{Lattice,LatticeD}.
The covariant spectator quark model 
defines a tool 
that can be used to extrapolate 
the lattice results both to the $Q^2 \to 0$ 
and to the physical pion mass limits.
Another strength of this formalism is the possibility of
its extension to the strange quark sector,  
enabling the application 
of the model to the octet and decuplet 
baryon systems.
In particular, the application 
to the  $\Omega^{-}$ form factors \cite{Omega}
is very promising, since meson cloud 
effects (kaon cloud) are expected to be very small in that system.
Our predictions could then be compared 
with lattice simulations nowadays 
performed at the 
strange quark physical mass \cite{Aubin,Boinepalli09,Alexandrou09c}.

\vspace{0.3cm}
\noindent
{\bf Acknowledgments:}

G.~R.~thanks Ross Young, David Richard   
and Anthony Thomas for helpful discussions.
The authors thank also Peter Moron 
for helpful clarifications relative to
Ref.~\cite{Boinepalli09} 
and Constantia Alexandrou for sharing the lattice data from 
Refs.~\cite{Alexandrou07,Alexandrou07X}.
This work was partially supported by Jefferson Science Associates, 
LLC under U.S. DOE Contract No. DE-AC05-06OR23177.
G.~R.~was supported by the Portuguese Funda\c{c}\~ao para 
a Ci\^encia e Tecnologia (FCT) under Grant  
No.~SFRH/BPD/26886/2006. 
This work has been supported in part by the European Union
(HadronPhysics2 project ``Study of strongly interacting matter'').

%

\appendix

\section{Generic form to the current}
\label{apGeneric}

Consider the $\Delta$ 
electromagnetic current $J^\mu$
\be
J^\mu = \bar w_\alpha \Gamma^{\alpha \beta \mu } w_\beta. 
\ee
The most general form, consistent with
parity, G-parity \cite{Nozawa90}, gauge invariance
(current conservation) and
the properties
\ba
& &(\not \! P_\pm -M_\Delta) w_\alpha(P_\pm)=0 \nonumber \\
& &
\bar w_\alpha(P_+) P_+^\alpha=0, \hspace{.3cm} P_-^\beta w_\beta(P_-)=0 \nonumber \\
& &
\gamma^\alpha w_\alpha=0,
\label{eqRSprop}
\ea 
reads
\ba
\Gamma^{\alpha \beta \mu}&= &
  \left(a g^{\alpha \beta }
+ b q^\alpha q^\beta \right) \gamma^\mu   \nonumber \\
& &  + \left(c g^{\alpha \beta}  +
      d q^\alpha q^\beta \right) P^\mu  \nonumber \\
& & + e \left(g^{\alpha \mu } - \frac{q^\alpha q^\mu}{q^2} \right) q^\beta 
\nonumber \\
& &  + f \left(g^{\beta  \mu } - \frac{q^\beta  q^\mu}{q^2} \right) q^\alpha. 
\ea
The coefficients $a$, $b$, $c$, $d$, $e$ and $f$ 
are general functions of $Q^2$.
Time reversal invariance implies that
$\Gamma(P_+,P_-)$ 
must be symmetric in the permutation 
$P_+ \rightleftharpoons P_-$ (or $q \to -q$), and then we conclude that
\be
f=-e.
\label{eqSim}
\ee

Using Eq.~(\ref{eqSim}), one has
\ba
\Gamma^{\alpha \beta \mu}&= &
  \left(a g^{\alpha \beta } 
+ b q^\alpha q^\beta \right) \gamma^\mu + \nonumber \\
& &   \left(c g^{\alpha \beta}  +
      d q^\alpha q^\beta \right) P^\mu + \nonumber \\
& & + e \left(g^{\alpha \mu } q^\beta -   
              q^\alpha g^{\beta  \mu }  \right). 
\ea
To obtain the final form we use the identity 
due to Fearing \cite{Nozawa90}:
\ba
g^{\alpha \mu } q^\beta -   
              q^\alpha g^{\beta  \mu } =
\frac{4M_\Delta^2+ Q^2}{2M_\Delta} g^{\alpha \beta} \gamma^\mu
\nonumber \\
- g^{\alpha \beta} (P_+ + P_-)^\mu
+ \frac{q^\alpha q^\beta}{M_\Delta} \gamma^\mu,
\label{eqFearing}
\ea
and the Gordon decomposition
\be
\frac{(P_+ + P_-)^\mu}{2 M_\Delta} =
\gamma^\mu - \frac{i \sigma^{\mu \nu} q_\nu}{2M_\Delta},
\label{eqGordon}
\ee
which holds between Rarita-Schwinger states 
and solutions of the Dirac equation.
Finally, one has
\ba
\Gamma^{\alpha \beta \mu}&= &
-\left[
F_1^\ast  g^{\alpha \beta} + F_3^\ast 
\frac{q^\alpha q^\beta}{4M_\Delta^2} \right]
\gamma^\mu \nonumber
\\
& &
-\left[
F_2^\ast g^{\alpha \beta} + F_4^\ast 
\frac{q^\alpha q^\beta}{4M_\Delta^2} \right]
\frac{i \sigma^{\mu \nu} q_\nu}{2M_\Delta},
\ea
where $F_i^\ast$ are linear combinations 
of $a,b,c$ and $e$.

%
%

\section{$\Delta$ form factors}
\label{apFF}

\subsection{Current $J_{S}^\mu$}
\label{apJS}

The current $J_{S}^\mu$ was written
in Ref.~\cite{DeltaFF} using the $\Delta$ 
S-state wave function of \cite{NDelta}:
\ba
& &
G_{E0}^S(Q^2)= 
\tilde g_\Delta
 {\cal I}_S  \\
& &
G_{M1}^S(Q^2)= 
\tilde f_\Delta{\cal I}_S  \\
& &
G_{E2}^S(Q^2) =
G_{M3}^S(Q^2) = 0,
\label{eqDeltaFF}
\ea
where
\be
{\cal I}_S=
\int_k \psi_S(P_+,k) \psi_S(P_-,k).
\ee
As $G_{E2}=G_{M3} \equiv 0$ there is no 
electric quadrupole neither 
magnetic octupole originated by
the $\Delta$ S-states.

%
%

\subsection{Current $J^\mu_{D3}$}
\label{apGD3}

Consider Eqs.~(\ref{eqPsiS}) and (\ref{eqPsiD3}) 
in a compact notation
\ba
& &\Psi_S= 
-\varepsilon_{P}^{\alpha \ast} w_\alpha \psi_S (T \cdot \xi^{1 \ast})
\\
& &
\Psi_{D3}=  
-3 \varepsilon_{P}^{\alpha \ast} W_{D3\, \alpha} \psi_{D3} 
(T \cdot \xi^{1 \ast}).
\ea
The transition between a $\Delta$ S and a $\Delta$ D3 
state can be decomposed in two processes:
\begin{itemize}
\item
S in the initial state (current $J_{S,D3}^\mu$)
\item
D3 in the initial state (current $J_{D3,S}^\mu$).
\end{itemize}

In the following we will 
use the properties of the spin S-states
(\ref{eqRSprop}) (S-state)
and also
\ba
& &
\not \! P_\pm W_{D3\, \beta}(P_\pm) =M_\Delta W_{D3\, \beta} (P_\pm)
\nonumber \\
& &
P_\pm^\beta W_{D3\, \beta}(P_\pm)=0, 
 \hspace{.3cm}
\gamma^\beta W_{D3\,\beta}(P_\pm)=0.
\label{eqW3prop}
\ea
Note that the last identity holds for spin states 
based in a spin 3/2 core \cite{NDelta,NDeltaD}.
Simple consequences of those relations are
\be
\bar W_{D3\, \alpha} \not \! q w_\beta=
\bar w_{\alpha} \not \! q W_{D3\, \beta}=0.
\ee

Because the S and D3-states have
the same core-spin, in general
\ba
\sum_{\lambda}\int_k 
\bar \Psi_{D3}(P_+,k) \Psi_S(P_-,k) \ne 0 
\nonumber \\
\sum_{\lambda}\int_k 
\bar \Psi_{S}(P_+,k) \Psi_{D3}(P_-,k) \ne 0,
\nonumber 
\ea
although the result becomes zero for $Q^2=0$.

\subsubsection{Transition S $\to$ D3}

The transition between an initial $\Delta$ S-state 
(momentum $P_-$) and a final $\Delta$ D3-state ($P_+$)
can be written as 
\be
J^\mu_{S,D3} =
3 \sum_{\lambda} \int_k 
\bar \Psi_{D3} (P_+,k) j_I^\mu \Psi_S (P_-,k),
\label{eqJSD3}
\ee
where the spin and polarization indices are suppressed 
for simplicity.
We can write (\ref{eqJSD3}) as
\be
J^\mu_{S,D3} =
3
\int_k
\left\{ 
\left[
\bar W_{D3\, \alpha} A^\mu w_\beta\right] 
\psi_{D3}^+ \psi_S^-
\right\}
\Delta^{\alpha \beta},
\label{eqJSD3b}
\ee
where $\psi_{D3}^+ = \psi_{D3}(P_+,k)$,
$\psi_{S}^- = \psi_{S}(P_-,k)$  and 
\ba
A^\mu &=&  (T \cdot \xi^{1 \ast})^\dagger (3 j_I^\mu) 
(T \cdot \xi^{1 \ast}) \nonumber \\
&=& 
\tilde e_\Delta \gamma^\mu + 
\tilde \kappa_\Delta
\frac{i \sigma^{\mu \nu} q_\nu}{2M_\Delta}. 
\label{eqA}
\ea
In the last equation we use
the definitions of $\tilde e_\Delta$
and $\tilde \kappa_\Delta$ from Eqs.~(\ref{eqEd})-(\ref{eqKd}).

Using the Gordon decomposition (\ref{eqGordon}), 
we write 
the current (\ref{eqJSD3b}) as
\be
J^\mu_{S,D3} = \tilde f_\Delta {\cal J}_{S,D3}^\mu
- \tilde \kappa_\Delta
\frac{(P_++P_-)^\mu}{2 M_\Delta} {\cal R}_{S,D3},
\ee
where $\tilde f_\Delta$ is defined by Eq.~(\ref{eqfDelta}), and 
\ba
& &
{\cal J}_{S,D3}^\mu
=3
\int_k
\left\{ 
\left[
\bar W_{D3\, \alpha} \gamma^\mu w_\beta\right] 
\psi_{D3}^+ \psi_S^-
\right\}
\Delta^{\alpha \beta} \\
& & 
{\cal R}_{S,D3}= 3
\int_k
\left\{ 
\left[
\bar W_{D3\,\alpha} w_\beta \right]
\psi_{D3}^+ \psi_S^- \right\}
 \Delta^{\alpha \beta}.
\ea
The expression for $\Delta^{\alpha \beta}$
is given by Eq.~(\ref{eqProp}).

To reduce the current $J^\mu_{S,D3}$
to a form
close to the standard one (\ref{eqJgen}), we work 
the spin algebra using the properties (\ref{eqRSprop}) and (\ref{eqW3prop}).
We also use the covariant result 
of  the integration in the azimuthal variable $\varphi$ 
\cite{NDeltaD}:
\be
\int_k {\cal D}^{\alpha \beta} 
\psi_{D3}^+\psi_S^- 
=
R^{\alpha \beta}(q,P_+)
\int_k 
b(\tilde k_+, \tilde q_+) \psi_{D3}^+\psi_S^-,
\label{eqIntPhi}
\ee
where
\ba
& &
b(\tilde k_+, \tilde q_+) =
\frac{3}{2} \frac{(\tilde k_+ \cdot \tilde q_+)^2}
{\tilde q_+^2}
-\frac{1}{2} \tilde k_+^2 
\label{eqBtilde}
\\
& &R^{\alpha \beta} (q,P_+)
=
\frac{\tilde q_+^\alpha  
\tilde q_+^\beta}{\tilde q_+^{2}}
-\frac{1}{3} \tilde g_+^{\alpha \beta}.
\label{eqRtilde}
\ea
The variable $\tilde g_+$ 
follows Eq.~(\ref{eqGtil}) and  $\tilde q_+$ 
comes from Eq.~(\ref{eqKtil}) with $k$ replaced by $q$.
In the limit $Q^2=0$ 
(initial and final state in the $\Delta$ rest frame)
$b(\tilde k, \tilde q)$ is reduced to ${\tilde k^2}Y_{20}(\hat k)$.

At the end, one has
\ba
{\cal J}_{S,D3}^\mu&=&
\bar w_\alpha 
\left[
\frac{1+2\tau}{1+\tau} \frac{q^\alpha q^\beta}{Q^2} \gamma^\mu
+ g^{\alpha \beta} \gamma^\mu  \right. \\
& &
\left.
- \frac{1}{1+\tau} \frac{q^\alpha}{M_\Delta}
g^{\beta \mu} 
+ 2 \frac{1}{1+ \tau} \frac{q^\alpha q^\beta}{Q^2}
\frac{P_-^\mu}{M_\Delta} \right] w_\beta \times {\cal I}_{D3}^+, 
\nonumber \\
{\cal R}_{S,D3}&=&
\bar w_\alpha 
\left[
\frac{3+2\tau}{1+\tau} \frac{q^\alpha q^\beta}{Q^2} 
+ g^{\alpha \beta} \right]w_\beta  \times {\cal I}_{D3}^+.
\ea
The common integral is defined by
\be
{\cal I}_{D3}^+= 
\int_k b(\tilde k_+,\tilde q_+) \psi_{D3}^+\psi_S^-.
\ee

\subsubsection{Transition D3 $\to$ S}
\label{apIntD3}

The transition between an initial $\Delta$ D3-state 
(momentum $P_-$) and a final $\Delta$ S-state ($P_+$)
can be written as 
\be
J^\mu_{D3,S} =
3 \sum_{\lambda} \int_k 
\bar \Psi_{S} (P_+,k) j_I^\mu \Psi_{D3} (P_-,k).
\label{eqJD3S}
\ee
Once again, we suppress the spin and polarization indices
for simplicity.
We can write (\ref{eqJD3S}) as
\be
J^\mu_{D3,S} = 3
\int_k
\left\{ 
\left[
\bar w_{\alpha} A^\mu W_{D3\, \beta}\right] 
\psi_{S}^+ \psi_{D3}^-
\right\}
\Delta^{\alpha \beta},
\ee
where $\psi_{D3}^- = \psi_{D3}(P_-,k)$,
$\psi_{S}^+ = \psi_{S}(P_+,k)$.
The operator $A^\mu$ is given by Eq.~(\ref{eqA}).

Using again the Gordon decomposition (\ref{eqGordon}),
we obtain
\be
J^\mu_{D3,S} = \tilde f_\Delta {\cal J}_{D3,S}^\mu
- \tilde \kappa_\Delta 
\frac{(P_+ +P_-)^\mu}{2 M_\Delta} {\cal R}_{D3,S},
\label{eqJD3Sb}
\ee
where $\tilde f_\Delta$ is defined by Eq.~(\ref{eqfDelta}), and 
\ba
& &
{\cal J}_{D3,S}^\mu
=3
\int_k
\left\{ 
\left[
\bar w_{\alpha} \gamma^\mu W_{D3\, \beta}\right] 
\psi_{D3}^- \psi_S^+
\right\}
\Delta^{\alpha \beta} \\
& & 
{\cal R}_{D3,S}= 3
\int_k
\left\{ 
\left[
\bar w_{\alpha} W_{D3\, \beta} \right] 
\psi_{D3}^+ \psi_S^- \right\}
 \Delta^{\alpha \beta}.
\ea

To reduce the current (\ref{eqJD3Sb}) to a form
close to the standard one (\ref{eqJgen}), we work 
the spin algebra as before.
We use again the covariant result 
of  the integration in the azimuthal variable $\varphi$ 
from Eq.~(\ref{eqIntPhi}) with the obvious 
replacement
of $P_+ \to P_-$
(which implies $\tilde k_+ \to \tilde k_-$, 
$\tilde q_+ \to \tilde q_-$ and 
$\tilde g_+^{\alpha \beta} \to \tilde g_-^{\alpha \beta}$).
One obtains
\ba
{\cal J}_{D3,S}^\mu&=&
\bar w_\alpha 
\left[
\frac{1+2\tau}{1+\tau} \frac{q^\alpha q^\beta}{Q^2} \gamma^\mu
+ g^{\alpha \beta} \gamma^\mu  \right. \\
& &
\left.
+ \frac{1}{1+\tau} g^{\alpha \mu} \frac{q^\beta}{M_\Delta}
+ 2 \frac{1}{1+ \tau} \frac{q^\alpha q^\beta}{Q^2}
\frac{P_+^\mu}{M_\Delta} \right] w_\beta \times {\cal I}_{D3}^-, 
\nonumber \\
{\cal R}_{D3,S}&=&
\bar w_\alpha 
\left[
\frac{3+2\tau}{1+\tau} \frac{q^\alpha q^\beta}{Q^2} 
+ g^{\alpha \beta} \right]w_\beta \times {\cal I}_{D3}^-,
\ea
where
\be
{\cal I}_{D3}^-= 
\int_k b(\tilde k_-,\tilde q_-) \psi_{D3}^-\psi_S^+.
\ee

One can prove now that:
\be
{\cal I}_{D3}^+ = {\cal I}_{D3}^- \equiv {\cal I}_{D3}.
\ee
One starts with the two integrals with explicit 
arguments written in the Breit frame.
Then we conclude that the integrand of ${\cal I}_{D3}^+$ 
is a function of $z$ and the integrand of ${\cal I}_{D3}^-$ is
a function of $-z$.
Replacing $z \to -z$ we obtain identity in the 
Breit frame.
As the integrals are covariant, 
the result holds for any frame.

\subsubsection{Final result}

Adding the two currents
\be
J_{D3}^\mu =  J_{S,D3}^\mu + J_{D3,S}^\mu, 
\ee
we define two components ($a$ and $b$)
\be
J_{D3}^\mu = J_a^\mu + J_b^\mu,
\ee
where
\ba
J_a^\mu &=&  \tilde f_\Delta {\cal J}_{D3}^\mu 
\label{eqJa}\\
J_b^\mu &=&  - 
\tilde \kappa_\Delta {\cal R}_{D3} \frac{(P_++P_-)^\mu}{2M_\Delta},
\label{eqJb}
\ea
with
\ba
{\cal J}^\mu_{D3}&=&  
{\cal I}_{D3} 
\left[
\bar w_\alpha 
\left\{
2 \frac{1 + 2 \tau}{1+ \tau}
\frac{q^\alpha q^\beta}{Q^2} \gamma^\mu 
 \right. \right. \nonumber \\
& &
\left. \left.
- 2 g^{\alpha \beta} \gamma^\mu
- \frac{1}{1+ \tau} \frac{1}{M_\Delta}
\left(  q^\alpha  g^{\beta \mu} -q^\beta g^{\beta \mu} \right)
\right. \right. \nonumber \\
& &
\left. \left.
+4 \frac{1}{1+\tau}
 \frac{q^\alpha q^\beta}{Q^2} 
\frac{(P_+ + P_-)^\mu}{2M_\Delta}
\right\} w_\beta
\right], \\
{\cal R}_{D3}&=&
2 {\cal I}_{D3} 
\left[
\bar w_\alpha 
\left\{
\frac{3 + 2 \tau}{1+\tau} 
\frac{q^\alpha q^\beta}{Q^2}
+ g^{\alpha \beta}
\right\}
w_\beta
\right].
\label{eqRD3}
\ea
Considering the Fearing relation (\ref{eqFearing})
and the Gordon decomposition (\ref{eqGordon}) 
we can finally write 
\ba
{\cal J}^\mu_{D3}&=&  
\frac{2}{1+\tau}{\cal I}_{D3} \times \nonumber \\
& &
\bar w_\alpha 
\left\{
\left[
g^{\alpha \beta}+ \frac{3}{\tau}
\frac{q^\alpha q^\beta}{4 M_\Delta^2} \right] 
\gamma^\mu  \right. \nonumber \\
& &
+ 
\left. 
\left[
- g^{\alpha \beta}
- \frac{2}{\tau} 
\frac{q^\alpha q^\beta}{4 M_\Delta^2}
\right] \frac{i \sigma^{\mu \nu}q_\nu}{2M_\Delta} 
\right\} w_\beta.
\ea
From the previous relation and
the definition of the form factors $F_i^\ast$, we 
conclude that the form factors associated 
with $J_a^\mu$, from Eq.~(\ref{eqJa}), are
\ba
& &
F_1^{\ast a}(Q^2)= -
 \frac{2}{1+\tau} \tilde f_\Delta {\cal I}_{D3}  \\
& &
F_2^{\ast a}(Q^2)=
  \frac{2}{1+\tau} \tilde f_\Delta {\cal I}_{D3} \\
& &
F_3^{\ast a}(Q^2)= -  
\frac{6}{1+ \tau} \tilde f_\Delta \frac{{\cal I}_{D3}}{\tau}  \\
& &
F_4^{\ast a}(Q^2)= 
\frac{4}{1+ \tau} \tilde f_\Delta \frac{{\cal I}_{D3}}{\tau}.
\ea
To evaluate the multipole form factors we consider 
the transformations (\ref{eqGE0})-(\ref{eqGM3}).
First we note that $F_2^{\ast a} = -F_1^{\ast a}$
and $F_4^{\ast a}= - \sfrac{3}{2}  F_3^{\ast a}$. 
Then
\ba
& &
F_1^{\ast a}+ F_2^{\ast a}= 0 \nonumber \\
& &
F_1^{\ast a}-\tau F_2^{\ast a}= 
- 2 \tilde f_\Delta {\cal I}_{D3}  \nonumber \\
& &
F_3^{\ast a}+ F_4^{\ast a}= - 
\frac{2}{1+\tau}
 \tilde f_\Delta \frac{{\cal I}_{D3}}{\tau}
\nonumber \\
& &
F_3^{\ast a}-\tau F_4^{\ast a}= 
- \frac{2}{1+\tau} \left(3+ 2 \tau \right)
 \tilde f_\Delta \frac{{\cal I}_{D3}}{\tau}.
\nonumber
\ea 
It follows
\ba
G_{E0}^a&\equiv& 0  \\
G_{M1}^a&=& \frac{4}{5} \tilde f_\Delta {\cal I}_{D3} \\
G_{E2}^a&=& 3 \tilde f_\Delta \frac{{\cal I}_{D3}}{\tau} \\
G_{M3}^a&=& \tilde f_\Delta \frac{{\cal I}_{D3}}{\tau}. 
\ea

Take now the current $J_b^\mu$ from Eq.~(\ref{eqJb}).
Using the Gordon decomposition 
(\ref{eqGordon}) and the form of ${\cal R}_{D3}$ 
from Eq.~(\ref{eqRD3}) one concludes that
\ba
& &
F_1^{\ast b}(Q^2)=
2  \kappa_\Delta {\cal I}_{D3} \\
& &
F_2^{\ast b}(Q^2)= -F_1^{\ast b} (Q^2)\\
& &
F_3^{\ast b}(Q^2)=
 2 \frac{3+ 2\tau}{1+ \tau}
  \tilde \kappa_\Delta 
\frac{{\cal I}_{D3}}{\tau} \\
& &
F_4^{\ast b}(Q^2)=-  F_3^{\ast b}(Q^2).
\ea
From the previous equations it is easy to conclude that
\ba
& &
F_1^{\ast b}+ F_2^{\ast b}=0, \hspace{.2cm}
F_3^{\ast b}+ F_4^{\ast b}=0 \nonumber \\
& &
F_1^{\ast b}-\tau F_2^{\ast b}= 
2 (1+\tau) 
\tilde \kappa_\Delta 
{\cal I}_{D3}
\nonumber \\
& &
F_3^{\ast b}-\tau F_4^{\ast b}=
2 (3 + 2\tau) 
 \tilde \kappa_\Delta 
{\cal I}_{D3}. 
\nonumber
\ea
And then,
\ba
& &G_{E0}^b \equiv 0 \\
& &G_{M1}^b \equiv 0 \\
& &G_{E2}^b = -3  
(1+\tau) 
\tilde \kappa_\Delta \frac{{\cal I}_{D3}}{\tau} 
\\
& &G_{M3}^b \equiv 0.
\ea
The magnetic contributions $b$ vanishes for $Q^2=0$.

Finally we take the sum of the $a$ and $b$ components. 
Using the definition for $\tilde f_\Delta$, we are left with
\ba
G_{E0}^{D3}(Q^2) &=&  0  \\
G_{M1}^{D3}(Q^2) &=&  \frac{4}{5} \tilde f_\Delta {\cal I}_{D3} \\
G_{E2}^{D3}(Q^2) &=& 
3 \tilde g_\Delta 
\frac{{\cal I}_{D3}}{\tau} \\
G_{M3}^{D3}(Q^2) &=& \tilde f_\Delta \frac{{\cal I}_{D3}}{\tau}.
\ea
Note that the electric form factors include 
the {\it electric factor} $\tilde g_\Delta$ , while
the magnetic form factors include the factor $\tilde f_\Delta$.
To achieve this final form it was essential 
to include
the terms associated with ${\cal R}_{D3}$:
the result of the overlap between the spin functions 
associated with core spins 3/2 (S and D3).
Note that, although ${\cal I}_{D3}$ vanishes 
for $Q^2=0$,
$\sfrac{{\cal I}_{D3}}{\tau} \to \mbox{const}$ 
as $Q^2 \to 0$.

%

%
%

\subsection{Current $J^\mu_{D1}$}
\label{apGD1}

Consider Eqs.~(\ref{eqPsiS}) and (\ref{eqPsiD1}) 
in a compact notation
\ba
& &\Psi_S= 
-\varepsilon_{P}^{\alpha \ast} w_\alpha \psi_S (T \cdot \xi^{1 \ast})
\\
& &
\Psi_{D1}=  
-3 \varepsilon_{P}^{\alpha \ast} W_{D1\, \alpha} \psi_{D1} 
(T \cdot \xi^{1 \ast}).
\ea

The transition between a $\Delta$ S and a $\Delta$ D1 
state can be decomposed into two processes:
\begin{itemize}
\item
S in the initial state (current $J_{S,D1}^\mu$)
\item
D1 in the initial state (current $J_{D1,S}^\mu$).
\end{itemize}

In the next calculations we 
use the properties of the spin S-states
(\ref{eqRSprop}) (S-state) and D1-state $W_{D1\, \beta}$:
\ba
& &
\not \! P_\pm W_{D1\, \beta}(P_\pm) =M_\Delta W_{D1\, \beta}(P_\pm) \\
& &
P_\pm^\beta W_{D1\, \beta}(P_\pm)=0. 
\label{eqW1prop}
\ea
See Ref.~\cite{NDeltaD} for details.

\subsubsection{Transition S $\to$ D1}

The transition between an initial $\Delta$ S-state 
(momentum $P_-$) and a final $\Delta$ D1-state ($P_+$)
can be written as 
\be
J^\mu_{S,D1} =
3 \sum_{\lambda} \int_k 
\bar \Psi_{D1} (P_+,k) j_I^\mu \Psi_S (P_-,k).
\label{eqJSD1}
\ee
We suppress the spin and polarization indicies for 
simplicity.
We can write (\ref{eqJSD1}) as
\be
J^\mu_{S,D1} =
3 \int_k
\left\{ 
\left[
\bar W_{D1\, \alpha} A^\mu w_\beta\right] 
\psi_{D1}^+ \psi_S^-
\right\}
\Delta^{\alpha \beta},
\label{eqJSD1b}
\ee
where $\psi_{D1}^+ = \psi_{D1}(P_+,k)$,
$\psi_{S}^- = \psi_{D1}(P_-,k)$  and 
$A^\mu$ is given by Eq.~(\ref{eqA}).

The Gordon decomposition (\ref{eqGordon}) of
the current (\ref{eqJSD1b}) gives:
\ba
J^\mu_{S,D1} &=&
3 \tilde f_\Delta
\int_k 
\left\{ 
\left[
\bar W_{D1\, \alpha} \gamma^\mu w_\beta\right] 
\psi_{D1}^+ \psi_S^-
\right\}
\Delta^{\alpha \beta} \nonumber \\
& & 
- 3 \tilde f_\Delta {\cal R} \frac{(P_+ + P_-)^\mu}{2M_\Delta},
\ea
where $\tilde f_\Delta$ is defined by Eq.~(\ref{eqfDelta}), and
\be
{\cal R}= 
\int_k 
\left\{ 
\left[
\bar W_{D1\, \alpha} w_\beta\right] 
\psi_{D1}^+ \psi_S^-
\right\}
\Delta^{\alpha \beta}.
\ee
The orthogonality between the states $W_{D1 \, \alpha}$ and $w_\beta$ gives
\be
{\cal R} \equiv 0,
\ee
equivalently to 
\be
3 \sum_\lambda \int_k \bar \Psi_{D1}(P_+,k) \Psi_S(P_-,k) \equiv 0.
\ee

By working the spin algebra and 
using the covariant integration in the azimuthal variable $\varphi$ 
\cite{NDeltaD}:
\be
\int_k {\cal D}^{\alpha \beta} 
\psi_{D1}^+\psi_S^- 
=
R^{\alpha \beta}(q,P_+)
\int_k 
b(\tilde k_+, \tilde q_+) \psi_{D1}^+\psi_S^-,
\ee
with $b$ and $R$ defined respectively by 
(\ref{eqBtilde}) and (\ref{eqRtilde}).
At the end, we obtain
\ba
& &J^\mu_{S,D1}=
\frac{1}{1+ \tau} \tilde f_\Delta {\cal I}_{D1}^+
\label{eqJD1a}\\
& &
\times
\left[
\bar w_\alpha 
\left\{
\frac{q^\alpha}{M_\Delta} g^{\beta \mu} 
+ 2 \frac{q^\alpha q^\beta}{Q^2}  \gamma^\mu 
- 2 \frac{q^\alpha q^\beta}{Q^2}  \frac{P_-^\mu}{M_\Delta}
\right\} w_\beta
\right]. \nonumber 
\ea
where 
\be
{\cal I}_{D1}^+= 
\int_k b(\tilde k_+,\tilde q_+) \psi_{D1}^+\psi_S^-.
\ee

\subsubsection{Transition D1 $\to$ S}

We consider now the transition from D1 to S:
\be
J^\mu_{D1,S} =
3 \sum_{\lambda} \int_k 
\bar \Psi_{S} (P_+,k) j_I^\mu \Psi_{D1} (P_-,k).
\label{eqJD1S}
\ee
We conclude that
\be
J^\mu_{D1,S} =
3 \tilde f_\Delta
\sum_{\lambda} \int_k 
\left\{ 
\left[
\bar w_{\alpha} \gamma^\mu W_{D1\, \beta}\right] 
\psi_{D1}^- \psi_S^+
\right\}
\Delta^{\alpha \beta}.
\ee
where the orthogonality was used again.

Working through the spin algebra, 

\be
\int_k {\cal D}^{\alpha \beta} 
\psi_{D1}^-\psi_S^+ 
=
R^{\alpha \beta}(q,P_-)
\int_k 
b(\tilde k_-, \tilde q_-) \psi_{D1}^+\psi_S^-,
\ee
we are left with
\ba
& &J^\mu_{D1,S}=
\frac{1}{1+ \tau} \tilde f_\Delta {\cal I}_{D1}^-
\label{eqJD1b}\\
& &
\times
\left[
\bar w_\alpha 
\left\{
-g^{\alpha \mu} \frac{q^\beta}{M_\Delta} 
+ 2 \frac{q^\alpha q^\beta}{Q^2}  \gamma^\mu 
- 2 \frac{q^\alpha q^\beta}{Q^2}  \frac{P_+^\mu}{M_\Delta}
\right\} w_\beta
\right], \nonumber 
\ea
where 
\be
{\cal I}_{D1}^-= 
\int_k b(\tilde k_-,\tilde q_-) \psi_{D1}^-\psi_S^+.
\ee

Similarly to Appendix \ref{apIntD3}, we can prove that
\be
{\cal I}_{D1}^+={\cal I}_{D1}^- \equiv  {\cal I}_{D1}.
\ee

\subsubsection{Final result}

Adding the two currents (\ref{eqJD1a}) and (\ref{eqJD1b})
\be
J_{D1}^\mu = J_{S,D1}^\mu + J_{D1,S}^\mu,
\ee
we obtain
\ba
J^\mu_{D1} &=&
{\tilde f_\Delta} 
\frac{1}{1+ \tau} 
{\cal I}_{D1}
\\
& &
\times
\bar w_\alpha 
\left\{
\frac{1}{M_\Delta} 
\left(q^\alpha g^{\alpha \mu} -q^\beta  g^{\alpha \mu} \right)
\right.
\nonumber \\
& & 
\left.
+ 4 \frac{q^\alpha q^\beta}{Q^2} \gamma^\mu
-  4 \frac{q^\alpha q^\beta}{Q^2} \frac{(P_++P_-)^\mu}{2M_\Delta}
\right\} w_\beta
\nonumber 
\ea
Using the Fearing-Nozawa relation (\ref{eqFearing})
and the Gordon decomposition (\ref{eqGordon}), we are left with
\ba
J_{D1}^\mu &=&
2 \tilde f_\Delta \frac{1}{1+ \tau} 
{\cal I}_{D1} \times \\
& &
\bar w_\alpha 
\left\{
\left[
\tau g^{\alpha \beta} + 2 
\frac{q^\alpha q^\beta}{4M_\Delta^2} \right] \gamma^\mu 
\right. \nonumber \\
& &
+ 
\left.
\left[
g^{\alpha \beta}
+ \frac{2}{\tau} 
\frac{q^\alpha q^\beta}{4M_\Delta^2} 
\right] 
\frac{i \sigma^{\mu \nu}q_\nu}{2 M_\Delta} 
\right\} w_\beta \nonumber
\ea
Comparing the previous equation with 
Eq.~(\ref{eqJen}), we conclude that
\ba
F_1^\ast &=& 
- \frac{2}{1+\tau}
\tilde f_\Delta  {\cal I}_{D1} \tau \\
F_2^\ast &=&  - \frac{2}{1+\tau} \tilde f_\Delta {\cal I}_{D1} \\
F_3^\ast &=& 
- 2 \frac{2}{1+\tau} \tilde f_\Delta {\cal I}_{D1} \\
F_4^\ast &=& - 
 2 \frac{2}{1+\tau} \tilde f_\Delta 
\frac{ {\cal I}_{D1}}{\tau}.
\ea

Note that, although $F_4^\ast$ 
includes a factor $1/\tau=4 M_\Delta^2/Q^2$, it is not singular.
By definition of  $b(\tilde q,\tilde k)$ 
the integral ${\cal I}_{D1}$ vanishes 
as $Q^2 \to 0$ canceling the divergence in $1/\tau$.
The condition ${\cal I}_{D1} =0$ when $Q^2=0$ 
is the result of the orthogonality 
between the states $L=0$ and $L=2$.

Using the simple relations
\ba
& &
F_1^\ast(Q^2)= \tau F_2^\ast(Q^2)  \\
& &
F_3^\ast(Q^2)= \tau F_4^\ast (Q^2), 
\ea
we can write
\ba
& &
F_1^\ast (Q^2) + F_2^\ast(Q^2)= -2 \tilde f_\Delta {\cal I}_{D1}
\nonumber \\
& &
F_1^\ast (Q^2)- \tau F_2^\ast(Q^2)=0 \nonumber \\
& &
F_3^\ast (Q^2) + F_4^\ast(Q^2)= -4
\tilde f_\Delta \frac{{\cal I}_{D1}}{\tau} \nonumber \\
& &
F_3^\ast (Q^2)- \tau F_4^\ast(Q^2)=0.
\nonumber  
\ea

Finally
\ba
& & 
G_{E0}^{D1}(Q^2) \equiv 0 \\
& & 
G_{M1}^{D1}(Q^2)= -
 \frac{2}{5} 
\tilde f_\Delta {\cal I}_{D1} \\
& & 
G_{E2}^{D1}(Q^2) \equiv 0 \\
& &
G_{M3}^{D1}(Q^2)=  2 \tilde f_\Delta \frac{{\cal I}_{D1}}{\tau}. 
\ea
In the limit $Q^2=0$ only $G_{M3}$ is different from zero.

\subsection{All contributions}

Each of the four  $\Delta$ form factors $G_\alpha$ 
($\alpha=E0, M1, E2, M3$) is 
then the result of the three contributions calculated above:   
\be
G_\alpha(Q^2)= N^2 \left[
G_\alpha^S(Q^2) + a  G_\alpha^{D3}(Q^2) + b 
G_\alpha^{D1}(Q^2) \right].
\ee

The final result becomes
\ba
G_{E0}(Q^2) &=&  N^2 \tilde g_\Delta {\cal I}_S  
\nonumber  \\
G_{M1}(Q^2) &=&  
N^2 \tilde f_\Delta {\cal I}_S 
+\frac{4}{5} (aN^2) \tilde f_\Delta {\cal I}_{D3}
-\frac{2}{5} (bN^2) \tilde f_\Delta {\cal I}_{D1}
 \nonumber \\
G_{E2}(Q^2) &=& 
3 (aN^2) \tilde g_\Delta  \frac{{\cal I}_{D3}}{\tau} \nonumber \\
G_{M3}(Q^2) &=& 
\tilde f_\Delta N^2\left[ 
a \frac{{\cal I}_{D3}}{\tau} +
2  b \frac{{\cal I}_{D1}}{\tau} \right].
\label{eqGM3b}
\ea

\section{Proof that ${\cal I}_D \sim Q^2$}
\label{apIQ2}

Consider the overlap integral
\be
{\cal I}_D =\int_k b(\tilde k_+,\tilde q_+) \psi_D(P_+,k) \psi_S(P_-,k),
\ee
where $D$ represents an arbitrary (D3 or D1) D-state.
The functions $\psi_D$ and $\psi_S$ 
are respectively the D-state and the S-state scalar wave functions
written in terms of the dimensionless variable $\chi$, as defined in Eq.~(\ref{eqChi}). 
That is the only assumption made in this appendix.
The function $b(\tilde k_+,\tilde q_+)$ is defined by Eq.~(\ref{eqB}).

Although the integral ${\cal I}_D$ is covariant the analysis 
can be considerably simplified in a particular frame.
We consider then the rest frame of the final baryon 
(mass $M$):
\ba
& &P_+=(M,0,0,0) \nonumber \\
& &P_-=(E,0,0,-|{\bf q}|) \nonumber \\
& &q=(\omega,0,0,|{\bf q}|),
\ea
where
\ba
& &
E= \sqrt{M^2+|{\bf q}|^2}= \frac{2M^2+Q^2}{2M} \nonumber \\
& &
|{\bf q}|=\frac{\sqrt{(4M^2+Q^2)Q^2}}{2M}= \frac{EQ}{M}
\nonumber  \\
& &
\omega= \frac{P_+ \cdot q}{M}= -\frac{Q^2}{2M}.
\ea
In this frame we can write 
$\tilde k_+= (0,k_x,k_y,k_z)$ and $\tilde q_+=(0,0,0,|{\bf q}|)$.
In the final baryon rest frame $\tilde k_+$ and $\tilde q_+$
has only spacial components.

It is now trivial to conclude 
($\tilde k_+ \cdot \tilde q_+=-|{\bf q}| k_z$) that
\ba
b(\tilde k_+,\tilde q_+) =
-\frac{1}{2}( 3 k_z^2- {\bf k}^2 ) = - {\bf k}^2 Y_{20}(z),
\ea
where we used $k_z= |{\bf k}| z$, with $z=\cos \theta$.
In the final baryon rest frame
there is no $Q^2$ dependence in $b$, and only the scalar wave functions
depend on $Q^2$.

The momentum dependence of the scalar wave 
appears through the functions $P_\pm \cdot k$,
which become in the final baryon rest frame:
\ba
& &
P_+ \cdot k = M E_s \\
& &
P_- \cdot k = E E_s + |{\bf q}| k_z = E E_s + \frac{E Q}{M} \mbox{k} z.
\ea
In the last equation we used $\mbox{k} =|{\bf k}|$.
The previous equations show that the angular dependence 
appears only in the S-state wave function.
To perform the angular  integration (variable $z$) for 
${\cal I}_D$ we need only to consider the factor 
\be
I= \int dz Y_{20}(z) \psi_S(P_-,k),
\label{eqIz}
\ee
where it is implicit that the integration is in the interval $[-1,1]$.

To proceed, we need to use the momentum dependence of $\psi_S$.
By taking the generic form
\be
\psi_S (P_-,k) = \frac{N_S}{D^n},
\label{eqPsiSp}
\ee
where $n$ is a integer ($n \ge 2$),
$N_S$ is some normalization constant, and
\ba
D= \alpha  + 2 \frac{P_- \cdot k}{M m_s} -2 
 = \beta + 2  \frac{E Q}{M^2} \mbox{k} z,
\ea
where $\beta=  (\alpha -2) +  2 \frac{E E_s}{M m_s}$.

Factorizing $D$ according with 
\be
D= \beta \left[
1 + 2 \frac{EQ}{M^2} \frac{\mbox{k}}{\beta} z
\right],
\ee
we can write
\ba
& &
\hspace{-1.2cm}
\frac{1}{D^n}=  
\frac{1}{\beta^n} \times \nonumber \\
& &
\hspace{-1.2cm}
\left[
a_0 + a_1 \left(\frac{EQ}{M^2} \frac{\mbox{k}}{\beta} \right) z +
a_2 \left(\frac{EQ}{M^2} \frac{\mbox{k}}{\beta} \right)^2 z^2 + ...
\right], 
\ea
where 
\be
a_j= 2^j \left(\begin{array}{c} -n  \cr j \cr \end{array}\right)= 2^j 
\frac{(-n)(-n-1)..(-n-j+1)}{j(j-1)..1},
\ee
with $j=0,1,...$ 
For $n=3$, for example: $a_0=1$, $a_1=-6$, $a_2=24$, etc.

The angular integration (\ref{eqIz}) is now 
considerably simplified
\ba
& &
\int dz 
\left\{ {\kern+4pt}
Y_{20}(z) \times \frac{}{} \right. \nonumber \\ 
& &
\left[
a_0 + a_1 \left(\frac{EQ}{M^2} \frac{\mbox{k}}{\beta} \right) z +
a_2 \left(\frac{EQ}{M^2} \frac{\mbox{k}}{\beta} \right)^2 z^2 + ...
\right]
\left. \frac{}{}
\right\}  \nonumber \\
& &= a_2 
\left(\frac{EQ}{M^2} \right)^2
\left[\int dz z^2Y_{20}(z) \right]  + {\cal O}(E^4 Q^4),
\ea
where the integral of the first term 
is zero $[\int Y_{20}(z)=0]$, the second term vanishes 
because is the integral of an odd function (in a symmetric interval),
and the same argument holds for the forth them (coefficient $a_3$).
The next nonzero contribution appears only with the fifth term 
(coefficient $a_4$). 
As $\int dz \left\{z^2Y_{20}(z)\right\}= \sfrac{4}{15}$, the previous result 
can be written as
\be
I=\frac{4}{15} \frac{a_2}{\beta^n}   
\left(\frac{E}{M^2} \frac{\mbox{k}}{\beta} \right)^2 Q^2 +
{\cal O}(E^4 Q^4).
\ee
We still have to perform
the integration in k. 
However, because 
$\sfrac{\mbox{k}}{\beta} 
\simeq \sfrac{1}{2} \sfrac{M}{E} m_s=\mbox{const}$ as 
k$\to \infty$, all integrals are well defined
providing that the original integral is well defined.
This means that integration in k preserves 
the decomposition presented above.

For small $Q^2$ 
we need only to consider the leading dependence in $Q^2$ 
of each term.
Using $\beta_0 = (\alpha-2) +  2 \frac{E_s}{m_s}$,  
one has
\ba
I&=& \frac{4}{15}
\frac{a_2}{\beta_0^n}   
\left(\frac{\mbox{k}}{M \beta_0} \right)^2 Q^2 +
{\cal O}(Q^4) \sim Q^2 
\ea
Therefore
\be
{\cal I}_D \sim Q^2,
\ee
which finishes the proof.

Note that this derivation is independent of $n$.
The only constraint to be imposed is that 
the power in $\psi_S$ is sufficient to assure 
the convergence of the integral (in k). 
An alternative derivation is possible 
in a different frame,
using also the properties of $\psi_D$.
For instance, in the initial baryon rest frame
all angular dependence is in $\psi_D$.
Considering  $\psi_D$ with the same form of $\psi_S$ from 
Eq.~(\ref{eqPsiSp}), we obtain the same limit for 
small $Q^2$ using the procedure described before, 
as it should be.

\end{document}